\documentclass[superscriptaddress,amsmath,amssymb, aps, prl,longbibliography,twocolumn, footinbib]{revtex4-2}

\usepackage{upgreek}
\usepackage{color}
\usepackage{graphicx}
\usepackage{nccmath}
\usepackage{bm}
\usepackage{hyperref}
\hypersetup{colorlinks,breaklinks,
            urlcolor=[rgb]{0,0,0.64},
            linkcolor=[rgb]{0,0,0.64},
            citecolor=[rgb]{0,0,0.64},
            filecolor=[rgb]{0,0,0.64}}
\usepackage{dcolumn}

\newcommand{\beginsupplement} {
    \setcounter{table}{0}
    \renewcommand{\thetable}{S\arabic{table}}
    \setcounter{figure}{0}
    \renewcommand{\thefigure}{S\arabic{figure}}
    \setcounter{equation}{0}
    \renewcommand{\theequation}{S\arabic{equation}}
}

\newcommand{\MIT}{Massachusetts Institute of Technology, Department of Physics, Cambridge, Massachusetts 02139, USA.}
\newcommand{\Cal}{University of California at Berkeley, Department of Chemistry, Berkeley, California 94720, USA.}
\newcommand{\UCLA}{University of California at Los Angeles, Department of Physics and Astronomy, Los Angeles, California 90095, USA.}
\newcommand{\Geballe}{Geballe Laboratory for Advanced Materials, Stanford University, Stanford, California 94305, USA.}
\newcommand{\StanfordAP}{Department of Applied Physics, Stanford University, Stanford, California 94305, USA.}
\newcommand{\StanfordMSE}{Department of Materials Science and Engineering, Stanford University, Stanford, California 94305, USA.}
\newcommand{\SIMES}{SIMES, SLAC National Accelerator Laboratory, Menlo Park, California 94025, USA.}
\newcommand{\SLAC}{SLAC National Accelerator Laboratory, Menlo Park, California 94025, USA.}
\newcommand{\Harvard}{Department of Physics, Harvard University, Cambridge, Massachusetts 02138, USA.}
\newcommand{\ETH}{Institute for Theoretical Physics, ETH Zurich, 8093 Zurich, Switzerland.}

\begin{document}

\title{Role of equilibrium fluctuations in light-induced order}

\author{Alfred~Zong}
\thanks{These authors contributed equally to this work.}
\affiliation{\MIT}
\affiliation{\Cal}
\author{Pavel~E.~Dolgirev}
\thanks{These authors contributed equally to this work.}
\affiliation{\Harvard}
\author{Anshul~Kogar}
\thanks{These authors contributed equally to this work.}
\affiliation{\MIT}
\affiliation{\UCLA}
\author{Yifan~Su}
\affiliation{\MIT}
\author{Xiaozhe~Shen}
\affiliation{\SLAC}
\author{Joshua~A.~W.~Straquadine}
\affiliation{\StanfordAP}
\affiliation{\SIMES}
\affiliation{\Geballe}
\author{Xirui~Wang}
\affiliation{\MIT}
\author{Duan~Luo}
\affiliation{\SLAC}
\author{Michael~E.~Kozina}
\affiliation{\SLAC}
\author{Alexander~H.~Reid}
\affiliation{\SLAC}
\author{Renkai~Li}
\affiliation{\SLAC}
\author{Jie~Yang}
\affiliation{\SLAC}
\author{Stephen~P.~Weathersby}
\affiliation{\SLAC}
\author{Suji~Park}
\affiliation{\SLAC}
\affiliation{\StanfordMSE}
\author{Edbert~J.~Sie}
\affiliation{\SIMES}
\affiliation{\Geballe}
\author{Pablo~Jarillo-Herrero}
\affiliation{\MIT}
\author{Ian~R.~Fisher}
\affiliation{\StanfordAP}
\affiliation{\SIMES}
\affiliation{\Geballe}
\author{Xijie~Wang}
\affiliation{\SLAC}
\author{Eugene~Demler}
\affiliation{\Harvard}
\affiliation{\ETH}
\author{Nuh~Gedik}\email[Correspondence to: ]{gedik@mit.edu}
\affiliation{\MIT}

\date{September 9, 2021}

\begin{abstract}
Engineering novel states of matter with light is at the forefront of materials research. An intensely studied direction is to realize broken-symmetry phases that are ``hidden'' under equilibrium conditions but can be unleashed by an ultrashort laser pulse. Despite a plethora of experimental discoveries, the nature of these orders and how they transiently appear remain unclear. To this end, we investigate a nonequilibrium charge density wave (CDW) in rare-earth tritellurides, which is suppressed in equilibrium but emerges after photoexcitation. Using a pump-pump-probe protocol implemented in ultrafast electron diffraction, we demonstrate that the light-induced CDW consists solely of order parameter fluctuations, which bear striking similarities to critical fluctuations in equilibrium despite differences in the length scale. By calculating the dynamics of CDW fluctuations in a nonperturbative model, we further show that the strength of the light-induced order is governed by the amplitude of equilibrium fluctuations. These findings highlight photoinduced fluctuations as an important ingredient for the emergence of transient orders out of equilibrium. Our results further suggest that materials with strong fluctuations in equilibrium are promising platforms to host ``hidden'' orders after laser excitation.
\end{abstract}

\maketitle

In a symmetry-breaking phase transition, fluctuations of the order parameter provide important information about the way an ordered state develops. Near the transition temperature $T_c$, fluctuations exhibit a diverging correlation length and correlation time, whose critical exponents define the underlying universality class. In contrast to the equilibrium situation, the role of order parameter fluctuations remains unclear if a phase transition proceeds under nonequilibrium conditions. Of particular interest are transitions instigated by an intense laser pulse, which has led to discoveries of many ``hidden'' orders that are not accessible in thermal equilibrium, such as light-induced superconductivity \cite{Kaiser2017,Mitrano2016,Buzzi2020}, charge or spin density waves \cite{Kogar2020,Zhou2021,Han2015,Kim2012}, and ferroelectricity \cite{Li2019,Nova2019}. These out-of-equilibrium orders are often short-lived, raising the question of whether they exist in the form of fluctuations and if so, how they are related to fluctuations in equilibrium. 

Empirically, several material classes that host transient states also display strong equilibrium fluctuations of the associated order \cite{Fausti2011,Kaiser2014,Hu2014,Nicoletti2014,Buzzi2020,Li2019,Nova2019,Borroni2017}. In underdoped cuprates where light-induced superconductivity was discovered \cite{Fausti2011,Kaiser2014,Hu2014,Nicoletti2014}, pronounced superconducting fluctuations are expected due to the small phase stiffness and poor screening \cite{Emery1995}. In $\kappa$-type organic salts where light-induced superconductivity was observed above $T_c$, Nernst effect measurements also pointed towards large fluctuations due to a nearby Mott criticality \cite{Nam2007,Kagawa2009,Buzzi2020}. In cases where equilibrium fluctuations do not yield an ordered state at finite temperature, such as in the quantum paraelectric phase of SrTiO$_3$, a terahertz pulse can induce a ferroelectric state in a metastable fashion \cite{Muller1979,Li2019,Nova2019}. These observations suggest that photoinduced orders may be a special manifestation of equilibrium fluctuations, but experimental evidence is lacking to formally establish a link between the two entities.

\begin{figure}[htb!]
	\centering
	\includegraphics[width=1\columnwidth]{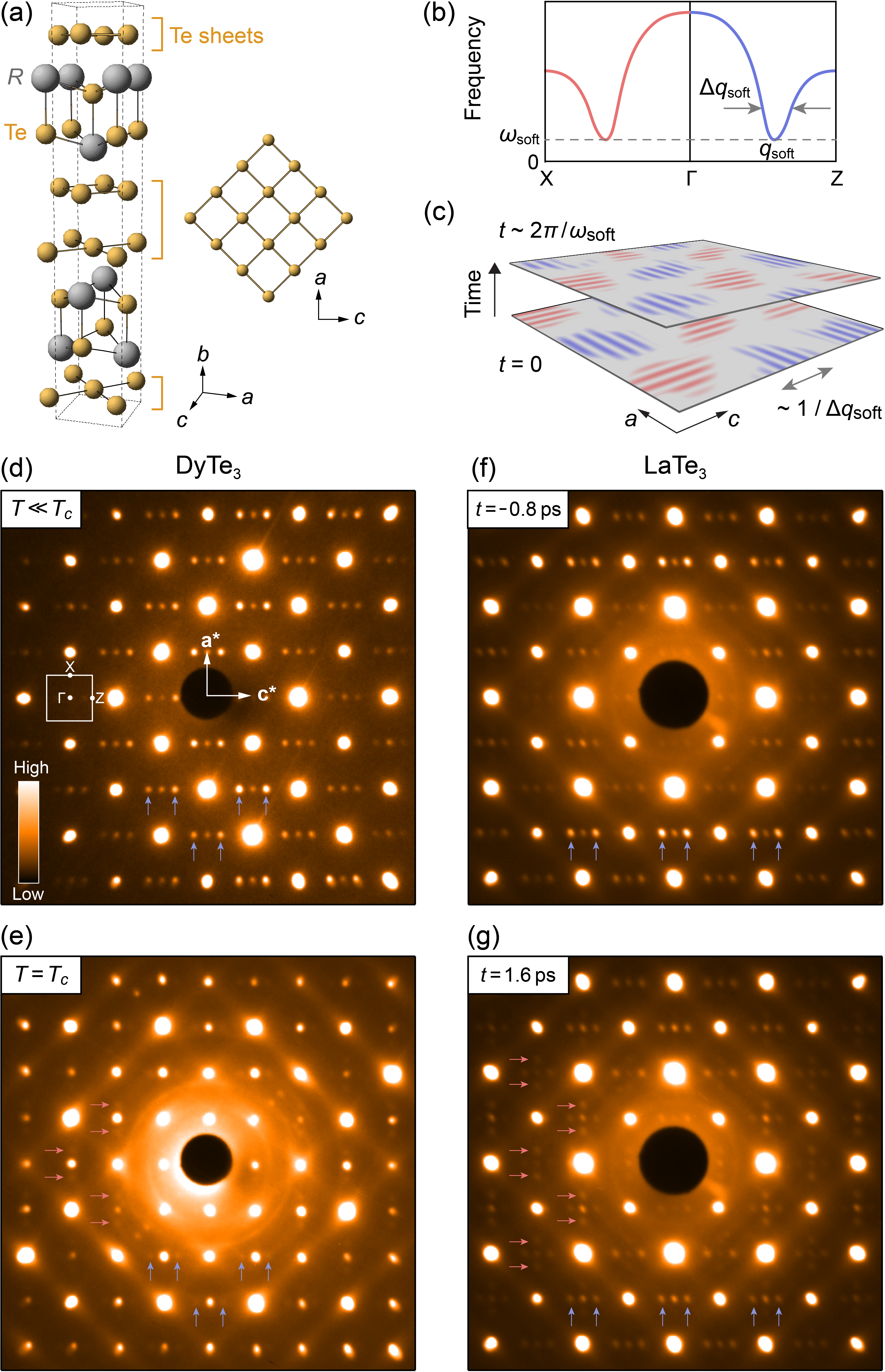}
	\caption{Competing charge density waves in rare-earth tritellurides. (a)~\emph{Left}: schematic of the layered structure of $R$Te$_3$, where dashed lines indicate the primary unit cell. \emph{Right}: Enlarged view of the nearly square-shaped Te sheets that host the CDW instabilities. (b)~Schematic phonon dispersion right above $T_c$ along $\Gamma$-X and $\Gamma$-Z, featuring two Kohn anomalies at $q_\text{soft}$. (c)~Schematic of fluctuating CDWs right above $T_c$. (d)(e)~Equilibrium electron diffractions of DyTe$_3$ ($T_c=306(3)$~K \cite{RuThesis}) taken at 100~K (d) and 307~K (e). (f)(g)~Time-resolved diffractions of LaTe$_3$ before (f) and 1.6~ps after (g) photoexcitation by an 80-fs, 800-nm laser pulse with an incident fluence of 2.1~mJ/cm$^2$, measured at 307~K. Blue and red arrows indicate the CDW peaks along the $c$- and $a$-axis, respectively. Difference in intensities of lattice Bragg peaks in (d) and (e) results from slight sample drift and tilt during the warm-up process.}
    \label{fig:intro}
\end{figure}

Here, through a side-by-side comparison, we show that a newly-discovered photoinduced charge density wave (CDW) \cite{Kogar2020,Zhou2021} shares the key characteristics of the CDW fluctuations at $T_c$ even though the former does not have a diverging correlation length. The comparison was enabled by a pump-pump-probe scheme with ultrafast electron diffraction, which gives a direct measurement of fluctuations through diffuse scatterings. Using a nonperturbative calculation, we further demonstrate that the intensity of the photoinduced CDW peak increases with the strength of the CDW fluctuations in equilibrium. The positive correlation suggests that photoinduced ``hidden'' state is more likely found in systems with significant equilibrium fluctuations, paving the way forward as we search for novel nonequilibrium orders.

The charge density wave is hosted by the rare-earth tritelluride ($R$Te$_3$) family. All members possess a layered structure and the CDW instability is found in the nearly square-shaped Te sheets [Fig.~\ref{fig:intro}(a)]. The quasi-two-dimensional nature of the crystals leads to a much reduced $T_c$ compared to the mean-field transition temperature. This gives rise to significant CDW fluctuations above $T_c$, as evidenced by Raman spectroscopy \cite{Eiter2013} and inelastic X-ray scattering \cite{Maschek2018}. The near-$C_4$ symmetry of the Te sheets leads to two competing CDWs: The dominant one has a modulation along the $c$-axis while the subdominant one along the orthogonal $a$-axis \cite{Ru2008}. Here, we focus on LaTe$_3$ ($T_c\approx670$~K) and DyTe$_3$ ($T_c=306(3)$~K) \cite{Hu2014b,Ru2008}. They share nearly identical properties except for the different transition temperatures \cite{Note1}. Hence, under similar experimental conditions, we have access to CDW fluctuations in the critical regime near $T_c$ (DyTe$_3$) as well as a state with only the dominant $c$-axis CDW (LaTe$_3$).

Figure~\ref{fig:intro}(d)(e) shows the equilibrium electron diffraction patterns of DyTe$_3$ in the $(H,0,L)$ plane, taken below and near $T_c$ (see \cite{Remark1} for experimental details). At 100~K, pairs of CDW satellite peaks are found along the $c$-axis at a wavevector $q_c=0.294(1)c^*$ (blue arrows), but no satellite peaks are observed along the orthogonal $a$-axis \cite{Note2}. When the sample is heated to $T_c$, the $c$-axis peaks significantly weaken but remain visible [Fig.~\ref{fig:intro}(e)]; in the meantime, diffuse spots arise along the $a$-axis (red arrows). Notably, the diffraction pattern appears symmetric between the $c$- and $a$-axis, as highlighted by three observations: (i)~brighter $(H\pm q_a,0,L)$ satellites are found along the $c$-axis than along the $a$-axis; vice versa for the $(H,0,L\pm q_c)$ peaks; (ii)~the CDW wavevectors are similar, $q_a\approx q_c$; (iii)~the satellite intensities are comparable for the two CDWs. Transverse atomic displacements associated with both CDWs account for the intensity pattern in (i) \cite{Remark1}. Observations (ii) and (iii) preclude the possibility of a long-range CDW along the $a$-axis that is known to occur in DyTe$_3$ at 68~K $\ll T_c$ \cite{Maschek2018} because this low-temperature $a$-axis peak has a markedly different wavevector and a much weaker diffraction intensity compared to its $c$-axis counterpart \cite{Straquadine2019,Kogar2020}. The symmetric appearance of the diffuse spots in Fig.~\ref{fig:intro}(e) is a signature unique to the critical regime near $T_c$. Below $T_c$, such symmetry is broken by the long-range $c$-axis CDW. At temperatures significantly exceeding $T_c$, fluctuations are weak, rendering any diffuse scattering invisible under the background intensity.

We now turn to LaTe$_3$ and study the behavior of the CDWs out of equilibrium. Figure~\ref{fig:intro}(f)(g) show the electron diffraction patterns taken 0.8~ps before and 1.6~ps after the incidence of an 80-fs, 800-nm laser pulse. After photoexcitation, the long-range CDW order along the $c$-axis is suppressed (blue arrows) while new peaks appear along the $a$-axis (red arrows), whose intensity increases monotonically with pump laser fluence \cite{Kogar2020,Remark1}. Remarkably, the CDW superlattice spots in this transient snapshot of the photoexcited state are visually indistinguishable from those in the equilibrium diffraction pattern recorded at $T_c$ in DyTe$_3$ [Fig.~\ref{fig:intro}(e)(g)]. In particular, the transient CDW satellites along both axes share a similar intensity and wavevector, hinting at a restored symmetry between the two CDWs.

The similarity between Fig.~\ref{fig:intro}(e) and \ref{fig:intro}(g) allows us to interpret the light-induced CDW state using an equilibrium picture close to $T_c$. In momentum space, the diffuse satellite peaks are indicative of the population of transient soft phonons along the $a^*$- and $c^*$-axis [Fig.~\ref{fig:intro}(b)]. In real space, this critical regime is characterized by short-range CDW patches in both directions [Fig.~\ref{fig:intro}(c)], with the correlation length inversely proportional to the momentum width of the Kohn anomaly \cite{Remark1}. From inelastic X-ray measurements \cite{Maschek2018}, the phonon energies at $q_a$ and $q_c$ are approximately 1 to 2~meV, corresponding to a fluctuating timescale of 2 to 4~ps for these CDW patches. A similar timescale is observed as the lifetime of the light-induced $a$-axis CDW [Fig.~\ref{fig:kink}(a)]. This energy-time correspondence suggests that the light-induced $a$-axis CDW is indistinguishable from a soft phonon at the corresponding wavevector, confirming the intimate link between the photoexcited state and the critical regime near $T_c$.

The comparison between the photoexcited and the critical state suggests that the photoinduced $a$-axis CDW in LaTe$_3$ does not have long-range order and remains fluctuating. While the statement can be rigorously proven by simple theoretical arguments \cite{Remark1}, here we give an estimate of the \emph{finite} correlation length of the $a$-axis CDW. Based on the diffraction peak width $w$ [Fig.~\ref{fig:intro}(g)], which is limited by instrumental resolution, the correlation length has a lower bound of $1/w\sim3.5$~nm, or 8 crystallographic unit cells (u.c.). Given the approximate CDW lifetime $\tau$ of 4~ps [Fig.~\ref{fig:kink}(a)], the correlation length is at most $v\tau\sim10$~nm (23~u.c.), where $v=2500~$m/s is the speed of sound along the $a$-axis \cite{Note3}. This upper bound is a testament that each fluctuating patch cannot establish phase coherence with its neighbors at a speed faster than phonon propagation. Compared to the correlation length of the dominant $c$-axis CDW in equilibrium, which is estimated to be at least 1.8~$\upmu$m within Te planes \cite{Ru2008}, the particularly small value of $v\tau$ hence confirms the absence of long-range order along the $a$-axis and suggests that the light-induced CDW consists entirely of short-range fluctuations.

\begin{figure}[htb!]
	\centering
	\includegraphics[width=1\columnwidth]{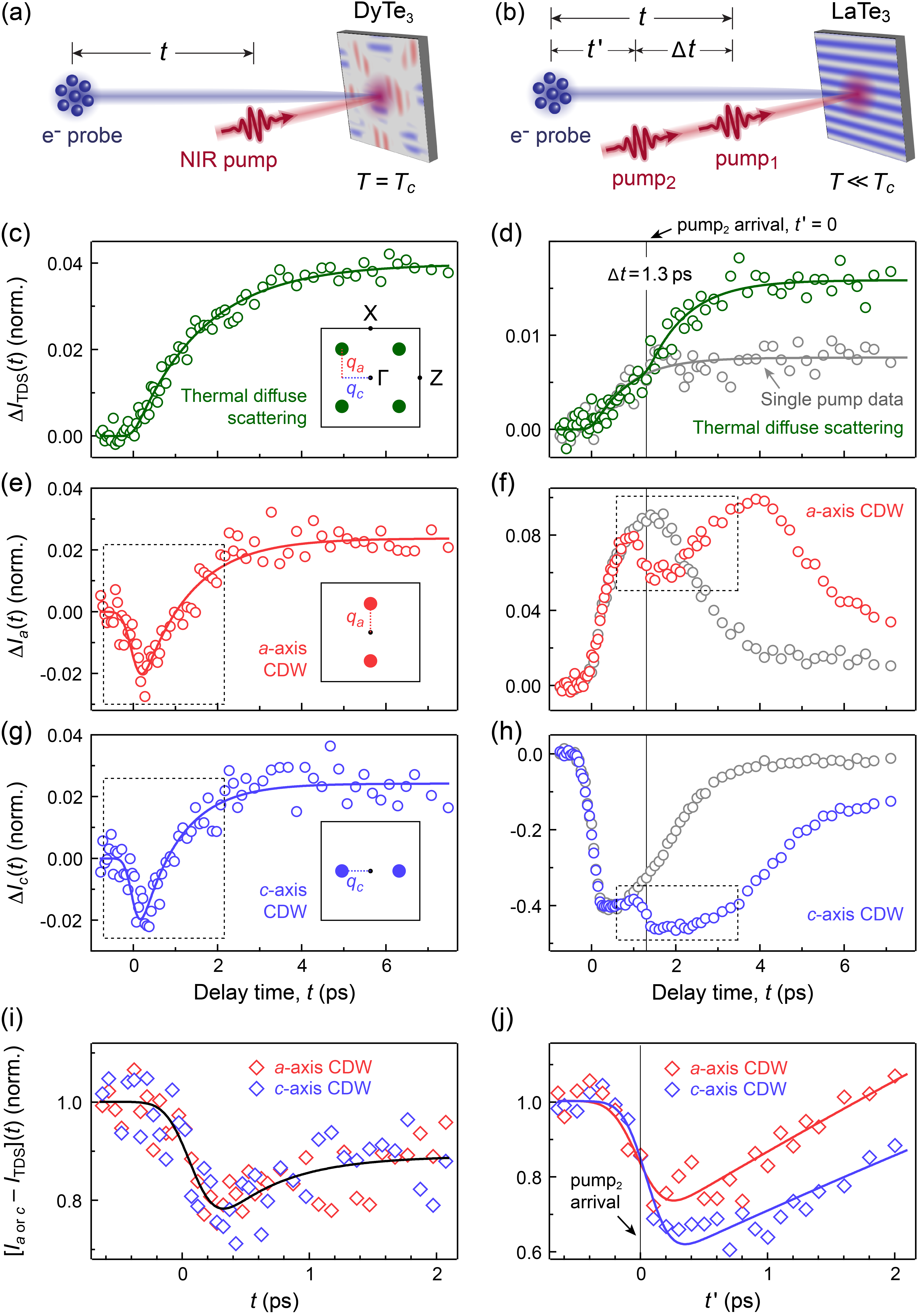}
	\caption{Response of CDW fluctuations to photoexcitation. (a)(b)~Schematic setups for DyTe$_3$ and LaTe$_3$. Both samples were kept at $T=307$~K. The incident fluence was 3.3~mJ/cm$^2$ in (a) and 1.0~mJ/cm$^2$ for each pump in (b). (c)--(h)~Changes in the integrated intensities for thermal diffuse scattering [$I_\text{TDS}$, (c)(d)], $a$-axis diffuse CDW peak [$I_a$, (e)(f)], and $c$-axis diffuse CDW peak [$I_c$, (g)(h)]. Integration areas are marked by solid circles in the insets. Traces are normalized by the average value of $I_c$ before photoexcitation. In (d)(f)(h), vertical lines indicate the arrival time of pump$_2$ at $\Delta t=1.3$~ps. For reference, dynamics in the absence of pump$_2$ is shown in gray. (i)(j)~Enlarged view of dashed rectangles in (e)--(h) after subtracting the respective thermal diffuse background [$I_\text{TDS}$ in (c)(d)]. In (i), traces are normalized by the average values at $t<0$. In (j), traces are plotted as a function of the relative delay between the probe pulse and pump$_2$ ($t'$), where intensities are normalized by the average value in the interval $t'\in[-0.7~\text{ps},-0.2~\text{ps}]$. A slightly larger reduction along the $c$-axis is attributed to a mismatch between pumped and probed volumes, leading to additional melting of residual long-range CDW by the second pulse. Solid curves in (c)--(j) are fits to a phenomenological model in Eq.~\eqref{eq:fit_erfexp}. The black fitted curve in (i) uses the averaged data along the $a$- and $c$-axes.}
\label{fig:double}
\end{figure}

An almost square-symmetric diffraction pattern after photoexcitation and at equilibrium $T_c$ is suggestive of a close connection between the two states. To further elucidate their relationship, we investigate their response to an external perturbation. By comparing the respective dynamics of the order parameter fluctuations, we can gain some crucial insights into the similarities and differences between the two regimes. To this end, we apply a second laser pulse to LaTe$_3$ right after the emergence of the $a$-axis satellite peak and record the intensity evolution of the CDW fluctuations along both axes. As a reference, we also photoexcite DyTe$_3$ at its CDW transition temperature, where fluctuations of both density waves abound.

We first examine the laser-induced response in DyTe$_3$ at its $T_c$ [Fig.~\ref{fig:double}(a)]. After photoexcitation, the diffuse satellite spots display an initial dip in intensity followed by a fast recovery, a trend perfectly mirrored in both axes [Fig.~\ref{fig:double}(e) and (g)]. These dynamics are in stark contrast to diffuse scattering intensities at other momenta away from Bragg or CDW peaks, where only a single-exponential rise is observed [Fig.~\ref{fig:double}(c)]. The dip can be understood in two equivalent ways. From the phonon perspective, it represents a transient stiffening of the soft mode \cite{Otto2019b,Remark1}. As electrons are excited to high energy, there is a transient reduction in the electronic band occupation near the Fermi energy that interacts with the lattice ions. This reduction leads to an increase in the renormalized phonon frequency and hence a decrease in the phonon population, as suggested by the equipartition theorem. An alternative viewpoint is based on the classical description of phonons as atomic displacements in real space. In each frame diffracted from a single electron pulse, we capture a snapshot of the system, such as the one depicted in Fig.~\ref{fig:intro}(c). The dip hence indicates a smaller lattice distortion amplitude in the fluctuating CDW patches, averaged over space and over all snapshots at the same pump-probe delay. The second perspective naturally connects the photoinduced melting of fluctuating CDWs to the melting of a long-range CDW. Locally, there is minimal distinction between the two processes and both occur over $\sim0.4$~ps, a timescale dictated by the phonon period associated with the CDW distortion \cite{Hellmann2012,Zong2019b}. In Fig.~\ref{fig:double}(e)(g), we observe that the intensities quickly rise after the dip, indicating an increased phonon population from laser-induced heating. After subtracting the thermal diffuse contribution, the dip only partially recovers [Fig.~\ref{fig:double}(i)], suggesting an elevated lattice temperature above $T_c$, where the Kohn anomaly becomes less pronounced.

Next, we study the dynamics in the photoexcited state of LaTe$_3$. As illustrated in Fig.~\ref{fig:double}(b), we use the first laser pulse to bring the material into a nonequilibrium state, where we have observed a symmetric appearance of diffuse satellite spots along both $a$- and $c$-axes. We then apply a second pulse to perturb this transient state and look at the response of the two competing CDW fluctuations. In the experiment, the two pump pulses share the same \emph{incident} fluence. To assess the \emph{absorbed} fluence, we note that the maximum value attained in thermal diffuse scattering doubles after the second pulse [Fig.~\ref{fig:double}(d)]. This observation affirms that energy absorption is minimally affected by the presence of excited carriers after the first pulse. We now move on to analyze the CDW peaks, shown in Fig.~\ref{fig:double}(f) and (h). Unlike their distinct behavior upon the initial photoexcitation, the intensity evolution of the peaks along both axes share almost identical trends after the second pulse. For a direct comparison between the two orders, we zoom in to their dynamics right after the second pulse and plot them together in Fig.~\ref{fig:double}(j), where intensities from thermal diffuse scattering have been subtracted using the same procedure applied to DyTe$_3$. Similar to the fluctuating CDWs in DyTe$_3$ near $T_c$, the two diffuse peaks in LaTe$_3$ feature a transient reduction in the fluctuation amplitude, followed by a recovery that lasts for more than 2~ps. Unlike DyTe$_3$, the satellite intensities in LaTe$_3$ are fully recovered compared to their values just before the second pulse, suggesting the nonthermal nature of these density wave fluctuations.

\begin{figure}[htb!]
	\centering
	\includegraphics[width=1\columnwidth]{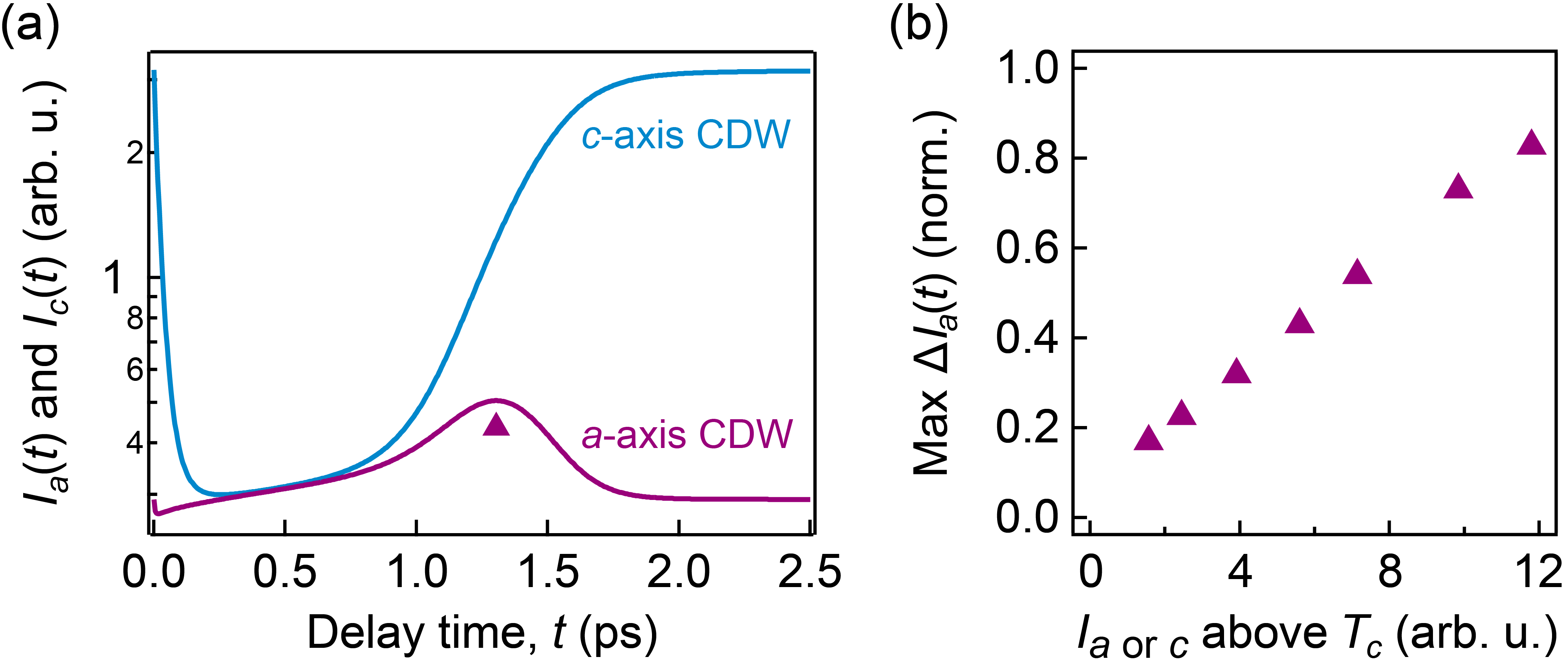}
	\caption{Simulated dynamics of photoinduced CDW and its relation to equilibrium fluctuations. (a)~Evolution of integrated intensities of the $c$-axis (blue) and $a$-axis (red) CDW peak upon photoexciting the unidirectional CDW state. Triangle marks the maximum intensity of the light-induced CDW. The nonzero value of the $a$-axis peak at $t=0$ originates from thermal fluctuations. (b)~Maximum intensity change of the photoinduced $a$-axis CDW peak [$\Delta I_a(t)$] as a function of equilibrium diffuse intensity at a fixed temperature above $T_c$, the latter of which quantifies thermal fluctuations and is indistinguishable between the two axes. $\Delta I_a(t)$ is normalized by $I_a(t=0)$ (see Fig.~\ref{fig:stiff}).}
\label{fig:sim}
\end{figure}

The similarities between the excited state in LaTe$_3$ and the critical state in DyTe$_3$ -- both in their diffraction snapshots (Fig.~\ref{fig:intro}) and in their photoinduced dynamics (Fig.~\ref{fig:double}) -- suggest that the light-induced CDW is a special manifestation of critical fluctuations. While the equilibrium fluctuations near $T_c$ are thermal and follow the scaling relations prescribed by the theory of renormalization group \cite{Goldenfeld1992}, the light-induced fluctuations may not conform to a thermodynamic distribution \cite{Dolgirev2020b}. To understand how the strength of equilibrium fluctuations affect the appearance of the light-induced CDW, we developed a time-dependent Ginzburg-Landau formalism within the Gaussian approximation (see \cite{Remark1} for derivation). This approach gives a nonperturbative solution to the light-induced dynamics, yielding quantities that have a one-to-one correspondence to the observables in our time-resolved diffraction experiments. Unlike $N$-temperature models \cite{Dolgirev2020a}, here we do not need to artificially assign a temperature to each degree of freedom in the system.

To assess the validity of the model, we first calculate intensity evolution of $a$- and $c$-axis CDW peaks after photoexcitation [Fig.~\ref{fig:sim}(a)]. The simulated trends successfully reproduce the experimental observations (Fig.~\ref{fig:kink}). The transient enhancement of intensity along the $a$-axis  is solely the result of CDW fluctuations without long-range order \cite{Remark1}. In Fig.~\ref{fig:sim}(b), as we reduce the order parameter stiffness to increase the amplitude of equilibrium fluctuations above $T_c$, the strength of the transient CDW order also increases under identical photoexcitation conditions. This positive correlation suggests that strong fluctuations in equilibrium constitute an important factor for observing light-induced ordering phenomena out of equilibrium.

Despite the similarities between the light-induced CDW and the critical fluctuations, there exist important differences \cite{Remark1}. For example, the transient lattice temperature of LaTe$_3$ stays far below its equilibrium $T_c$, and there is no change in the in-plane lattice anisotropy after photoexcitation, distinct from the evolution of $a$ and $c$ lattice parameters across $T_c$ \cite{Kogar2020,Ru2008}. Importantly, the light-induced CDW has a finite correlation length for all time delays but at the critical point in equilibrium, correlation length diverges with fluctuations occurring at all length scales. Hence, strictly speaking, the photoexcited state is not truly critical as described in a thermodynamic transition.

By leveraging the symmetry between two competing CDWs in $R$Te$_3$, we have elicited the correspondence between a photoinduced order and critical fluctuations in equilibrium. The parallels provide a nonthermal pathway to access hidden symmetries of a system even if $T_c$ is unattainable under equilibrium condition. The similarities also hint at the existence of universal scaling laws that govern the dynamics of a highly nonequilibrium system \cite{Dolgirev2020b}, which have been detected in scattering experiments with high momentum resolution and an extended time delay \cite{Laulhe2017,Vogelgesang2018,Mitrano2019}. Furthermore, our results offer a generic mechanism for the creation of photoinduced states, which can emerge as order parameter fluctuations in the absence of long-range order. This insight suggests that one should look for material classes that exhibit strong order parameter fluctuations in equilibrium in order to look for ``hidden'' states out of equilibrium. Experimental signatures for such strong fluctuations depend on the order parameter, ranging from diffuse peaks in a charge or spin density wave system to Nernst effect in a superconductor \cite{Remark1}. We expect the connection between equilibrium fluctuations and out-of-equilibrium ordering to hold regardless of microscopic details, providing a guiding principle in our search for other light-induced states.

\begin{acknowledgments}
We thank Mariano~Trigo and Yu~He for helpful discussions. This work was mainly funded by the U.S. Department of Energy, BES DMSE (data taking and analysis) and the Gordon and Betty Moore Foundation's EPiQS Initiative grant GBMF9459 (modeling and manuscript writing). We acknowledge support from the U.S. Department of Energy BES SUF Division Accelerator \& Detector R\&D program, the LCLS Facility, and SLAC under contract No.'s DE-AC02-05-CH11231 and DE-AC02-76SF00515 (MeV UED at SLAC). Sample growth and characterization work at Stanford was supported by the U.S. Department of Energy, Office of Basic Energy Sciences, under contract number DEAC02-76SF00515. A.Z. acknowledges support from the Miller Institute for Basic Research in Science. This research was partly supported by the Army Research Office through Grant No. W911NF1810316, and the Gordon and Betty Moore Foundation EPiQS Initiative through grant GBMF9643 to P.J.-H. (sample preparation and characterization). This work made use of the Materials Research Science and Engineering Center Shared Experimental Facilities supported by the National Science Foundation (NSF) (Grant No. DMR-0819762). This work was performed in part at the Harvard University Center for Nanoscale Systems (CNS), a member of the National Nanotechnology Coordinated Infrastructure Network (NNCI), which is supported by the National Science Foundation under NSF ECCS Award No. 1541959. P.E.D. and E.D. were supported by Harvard-MIT CUA, AFOSR-MURI: Photonic Quantum Matter award FA95501610323, Harvard Quantum Initiative.
\end{acknowledgments}

\footnotetext[1]{Unlike LaTe$_3$ that only develops a unidirectional CDW under ambient pressure, there are two CDW transitions in DyTe$_3$ \cite{RuThesis}. Starting from its normal metallic state, DyTe$_3$ first develops a unidirectional CDW at $T_{c1}$; the second transition into a bidirectional CDW state occurs at a lower temperature $T_{c2}$. Here, we are only concerned with the high-temperature transition and we denote $T_{c1}$ by $T_c$ for brevity.}

\footnotetext[2]{The image was taken above $T_{c2}=68$~K for DyTe$_3$ \cite{Maschek2018}.}

\footnotetext[3]{The speed of sound is deduced from the phonon dispersion in DyTe$_3$ calculated by density functional perturbation theory and verified by inelastic X-ray scattering \cite{Maschek2018}. The speed of sound associated with the phason excitation may be an alternative choice for this correlation length estimate. The phason dispersion is unavailable for $R$Te$_3$ but we take note of values in other incommensurate CDWs. The phason speed ranges from $4\times10^2$~m/s in 1$T$-TaS$_2$ \cite{Minor1989} to $2\times10^4$~m/s in K$_{0.3}$MoO$_3$ \cite{Pouget1991}, hence not changing the conclusion that the largest possible correlation length of the transient CDW in LaTe$_3$ is still orders of magnitude smaller compared to its dominant CDW in equilibrium.}

\newcommand{\noopsort}[1]{} \newcommand{\printfirst}[2]{#1}
  \newcommand{\singleletter}[1]{#1} \newcommand{\switchargs}[2]{#2#1}

\cleardoublepage
\onecolumngrid
\begin{large}
\begin{center}
\textbf{Supplemental Material to ``Role of equilibrium fluctuations in light-induced order''}
\end{center}
\end{large}
\vspace{0.5cm}
\twocolumngrid

\beginsupplement

\section{Experimental details}\label{sec:method}

\subsection{Sample preparation}

Single crystals of LaTe$_3$ and DyTe$_3$ were grown by slow cooling of a binary melt \cite{Ru2006}, and then mechanically exfoliated to a typical size of approximately $150\,\upmu\text{m}\times150\,\upmu\text{m}\times60$\,nm. Via an all-dry viscoelastic stamping method \cite{Bie2021}, thin flakes were transferred to a commercial 10-nm-thick silicon nitride window (SiMPore), which was mounted on a copper holder for ultrafast electron diffraction measurements. All preparations were performed in argon or nitrogen gas as $R$Te$_3$ compounds are readily oxidized under ambient conditions \cite{Ru2006}.

\subsection{Single- and double-pump ultrafast electron diffraction}

The ultrafast electron diffraction experiments were carried out at the MeV UED beamline at SLAC National Accelerator Laboratory \cite{Weathersby2015,Shen2018}. The 800-nm (1.55-eV), 80-fs pump pulse from a commercial Ti:sapphire regenerative amplifier laser (Vitara and Legend Elite HE, Coherent Inc.) was split into the pump and probe arms. The probe arm was frequency tripled and focused to a copper photocathode, and 3.1~MeV (or 4.2~MeV) electron bunches were generated by radiofrequency photoinjetors at a repetition rate of 180~Hz (or 360~Hz). The electron beam was normally incident on the sample with a $90\times90$~$\upmu$m$^2$ spot size, measured at full-width at half maximum (FWHM). The diffraction pattern was imaged by a phosphor screen and recorded by an electron-multiplying charge-coupled device (EMCCD, Andor iXon Ultra 888). A circular through hole in the center of the phosphor screen allowed the passage of undiffracted electron beam to prevent camera saturation. 

The pump arm goes through a linear translation stage (Delay~1) before being split into two paths: a fixed path and a variable path. The variable path has an additional linear stage (Delay~2) that adjusts the relative optical delay between the two pump paths. Each pump path can be individually blocked and its pulse energy can be separately tuned. The two pump paths were recombined before entering the vacuum chamber, and the pump was then focused to an area larger than $350\times350$~$\upmu$m$^2$ (FWHM) in the sample at an incidence angle around $5^{\circ}$ from sample normal. For single-pump measurements, one of the two pump paths was blocked and Delay~1 was scanned. For double-pump measurements, both pump paths were unblocked. Delay~1 was varied with a fixed value of Delay~2 that set the relative arrival time between the two pump pulses. The relative time delay and incident fluence of the two excitation pulses were calibrated by illuminating them separately on the same sample and observing the initial intensity drop of the CDW peaks.

\subsection{Time trace fitting}

In Figs.~\ref{fig:double} and \ref{fig:roi}, the time evolution of diffraction intensities was fit to a phenomenological model to describe the photoinduced changes \cite{Hellmann2012,Moore2016}
\begin{align}\label{eq:fit_erfexp}
\Delta I(t) = \Bigg[\frac{1}{2}&\left(1+\textrm{Erf}\left(2\sqrt{2}(t-t_0 )/w\right) \right)\notag\\
\cdot&\left(I_{\infty}+I_0 e^{-(t-t_0)/\tau}\right)\Bigg] *g(w_0,t).
\end{align}
In this model, $w$ represents the intrinsic system response time to photoexcitation; $I_0$ represents the maximum intensity change; $I_{\infty}$ denotes the value of $\Delta I$ at long time delays; $\tau$ is the characteristic relaxation time to the quasi-equilibrium; $t_0$ is associated with the relative arrival time of pump and probe pulses, and it is the time delay when $\Delta I$ reaches $(I_0+I_\infty)/2$. The effect of the finite pulse width in both pump and probe branches is taken into account by convolving the terms in the square brackets with a normalized Gaussian pump-probe cross-correlation function $g(w_0,t)$, where $w_0$ denotes the FWHM. In the curve fits, $w_0$ takes the value of 300~fs (or 230~fs) for the 3.1~MeV (or 4.2~MeV) electrons used. If the time trace can be described by a single-exponential evolution, such as the thermal diffuse intensity in Figs.~\ref{fig:double}(c)(d) and \ref{fig:roi}(d), Eq.~\eqref{eq:fit_erfexp} is simplified by setting $I_\infty=-I_0$ and $w\rightarrow0^+$.

\subsection{Transient lattice temperature}

An important observation made in the main text is the close resemblance between photoexcited LaTe$_3$ at 307~K and equilibrium DyTe$_3$ near its $T_c$. To rule out the trivial scenario where the laser pulse transiently heats LaTe$_3$ to its $T_c$ around 670~K \cite{Hu2014b}, here we estimate the transient increase of its lattice temperature. Assuming no heat dissipation out of the probed volume, an upper bound of the temperature rise is given by 
\begin{align}
    \Delta T = \frac{(1-R)F}{dc},
\end{align}
where $R=0.59$ is the room temperature reflectivity at 800~nm \cite{Sacchetti2006}, $F$ is the incident fluence, $d=44$~nm is the penetration depth at which the intensity of the 800-nm light decays to $1/e$ of its original value \cite{Zong2019b}, and $c=1.3$~J/(cm$^3\cdot$K) is the room temperature specific heat capacity \cite{Ru2006,Ramsey1965}. In this work, a maximum incident fluence of 2.1~mJ/cm$^2$ was used to photoexcite LaTe$_3$, leading to a maximum possible increase of $\Delta T=151$~K in the lattice temperature. Given the low repetition rate of 180~Hz or 360~Hz used in the experiment and given the equilibrium sample temperature of 307~K, we conclude that LaTe$_3$ is much below its CDW transition temperature throughout the temporal evolution.

\section{Diffraction intensity near a Kohn anomaly}\label{sec:kohn}

In Fig.~\ref{fig:intro}(c), the characteristic length scale of a fluctuating CDW domain is labeled as $1/\Delta q_\text{soft}$, where $\Delta q_\text{soft}$ is the width of the Kohn anomaly in the phonon dispersion. In this section, we formalize the relation between the momentum width of the Kohn anomaly and the corresponding width of the diffuse scattering peak, the latter of which measures the reciprocal of the fluctuating domain size.

\begin{figure*}[htb!]
	\centering
	\includegraphics[width=0.80\textwidth]{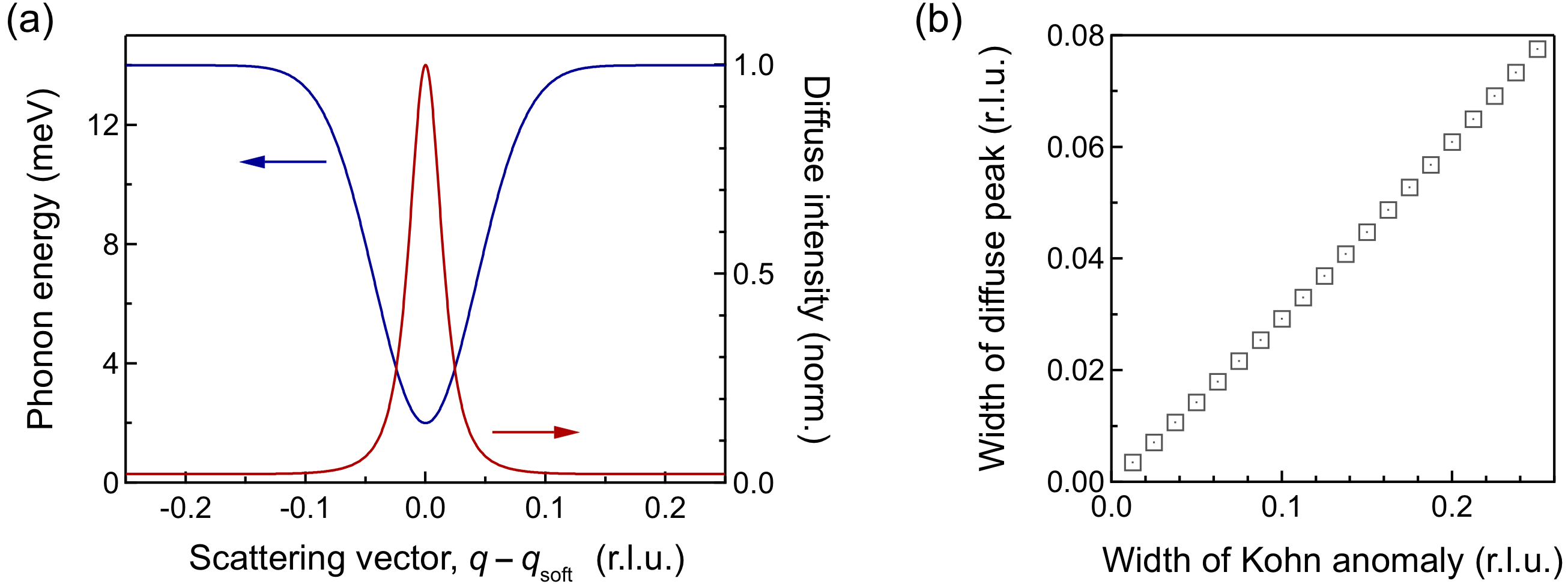}
	\caption{Widths of Kohn anomaly and thermal diffuse peak. (a)~Sketch of a Kohn anomaly in the phonon dispersion (blue curve, left axis), adopting experimentally observed parameters for DyTe$_3$ near its $T_c$ \cite{Maschek2018}. The calculated one-phonon diffuse intensity at $T=307$~K for this dispersion is superimposed (red curve, right axis). Intensity is normalized by its maximum value. (b)~FWHMs of the diffuse peak as a function of FWHMs of the Kohn anomaly, showing a quasi-linear relation. The FWHM of the diffuse peak is obtained by fitting it to a Lorentzian profile.}
\label{fig:kohn}
\end{figure*}

Under the kinematic approximation, the diffraction intensity at wavevector $\mathbf{k}$ is the sum of contributions from elastic scattering and inelastic scatterings with $n\geq 1$ phonons:
\begin{align}
	I(\mathbf{k}) = I_0 (\mathbf{k}) + \sum_{n\geq1} I_n(\mathbf{k}).
\end{align}
As the scattering cross sections with two or more phonons are relatively small, we are mostly concerned with $I_0$ and $I_1$, given by \cite{RenedeCotret2019}
\begin{align}
I_0(\mathbf{k}) &= I_c \left| \sum_\alpha f_\alpha(\mathbf{k}) e^{-W_\alpha(\mathbf{k})}e^{-i\mathbf{k}\cdot \mathbf{r}_\alpha}   \right|^2, \label{eq:I0}\\
I_1(\mathbf{k}) &= I_c \sum_j \frac{n_{j,\mathbf{q}}+1/2}{\omega_{j,\mathbf{q}}} \left|F_{1j}(\mathbf{k})\right|^2,\label{eq:I1}
\end{align}
where $I_c$ is some constant of proportionality, $\mathbf{k}$ is the total scattering wavevector, $\mathbf{q}\equiv \mathbf{k}-\mathbf{G}$ is the reduced crystal momentum, defined with respect to the closest reciprocal lattice vector $\mathbf{G}$ for a given $\mathbf{k}$. $\mathbf{r}_\alpha$ is the position of atom $\alpha$ in the unit cell, $W_\alpha(\mathbf{k})$ is the Debye-Waller factor, and $f_\alpha(\mathbf{k})$ is the atomic form factor. Index $j$ runs over the phonon branches and $n_{j,\mathbf{q}}$, $\omega_{j,\mathbf{q}}$ correspond to the population and frequency for the phonon at reduced momentum $\mathbf{q}$ in branch $j$. $\left|F_{1j}(\mathbf{k})\right|^2$ is the one-phonon structure factor, given by
\begin{align}
\left|F_{1j}(\mathbf{k})\right|^2 = \left| \sum_\alpha e^{-W_\alpha(\mathbf{k})}\frac{f_\alpha(\mathbf{k})}{\sqrt{m_\alpha}} \left(\mathbf{k}\cdot \mathbf{e}_{j,\alpha,\mathbf{q}}\right) \right|^2,
\end{align}
where $m_\alpha$ is the mass of atom $\alpha$ and $\mathbf{e}_{j,\alpha,\mathbf{q}}$ is the displacement polarization vector for atom $\alpha$ under the motion of phonon in branch $j$ with momentum $\mathbf{q}$. As we explain in Sec.~\ref{sec:transverse}, the transverse polarization of the CDWs in $R$Te$_3$ leads to a momentum-dependent satellite intensity in the diffraction pattern, which is accounted for by the $\mathbf{k}\cdot \mathbf{e}_{j,\alpha,\mathbf{q}}$ term. The Debye-Waller factor $W_\alpha(\mathbf{k})$ describes the intensity reduction in both $I_0$ and $I_1$, which results from the excitation of phonons in all branches and momenta. It reads
\begin{align}
	W_\alpha(\mathbf{k}) = \frac{1}{4m_\alpha} \sum_{j,\mathbf{q}} \left|u_{j,\mathbf{q}}\right|^2 \left|\mathbf{k}\cdot\mathbf{e}_{j,\alpha,\mathbf{q}}\right|^2.
\end{align}
Here, $u_{j,\mathbf{q}}$ is the vibration amplitude for a particular phonon mode,
\begin{align}
\left|u_{j,\mathbf{q}}\right|^2 = \frac{2\hbar\left(n_{j,\mathbf{q}}+1/2\right)}{N_c\, \omega_{j,\mathbf{q}}},
\end{align}
where $N_c$ is the number of unit cells. 

From Eq.~\eqref{eq:I1}, the presence of a Kohn anomaly -- a dip in the frequency $\omega_{j,\mathbf{q}}$ at a particular momentum $\mathbf{q}_\text{soft}$ for phonon branch $j$ -- would lead to a locally enhanced one-phonon scattering. The enhancement is primarily due to the $n_{j,\mathbf{q}}$ term in the numerator and the $\omega_{j,\mathbf{q}}$ term in the denominator. If we choose a scattering vector $\mathbf{k}$ that is aligned with the polarization vector $\mathbf{e}_{j,\alpha,\mathbf{q}}$ for the soft mode in question, we expect the one-phonon structure factor $\left|F_{1j}(\mathbf{k})\right|^2$ to be a slow-varying function of $\mathbf{q}$ within a Brillouin zone, so it is less affected by the Kohn anomaly. To see the relationship between $\omega_{j,\mathbf{q}}$ and $I_1(\mathbf{k})$, a schematic of a Kohn anomaly is sketched in Fig.~\ref{fig:kohn}(a) (blue curve), using experimentally determined parameters for DyTe$_3$ near its $T_c$ \cite{Maschek2018}. The corresponding $I_1$ from this soft phonon branch is shown in the red curve, where the contribution from $|F_{1j}|^2$ is neglected. Here, we use the Bose-Einstein distribution for the phonon population $n_{j,\mathbf{q}} =1/ \left(e^{\omega_{j,\mathbf{q}}/k_BT} - 1\right)$, where $T$ is set to 307~K, near the $T_c$ of DyTe$_3$. As expected, a diffuse peak develops at $q_\text{soft}$, whose width matches that of the Kohn anomaly up to a small factor. In Fig.~\ref{fig:kohn}(b), the width of the diffuse peak is plotted for a range of widths for the Kohn anomaly, showing a quasi-linear relationship. Therefore, it is indeed justified to use the width of the Kohn anomaly to estimate the width of the diffuse peak up to a small constant factor, which in turn indicates the spatial extent of the fluctuating CDW domains [Fig.~\ref{fig:intro}(b)(c)].

\begin{figure*}[htb!]
	\centering
	\includegraphics[width=0.78\textwidth]{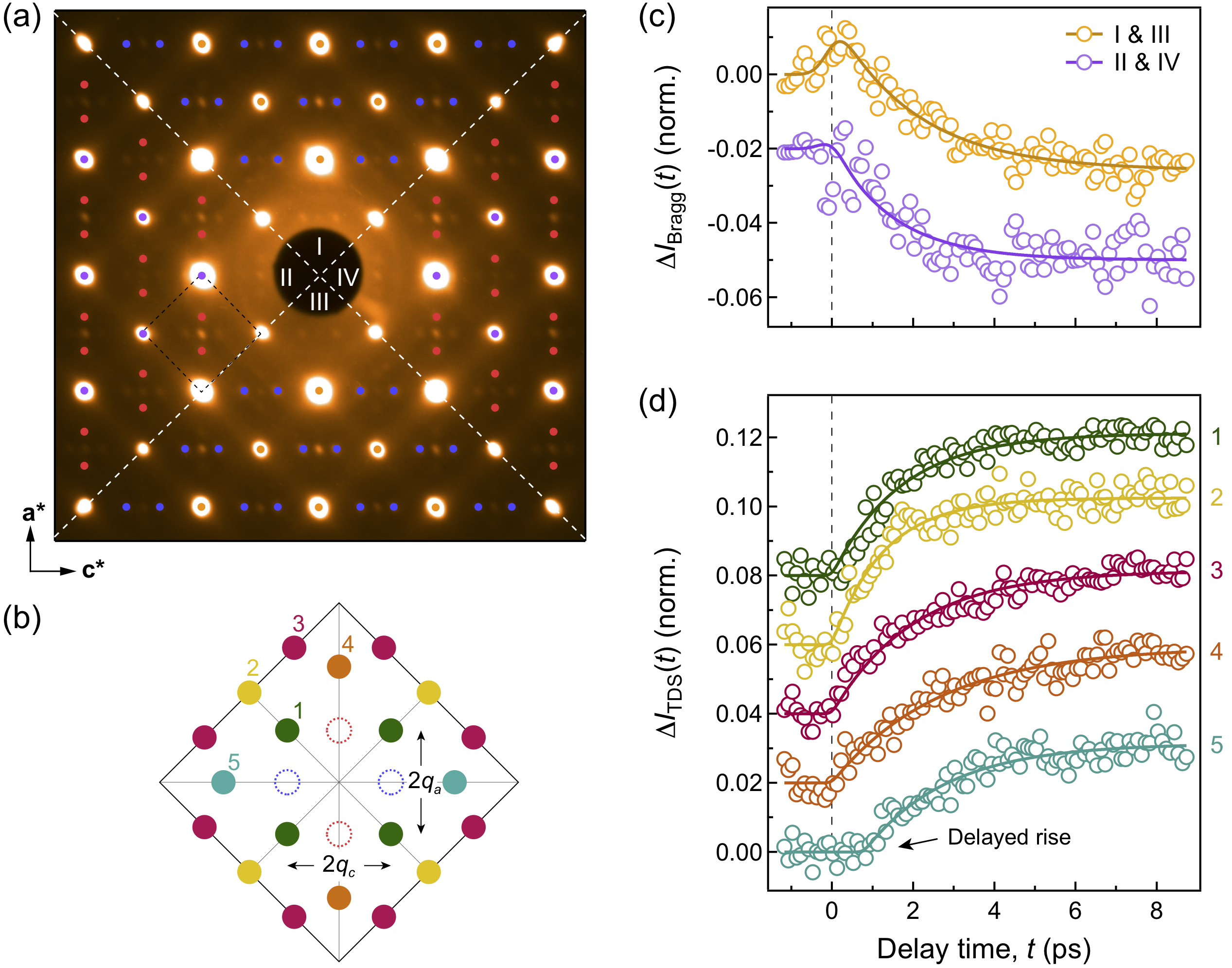}
	\caption{Evolution of lattice Bragg peaks and diffuse scatterings. (a)~The same diffraction pattern of LaTe$_3$ as in Fig.~\ref{fig:intro}(g), overlaid with circular ROIs that indicate the integration regions for CDW peak intensities $I_c$ (blue) and $I_a$ (red). ROIs are divided into quadrants I--IV (white dashed lines) due to the transverse nature of the lattice distortion. Yellow and purple ROIs on Bragg peaks correspond to integration areas for (c). (b)~An enlarged view of the dashed diamond in (a) centered around an $(H,0,L)$ order where $H+L$ is an odd integer. Color-coded circles are integration areas for diffuse scatterings ($I_\text{TDS}$) plotted in (d), where intensities in areas of the same color across multiple diffraction orders are averaged. Locations of CDW peaks are indicated by the dashed circles. (c)~Changes in the integrated intensity of Bragg peaks in quadrant I \& III (yellow) and quadrant II \& IV (purple). (d)~Evolutions of the change in the diffuse scattering intensity at locations 1 to 5, labeled in (b). Intensities in (c)(d) are normalized to their averaged value at $t<0$. Curves are vertically offset by 0.02 for clarity. For all panels, the incident laser fluence was 2.1~mJ/cm$^2$ and curves are fits to Eq.~\eqref{eq:fit_erfexp}.}
\label{fig:roi}
\end{figure*}

\section{Transverse atomic displacement and Bragg peak dynamics}\label{sec:transverse}

From the diffraction patterns in Fig.~\ref{fig:intro}(d)--(g), the $c$-axis CDW peaks at $(H,0,L\pm q_c)$ are most prominent when $|H|>|L|$. On the other hand, the $a$-axis satellites at $(H\pm q_a, 0,L)$ are the brightest for $|H|<|L|$. Here, we adopt the convention of $q_a\approx q_c \approx 2/7$, so $H+L$ is an odd integer. As we explain in the following, this momentum dependence suggests the transverse polarization of the atomic displacement underlying the periodic lattice distortion \cite{Maschek2015,Maschek2018}. For small distortion, the intensity of the CDW peak contains a factor of $\mathbf{u}\cdot\mathbf{k}$, where $\mathbf{u}$ is the atomic displacement and $\mathbf{k}$ is the scattering vector \cite{Dolgirev2020a}. For a transverse mode, $\mathbf{u}\perp \mathbf{q}$, where $\mathbf{q}$ is the CDW ordering vector. Hence, $\mathbf{u} \parallel \mathbf{a}^*$ for the $c$-axis CDW and $\mathbf{u} \parallel \mathbf{c}^*$ for the $a$-axis CDW. The maximum satellite peak intensity occurs when $\mathbf{k}\parallel \mathbf{u}$, leading to the momentum pattern observed. For this reason, we divide the diffraction pattern into four quadrants, shown in Fig.~\ref{fig:roi}(a). When plotting the intensity evolutions of CDW peaks -- $I_a(t)$ and $I_c(t)$ -- we only average over peaks in the corresponding quadrants to maximize the signal. The integration areas for $I_a(t)$ and $I_c(t)$ are marked by red and blue circles in Fig.~\ref{fig:roi}(a), respectively.

Another consequence of the transverse polarization is encoded in the intensity evolution of the Bragg peak, shown in Fig.~\ref{fig:roi}(c). The yellow markers correspond to quadrants I \& III while the purple markers correspond to quadrants II \& IV, as indicated by the color-coded regions of interest (ROIs) in Fig.~\ref{fig:roi}(a). While the yellow curve shows an initial increase in intensity, the purple curve only exhibits an intensity decay over $\sim2$~ps. The decay arises from the photoinduced Debye-Waller factor that redistributes the spectral weight from the Bragg peak to elsewhere in the momentum space as phonon branches are populated. The initial rise can only be accounted for by the transient suppression of the equilibrium $c$-axis CDW. This intensity gain of the Bragg peak also contains the factor $\mathbf{u}\cdot\mathbf{k}$ \cite{Dolgirev2020a}, so the effect is the most pronounced when $\mathbf{u}\parallel\mathbf{k}$. For the $c$-axis CDW, $\mathbf{u} \parallel \mathbf{a}^*$, so the intensity gain is mostly manifested in Bragg peaks from quadrants I \& III.

\section{Momentum-dependent diffuse scattering dynamics}

In Fig.~\ref{fig:double}(c)(d), we plotted the temporal evolutions of diffuse scattering at $\pm\mathbf{q}_a\pm\mathbf{q}_c$, showing a single-exponential rise after photoexcitation. In this section, we examine the momentum-dependence of the diffuse scattering dynamics. In Fig.~\ref{fig:roi}(b), we identified five representative momentum positions together with their symmetry-equivalent points within the diamond-shaped ROI. The intensity evolution of thermal diffuse scattering from each location is plotted in Fig.~\ref{fig:roi}(d). The short-time dynamics of all curves is well described by a single-exponential rise, indicating the transient excitation of incoherent phonons. In particular, for locations 1 to 4, the intensity rise starts immediately after photoexcitation and all curves plateau at a quasi-equilibrium value of a 4\% increase. For comparison, under the same excitation condition, the maximum intensity rise at $\pm \mathbf{q}_a$ is approximately 10\%, plotted in Fig.~\ref{fig:kink}(a). The time constants for the intensity rise vary among different momenta, ranging from 1 to 2.5~ps. These rise times encode information about the momentum-dependent electron-phonon and phonon-phonon coupling strengths \cite{Stern2018}, which are subject to future investigations.

\begin{figure*}[htb!]
	\centering
	\includegraphics[width=0.70\textwidth]{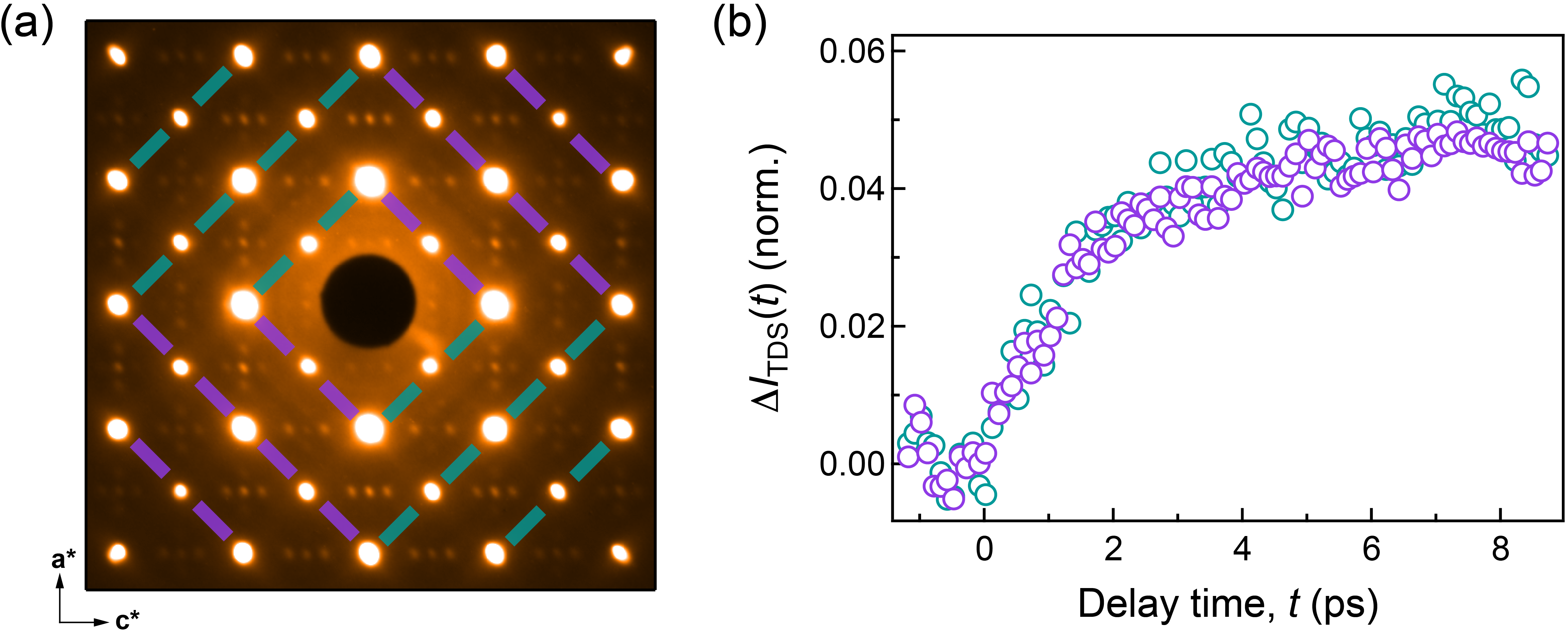}
	\caption{Diffuse scattering dynamics along different directions. (a)~Transient diffraction pattern of LaTe$_3$ following photoexcitation by a 2.1~mJ/cm$^2$ pulse, reproduced from Fig.~\ref{fig:intro}(g). It is overlaid with ROIs that cover the diffuse scattering signals of transverse acoustic phonons along the two diagonal directions (purple and green). (b)~Evolutions of the change in the diffuse scattering intensity for the two color-coded ROIs in (a), showing indistinguishable dynamics along the two directions. Intensities are normalized to their averaged value at $t < 0$.}
\label{fig:diff_acoustic}
\end{figure*}

The diffuse scattering dynamics at location~5 is markedly different from the others. First, there is a 0.8~ps delay of the rise, indicated by the arrow in Fig.~\ref{fig:roi}(d). Second, the intensity only increases by 3\% when it plateaus, distinctly smaller than the others. These features suggest that there is another dynamics apart from the increase in the phonon population. One possibility is that location 5 lies in close proximity to high-order $c$-axis CDW peaks \cite{Straquadine2019}. The photoinduced melting of these high-order peaks will both delay and offset the rise of diffuse scattering, explaining the anomaly at location 5. Another possibility is that the electron-phonon coupling vertex is anomalously small at this location, leading to a delayed rise of the phonon population.

In Fig.~\ref{fig:roi}(d), the ROIs for the diffuse intensity dynamics are taken in a four-fold symmetric manner, as illustrated in Fig.~\ref{fig:roi}(b). This is because the transient diffuse intensity change appears isotropic and does not distinguish between the $a$- and $c$-axis. We verify this lack of anisotropy in Fig.~\ref{fig:diff_acoustic}, where diffuse signals in two sets of ROIs along the two diagonal directions are analyzed. These ROIs correspond to the acoustic transverse phonon branches in the (001) diffraction plane, which give rise to the diamond-shaped diffuse scattering signals in Fig.~\ref{fig:intro}(e)(g). Upon photoexcitation, the diffuse scattering evolutions along the two orthogonal directions are perfectly overlapped [Fig.~\ref{fig:diff_acoustic}], reaffirming the direction-independent nature of the diffuse signals.

\section{Initial CDW dynamics after photoexcitation}\label{sec:kink}

\begin{figure}[htb!]
	\centering
	\includegraphics[width=0.4\textwidth]{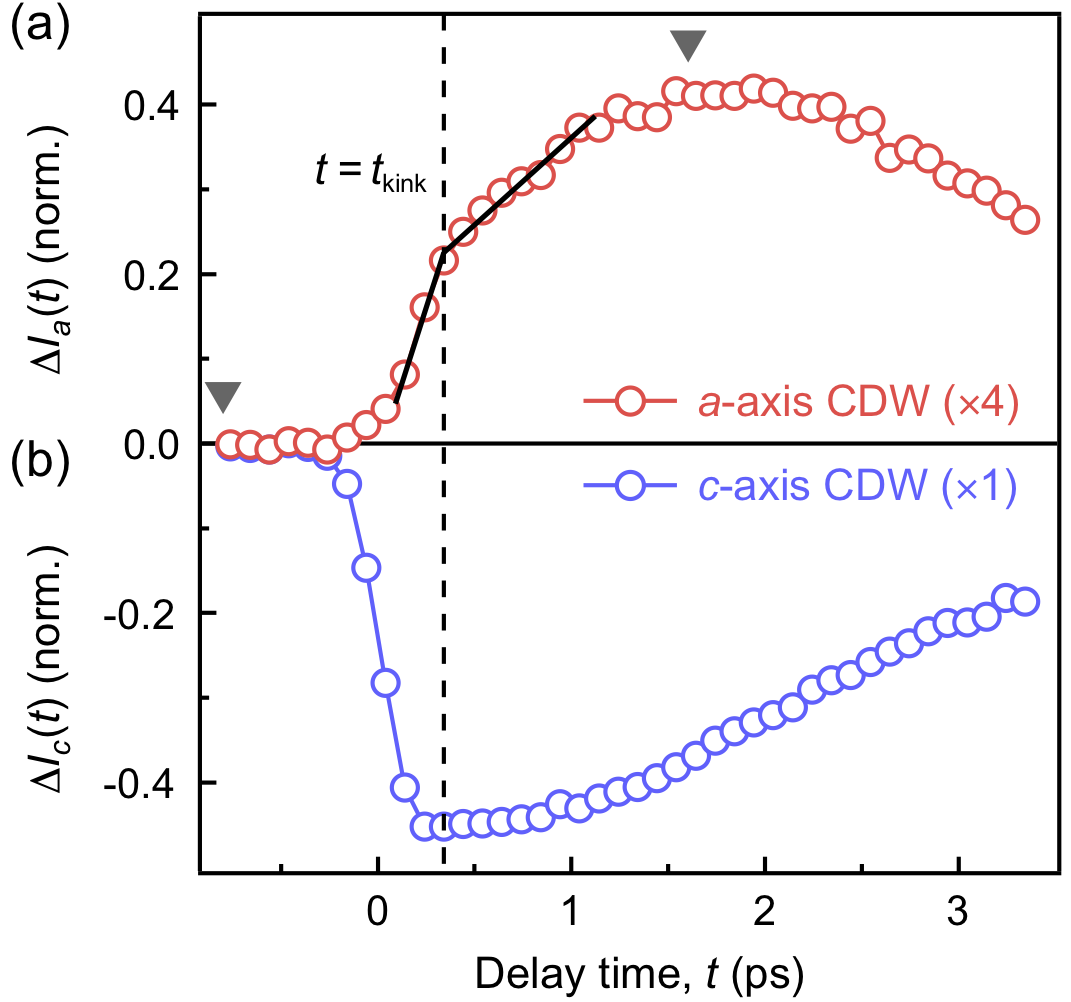}
	\caption{Photoinduced dynamics of $c$-axis and $a$-axis CDWs. (a)~Changes in the integrated intensities of the transient $a$-axis CDW peak, $I_a$. (b)~Similar plot for the equilibrium $c$-axis CDW peak, $I_c$. Peaks in multiple Brillouin zones are averaged for improved signal-to-noise ratio (see Sec.~\ref{sec:transverse}). Both traces are normalized by the average value of $I_c$ before photoexcitation. Triangles indicate the time points for the snapshots in Fig.~\ref{fig:intro}(f)(g). Dashed line marks the kink in the evolution of $I_a$ and solid lines are guides to the eye. Data was taken at 307~K with an incident pump laser fluence of 2.1~mJ/cm$^2$.}
\label{fig:kink}
\end{figure}

In Fig.~\ref{fig:double}(f)(h) of the main text, the initial rise of the light-induced $a$-axis CDW peak appears slower than the drop of the equilibrium $c$-axis CDW peak. Here, we elaborate on their initial fast dynamics and explain the origin of the discrepancy. As shown in Fig.~\ref{fig:kink}, within 350~fs after photoexcitation, the $c$-axis peak is maximally suppressed. The growth of the fluctuation along the $a$-axis is conspicuously slower, peaking at a pump-probe delay of $\sim2$~ps. In addition, its intensity evolution exhibits a kink at $t_\text{kink}=350$~fs (dashed line), corresponding to the time delay when the $c$-axis intensity reaches the minimum. This correspondence is not coincidental and hints at the competing relation between the two CDWs. The rapid growth of fluctuations before $t_\text{kink}$ is facilitated by the suppression of the equilibrium order. After $t_\text{kink}$, there is no further decrease of the $c$-axis CDW and the development of $a$-axis fluctuations slows down. From a microscopic point of view, a vanishing equilibrium CDW restores electronic densities of states to the Fermi level when the corresponding CDW gap closes, providing mobile carriers that can scatter the soft phonons associated with the CDW fluctuations. After $t_\text{kink}$, no more carriers are added to the Fermi level, so the fluctuations do not grow as quickly as before.

\section{Fluence-dependent slowing-down in the fluctuation dynamics}\label{sec:flu}

\begin{figure}[htb!]
	\centering
	\includegraphics[width=0.8\columnwidth]{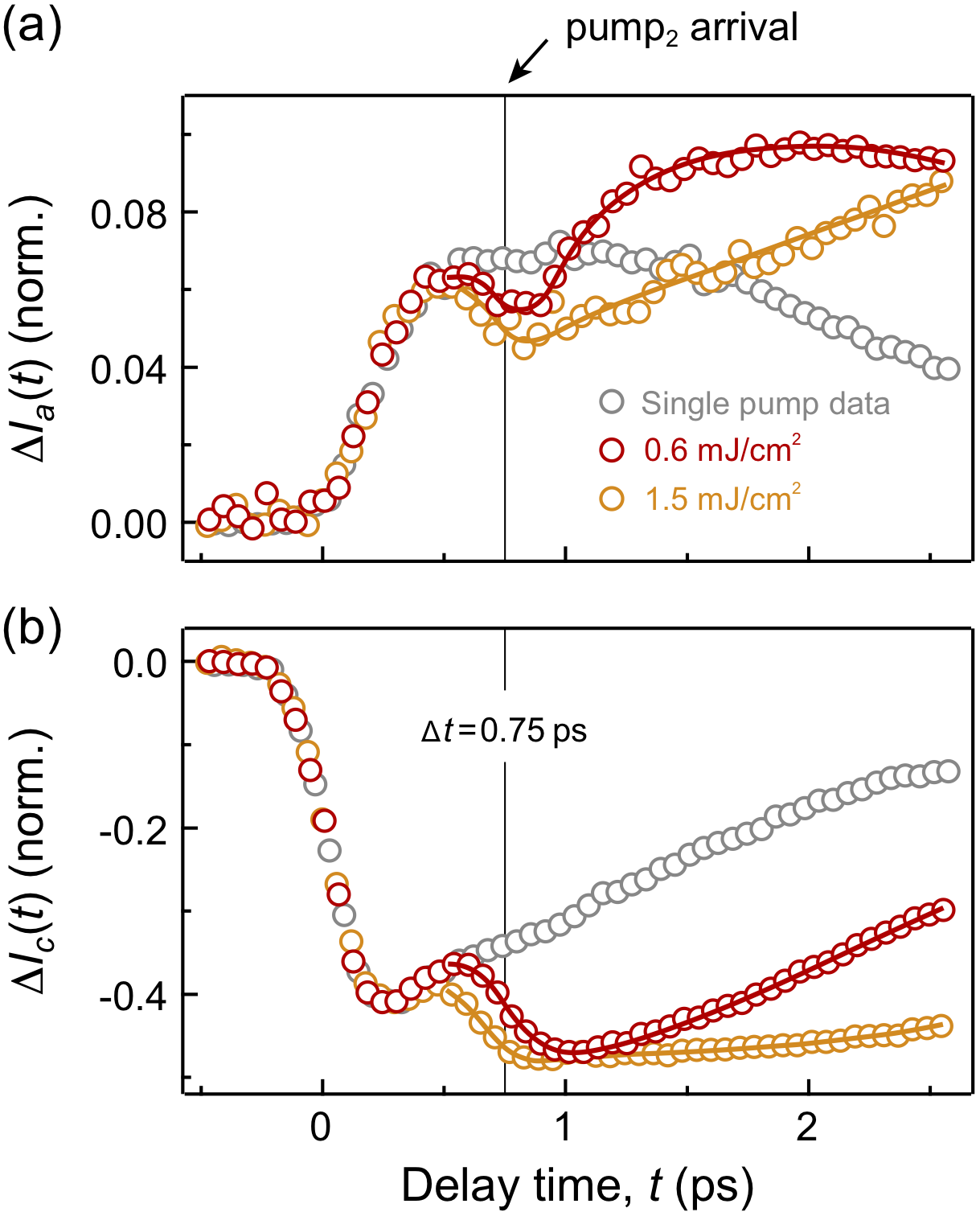}
	\caption{Fluence-dependent growth of fluctuating CDWs. (a)~Changes in the integrated intensities for the transient $a$-axis CDW peaks following photoexcitation by two pulses. (b)~Similar plots for the $c$-axis CDW. For reference, dynamics induced by a single pulse is shown in gray. The arrival time of the second pulse is marked by the vertical line. The incident fluence of pump$_1$ was 1.1~mJ/cm$^2$; various fluences of pump$_2$ were labeled. Traces are normalized by the average value of $I_c$ before photoexcitation. Solid curves are guides to the eye.}
\label{fig:flu}
\end{figure}

A close examination of Fig.~\ref{fig:double}(f) in the main text reveals that the growth of the $a$-axis fluctuation significantly slows down after the incidence of the second pump pulse. To understand the origin of this effect, we repeated the double-pump experiment with varying laser fluences of the second pulse (Fig.~\ref{fig:flu}). As more energy is injected into the system, both $a$- and $c$-axis fluctuations are found to slow down in their joint growth. The slower growth as a function of increasing fluence is captured by our nonperturbative model formulated in Sec.~\ref{sec:theory} [Fig.~\ref{fig:theory}(b)]. An important implication of this slowing-down is that the photoinduced fluctuation would appear to last longer with higher laser fluence \cite{Kogar2020}. Hence, one could in principle use tailored fluence to adjust the lifetime of light-induced fluctuating states.

\section{Nonperturbative calculation of photoinduced fluctuations}\label{sec:theory}

In this section, we explain the time-dependent $O(N)$ model and its nonperturbative solution to the light-induced dynamics of both $a$-axis and $c$-axis CDWs. Without assuming any thermal state of the excited system, our method yields the exact observable in an ultrafast electron diffraction measurement. The benefit of our approach is that it contains only a few model parameters, yet it captures all essential aspects of the equilibrium phase diagram of the material as well as the dynamics in the ultrafast regime. 

\subsection{Ginzburg-Landau free energy and equilibrium phase diagram}

We start by considering the simplest Ginzburg-Landau free energy functional that contains all necessary ingredients for the two competing CDWs,
\begin{align}
 {\cal F}[{\bm\psi}_c,{\bm\psi}_a] = {\cal F}_c[{\bm\psi}_c] + {\cal F}_a[{\bm\psi}_a] + {\cal F}_\text{int}[{\bm\psi}_c,{\bm\psi}_a],\label{eq:F_all}
\end{align}
where ${\cal F}_{i=a,c}[{\bm\psi}_i]$ is the usual Ginzburg-Landau potential
\begin{align}
 {\cal F}_{i}[{\bm\psi}_i] = \int d^d \mathbf{x}~\left[ \frac{r_i}{2} {\bm\psi}_{i}^2 + \frac{K_i}{2} (\nabla {\bm\psi}_{i})^2 + u_i ({\bm\psi}_{i}^2)^2 \right],\label{eq:F_i}
\end{align}
and ${\cal F}_\text{int}[{\bm\psi}_c,{\bm\psi}_a]$ describes the competing interaction
\begin{align}
 {\cal F}_\text{int}[{\bm\psi}_c,{\bm\psi}_a] = 2\eta \int d^d  \mathbf{x} ~ {\bm\psi}_{c}^2{\bm\psi}_{a}^2,\text{~where } \eta >0.
\end{align}
Here, $r_i$, $u_i$, $K_i$, and $\eta$ are the model parameters. ${\bm\psi}_{i=a,c}(\mathbf{x})$ is an $N$-component vector of real numbers, which is used to denote the order parameter for the symmetry-breaking transition \cite{Mazenko1985,Bray1994}
\begin{align}
{\bm\psi}_{i=a,c}(\mathbf{x}) = \left[\psi_{i,1} (\mathbf{x}),\dots, \psi_{i,N} (\mathbf{x})\right].
\end{align}
In our case, $N=2$ for incommensurate CDWs in $R$Te$_3$ and the spatial dimension is $d=3$.

\begin{figure*}[htb!]
	\centering
	\includegraphics[width=1\textwidth]{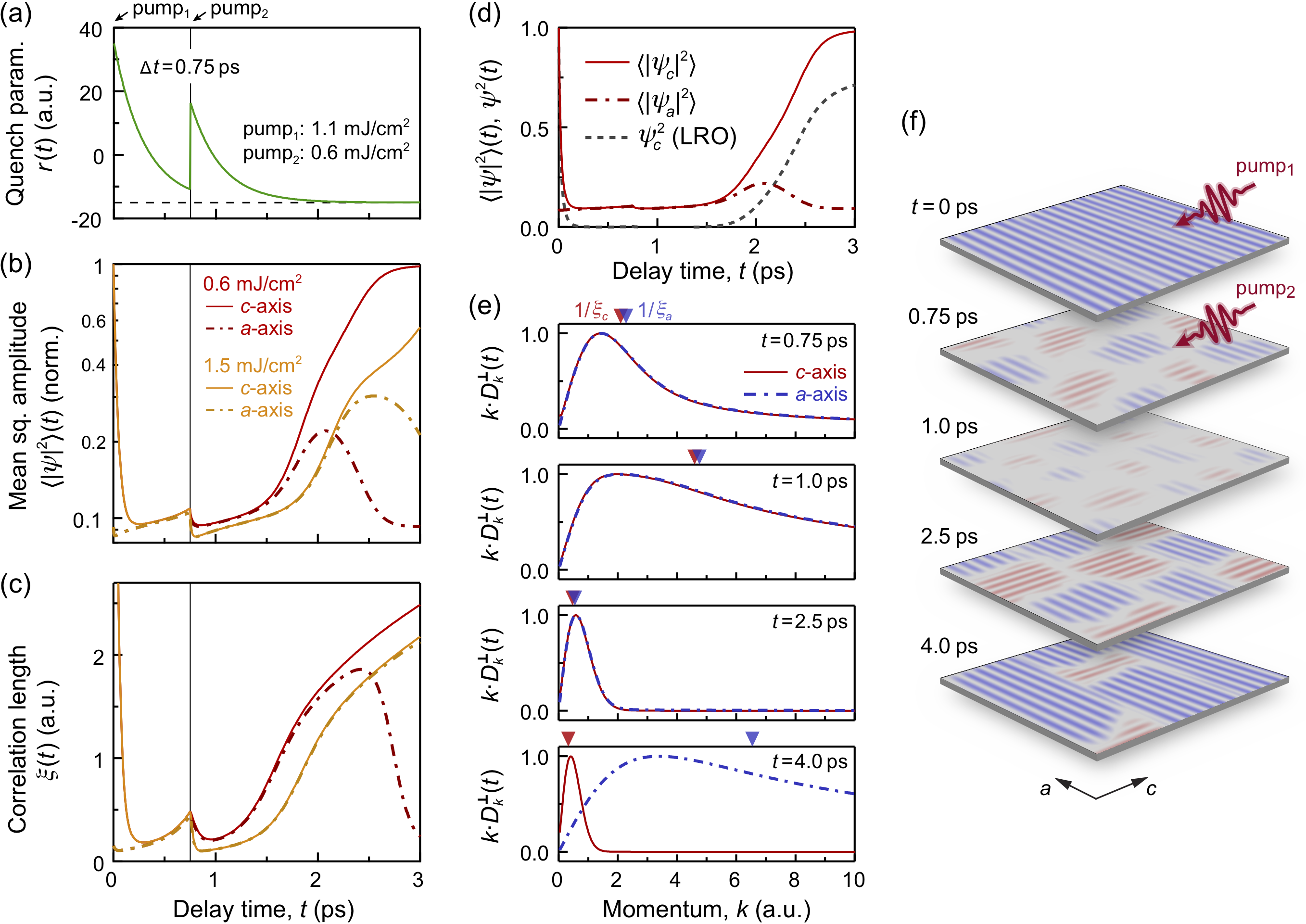}
	\caption{Simulated dynamics of $a$- and $c$-axis CDWs following a double-pump scheme. (a)~Quench protocol for the double-pump experiment, following Eq.~\eqref{eq:quench_double}. The simulated fluence ratio of the first and the second pulse adopts the experimental value of 1.1 and 0.6~mJ/cm$^2$. Horizontal dashed line represents the equilibrium value of $r\equiv(r_a+r_c)/2$. $\tau_\text{QP}$ is taken to be 0.3~ps, obtained from transient reflectivity measurements \cite{Zong2019a}. (b)(c)~Simulated mean square order parameter amplitude $\langle|\psi_{a,c}|^2\rangle(t)$ and correlation length $\xi_{a,c}(t)$ for two different pump$_2$ fluences, matching those in Fig.~\ref{fig:flu}. Values of $\langle|\psi_{a,c}|^2\rangle(t)$ are normalized by $\langle|\psi_c|^2\rangle(0)$. (d)~Evolution of the long-range order component $\psi_c^2(t)$ (gray dashed curve) of the mean square order parameter amplitude $\langle|\psi_c|^2\rangle(t)$. Red curves are the same as those in (b). All curves are normalized by $\langle|\psi_c|^2\rangle(0)$. There is no long-range order for the $a$-axis CDW: $\psi_a^2(t) = 0$ for all $t$. (e)~Momentum distributions of transverse fluctuations at indicated time delays; each curve is individually normalized by its maximum value. $D^\perp_{k,i}(t)$ is the transverse correlation function defined in Eq.~\eqref{eq:D_perp_def}. The extracted inverse correlation lengths, $1/\xi_{a,c}$, are indicated by the triangles. The simulation parameters used are the same as the red curves in (c). (f)~Schematic evolution of CDWs upon photoexcitation, illustrating the time-dependent correlation lengths in (e).}
\label{fig:theory}
\end{figure*}

The  mean-field  phase  diagram of the model in Eq.~\eqref{eq:F_all} agrees with the phenomenology of rare-earth tritellurides. Specifically, this model has two qualitatively different regimes \cite{Chaikin1995}: (i)~bicritical regime ($u_au_c< \eta^2$) corresponds to a situation when only the dominant order, $\psi_c$, can develop in equilibrium; (ii)~tetracritical regime ($u_au_c> \eta^2$) represents the case when the two order parameters can coexist. We focus on the former situation for LaTe$_3$ and further impose $u_a=u_c=u$ and $K_a=K_c=K$ because the anisotropy between the orthogonal CDWs is small. The equilibrium asymmetry between the two CDWs lies in a slightly smaller value of $r_c$ compared to $r_a$, so the CDW always occurs along the $c$-axis. 

\subsection{Photoexcitation protocol}

Photoexcitation is modeled by an impulsive change to the phenomenological parameters $r_a$ and $r_c$ in Eq.~\eqref{eq:F_i}. If we define $r\equiv (r_c+r_a)/2$ and $g\equiv(r_c-r_a)/2$, the quench protocol is
\begin{align}\label{eq:quench_double}
r(t) = r_0 &+ \Theta(t-t_1) e^{-(t-t_1)/\tau_\text{QP}}F_1 \notag\\
&+ \Theta(t-t_2)e^{-(t-t_2)/\tau_\text{QP}}F_2,
\end{align}
where $r_0$ is the equilibrium value; $t_{1,2}$ and $F_{1,2}$ denote the arrival times and fluences of the two excitation pulses; for single-pump measurement, we set $F_{2} =0$. $g$ is assumed to be constant throughout the temporal evolution to minimize the number of adjustable parameters and to mirror the equilibrium phase diagram of a bicritical regime. $\tau_\text{QP}$ represents the typical relaxation timescale of excited quasiparticles, which are responsible for the modification of the free energy potential. A representative $r(t)$ is shown in Fig.~\ref{fig:theory}(a) for a double-pump pulse sequence.

\subsection{Equations of motion for long-range order and fluctuations}

Here, we explain the equations of motion for the long-range order and order parameter fluctuations corresponding to the free energy functional $ {\cal F}[{\bm\psi}_c,{\bm\psi}_a]$ in Eq.~\eqref{eq:F_all}. To describe the photoinduced evolution, we assume overdamped order parameter dynamics (model-A \cite{Hohenberg1977}) and follow the formalism in ref.~\cite{Dolgirev2020b}. We neglected the coherent oscillatory dynamics associated with the phononic degree of freedom. The overdamped dynamics is expected to be reliable in describing long-time dynamics after a strong laser pulse, which results in a proliferation of low-energy, low-momenta order parameter collective modes. Adding the phonon contribution will better capture the short-time dynamics, especially the initial response time \cite{Zong2019b}. 

Without loss of generality, we assume that spontaneous symmetry breaking occurs along the first component of the order parameter, $\psi_{i,1}$. In the temporal evolution after photoexcitation, the presence of long-range order is hence represented by a nonzero expectation value 
\begin{align}
\psi_i(t) \equiv \langle\psi_{i,1}(t)\rangle. 
\end{align}
Irrespective of whether long-range order exists, the order parameter can also exhibit transverse and longitudinal fluctuations, represented by the transient equal-time two-point correlation functions \cite{Dolgirev2020b},
\begin{align}
D^\parallel_{\mathbf{k},i}(t) &\equiv \langle \psi_{i,1}(-\mathbf{k};t)\,\psi_{i,1}(\mathbf{k};t)  \rangle_c, \label{eq:D_para_def}\\
D^\perp_{\mathbf{k},i}(t) &\equiv \langle \psi_{i,\alpha\neq1}(-\mathbf{k};t)\,\psi_{i,\alpha\neq1}(\mathbf{k};t)  \rangle_c. \label{eq:D_perp_def}
\end{align}
Here, $D^\parallel_{\mathbf{k},i}(t)$ describes the Higgs modes while $D^\perp_{\mathbf{k},i}(t)$ represents the Goldstone modes. The evolutions of $\psi_i(t)$, $D^\parallel_{\mathbf{k},i}(t)$, and $D^\perp_{\mathbf{k},i}(t)$ can be numerically solved following their respective equations of motion:
\begin{align}
\frac{d \psi_i(t)}{dt} &= -\Gamma_i\,   r_{{\rm eff}, i}\, \psi_i \label{eq:dt_phi},\\
\frac{d D^{\perp}_{{\bf k},i}(t)}{dt} &= 2T\Gamma_i - 2\Gamma_i (K  {\bf k}^2 + r_{{\rm eff}, i})D^{\perp}_{{\bf k},i},\label{eq:dD_perp}\\
\frac{d D^{\parallel}_{{\bf k},i}(t)}{dt} &= 2T\Gamma_i - 2\Gamma_i (K  {\bf k}^2 + r_{{\rm eff}, i} + 8u\psi_i^2)D^{\parallel}_{{\bf k},i},\label{eq:dD_par}
\end{align}
where $T$ is the temperature of a phononic bath that provides thermalization of the excited system, and is assumed to be unchanged after photoexcitation. $\Gamma_i$ is a measure of the relaxation rate of the corresponding long-range order. Given the photoinduced symmetry between the two CDWs in LaTe$_3$ and DyTe$_3$, we assume that $\Gamma_a = \Gamma_c = \Gamma$ are the same for the two density waves. In principle, one is free to impose different $\Gamma$ values for different order parameters, leading to long-lasting metastable states that are not in the global free-energy minimum \cite{Sun2020}. $r_{{\rm eff}, i}$ is a self-consistent ``mass'' term, defined as
\begin{align}
r_{{\rm eff},a}(t)  &= 4u \Big[ \psi_a^2 + \int \frac{d^3 {\bf k}}{(2\pi)^3}( D^{\parallel}_{{\bf k},a} + (N-1)D^{\perp}_{{\bf k},a})\Big] \notag\\
&+ 4\eta\Big[ \psi_c^2 + \int \frac{d^3 {\bf k}}{(2\pi)^3}( D^{\parallel}_{{\bf k},c} + (N-1)D^{\perp}_{{\bf k},c})\Big]\notag\\
&+ r_a(t),
\end{align}
and a symmetric expression holds for $r_{{\rm eff},c}$.

\begin{figure*}[htb!]
	\centering
	\includegraphics[width=0.8\textwidth]{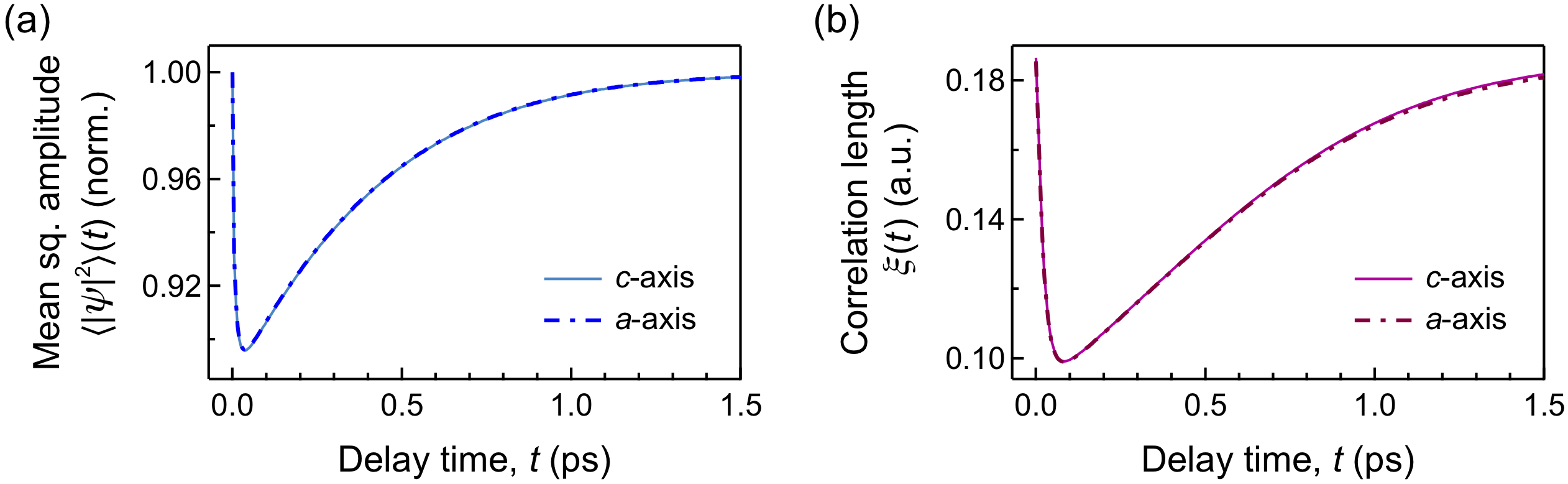}
	\caption{Simulated dynamics of fluctuating CDWs above {\em T}$_{\text{\em c}}$. (a)~Temporal evolutions of $\langle|\psi_{i}|^2\rangle(t)$ along the $a$- and $c$-axis when the initial state has no long-range order. The curves are normalized by values at $t=0$. (b)~Similar plots for $\xi_{i}(t)$. The CDW instabilities along the two axes show nearly identical dynamics.}
\label{fig:above_tc}
\end{figure*}

We make a few remarks on Eqs.~\eqref{eq:dt_phi}--\eqref{eq:dD_par}. In equilibrium, there is no long-range order for the $a$-axis CDW ($\psi_a=0$), so it remains zero throughout the time evolution as dictated by Eq.~\eqref{eq:dt_phi}. It also implies that the dynamics of $D^{\perp}_{{\bf k},a}(t)$ and $D^{\parallel}_{{\bf k},a}(t)$ are identical because there is no distinction between transverse and longitudinal fluctuations in the absence of any symmetry breaking along the $a$-axis.

\subsection{Intensity evolutions of \textit{c}- and \textit{a}-axis CDW peaks}

In electron diffraction, the integrated intensity surrounding the CDW peaks measures the mean square order parameter amplitude at each pump-probe delay $t$, averaged over the probed sample volume. We denote this observable by
\begin{align}
\langle |\psi_i|^2 \rangle (t) &\equiv \int d^3\mathbf{x}~\left[\langle\psi_{i,1}^2(\mathbf{x},t)\rangle + \langle\psi_{i,2}^2(\mathbf{x},t)\rangle\right] \\
 &=\underbrace{\psi_{i}^2(t)}_\text{LRO} + \underbrace{\int\frac{d^3\mathbf{k}}{(2\pi)^3} \left[D^\parallel_{\mathbf{k},i}(t) + D^\perp_{\mathbf{k},i}(t) \right]}_\text{Fluctuations},\label{eq:mspsi}
\end{align}
where the second line follows from the Fourier transform. As labeled in Eq.~\eqref{eq:mspsi}, the integrated intensity contains contributions from both the long-range order and the fluctuations. This statement is true for the dominant $c$-axis CDW. However, as explained above, Eq.~\eqref{eq:dt_phi} imposes that the long-range order is always absent in the subdominant $a$-axis CDW throughout the temporal evolution. Therefore, the diffraction intensity in the $a$-axis satellite is an exclusive marker for the order parameter fluctuations.

In Fig.~\ref{fig:sim}(a) of the main text, we presented the evolutions of $I_{i=a,c}(t)\equiv \langle |\psi_i|^2 \rangle (t)$ following a single pump. Here, we also explore their dynamics after a double-pump sequence, mimicking the experimental scheme in Fig.~\ref{fig:double}(f)(h). After the first photoexcitation event, $\langle |\psi_c|^2 \rangle (t)$ decreases significantly due to the loss of long-range order $\psi_c^2(t)$, which is shown separately in Fig.~\ref{fig:theory}(d) (gray dashed curve). As the long-range $c$-axis CDW is suppressed, short-range fluctuations of both CDWs start to develop in an indistinguishable manner. The effect of the second pump is to transiently suppress the codeveloping fluctuations, which swiftly rebound and grow over the next picosecond. As the fluence of the second pump increases, the joint growth of both $\langle |\psi_a|^2 \rangle (t)$ and $\langle |\psi_c|^2 \rangle (t)$ slows down, in agreement with experimental observations in Fig.~\ref{fig:flu}. The mean square amplitudes of the two CDWs start to diverge once the long-range order $\psi_c^2(t)$ starts to recover from zero [Fig.~\ref{fig:theory}(d)].

We can use the same formalism to study the dynamics when the initial state lacks any long-range $c$-axis CDW, such as in DyTe$_3$ near its $T_c$. Figure~\ref{fig:above_tc}(a) shows the evolutions of mean square order parameter amplitude for the fluctuating CDWs above $T_c$, mimicking the dynamics measured experimentally in Fig.~\ref{fig:double}(i). In particular, there is no observable distinction between the two competing orders because long-range order never sets in during the nonequilibrium evolution.

\subsection{Evolution of correlation lengths}

We can further infer the correlation length of fluctuating CDW patches from the transverse correlation function $D^\perp_{\mathbf{k},i}(t)$. In equilibrium, the corresponding transverse correlation length increases as one approaches $T_c$ from above and remains divergent in the symmetry-broken phase, as expected in a second-order phase transition \cite{Dolgirev2020b}. In Fig.~\ref{fig:theory}(e), $kD^\perp_{k,i}(t)$ are shown for several representative time delays, corresponding to the simulated dynamics in Fig.~\ref{fig:theory}(b). Right after the second pulse, the distribution shifts to higher momenta, indicating a reduction of the CDW correlation length in the real space ($t=1.0$~ps). As both CDWs develop in amplitude, their correlation lengths also grow in a symmetric manner ($t=2.5$~ps). Finally, as long-range order along the $c$-axis develops, the correlation functions of the two CDWs depart and the $a$-axis order loses its phase coherence ($t=4.0$~ps). 

\begin{figure*}[htb!]
	\centering
	\includegraphics[width=0.7\textwidth]{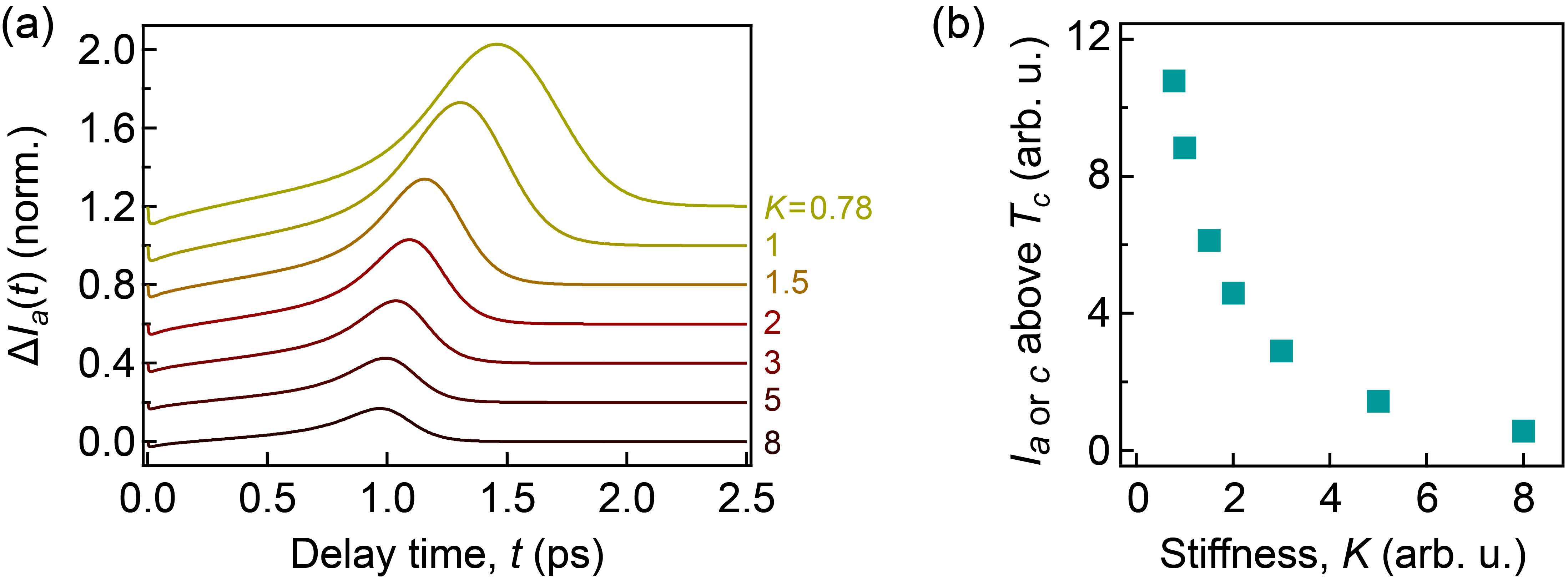}
	\caption{Variation of photoinduced and equilibrium fluctuations due to stiffness of the order parameter. (a)~Simulated change in the $a$-axis CDW peak intensity upon photoexciting the equilibrium long-range $c$-axis CDW. Values are normalized by pre-photoexcitation value and curves are vertically offset by 0.2 for clarity. Each curve corresponds to a stiffness constant, $K$. Here, $I_a(t)\equiv \langle |\psi_a|^2\rangle(t)$. The curves shown in Fig.~\ref{fig:sim}(a) are plotted with $K=1$. (b)~Calculated diffuse CDW peak intensity at a fixed temperature above $T_c$ for different stiffness parameters, $K$. As there is no long-range order, diffuse scatterings along $a$- and $c$-axis are indistinguishable.}
    \label{fig:stiff}
\end{figure*}

From the correlation function $D^\perp_{\mathbf{k},i}(t)$, we can extract a correlation length $\xi_i$, whose reciprocal is indicated by the triangles in Fig.~\ref{fig:theory}(e). In the simulation, $1/\xi_i$ is taken as the centroid of $D^\perp_{\mathbf{k},i}$ up to an ultraviolet cutoff, though the exact definition is not important. The qualitative change in the correlation lengths of the respective CDWs is summarized by the schematic in Fig.~\ref{fig:theory}(f); the quantitative change of $\xi_{a,c}(t)$ is captured in Fig.~\ref{fig:theory}(c). In general, photoexcitation decreases the correlation length; the amount of suppression positively scales with the fluence. The recovery of the correlation lengths are predicted to follow a scaling law at long time delays \cite{Dolgirev2020b}, hence a longer recovery time is expected for a larger suppression.

\subsection{Correlation between photoinduced {\em a}-axis CDW and equilibrium fluctuations}

To study how the equilibrium fluctuations affect the light-induced $a$-axis CDW, we vary the stiffness parameter, $K_i$ in Eq.~\eqref{eq:F_i}, to adjust the strength of the equilibrium fluctuations. Recall that due to the symmetry between the $a$- and $c$-axes, we set $K_a=K_c=K$, so fluctuations in both axes are equally affected. A small stiffness implies a large deviation from the mean-field behavior near $T_c$ and hence significant order parameter fluctuations are expected. Indeed, at a fixed temperature above $T_c$, Fig.~\ref{fig:stiff}(b) shows that smaller stiffness leads to a larger CDW diffuse scattering intensity in our calculation. Turning to the nonequilibrium situation, we plot in Fig.~\ref{fig:stiff}(a) the evolutions of diffraction intensity change for the light-induced $a$-axis CDW at several stiffness values at a fixed temperature below $T_c$. For comparison, these traces are normalized by their equilibrium values before photoexcitation, and are vertically displaced for clarity. As stiffness decreases, the light-induced order strengthens, as summarized in Fig.~\ref{fig:sim}(b). This relationship hints that the transient CDW order is closely tied to the degree of order parameter fluctuations in equilibrium, as we discussed in the main text.

\subsection{Codevelopment of competing orders}\label{sec:codevelop}

An important experimental observation is the codevelopment of competing orders in LaTe$_3$, best illustrated by the dynamics following the second pump pulse [Fig.~\ref{fig:double}(j)]. This is very well captured by our simulation in Fig.~\ref{fig:theory}(b)(c). The timespan of this stage of concurrent growth increases with the pump fluence (Fig.~\ref{fig:flu}). A corollary is that the lifetime of the transient CDW would appear to be prolonged by an increasing pulse fluence. Therefore, a long-lasting photoinduced order does not necessarily imply a protracted absence of the competing order; rather, both could experience a slowing-down in their dynamics.

The observed transient symmetry between the two CDWs also distinguishes our theoretical model from the previous work in ref.~\cite{Sun2020}, where a similar model-A-based approach has been utilized to investigate the nonequilibrium dynamics of two competing orders. In ref.~\cite{Sun2020}, it is assumed that the two orders have very different relaxation rates [$\Gamma_i$ in Eq.~\eqref{eq:dt_phi}], giving rise to the possibility of transient development of a metastable subdominant order at the cost of suppressing the dominant one. By contrast, here we encounter a regime where both order parameters have similar relaxation timescales, resulting in the intriguing effect of a concurrent development after photoexcitation.

\section{Fluctuations in the dominant CDW}\label{sec:c_fluctuate}

We argued in the main text that the photoinduced $a$-axis CDW lacks long-range order and consists solely of order parameter fluctuations. Given the symmetry between the $a$- and $c$-axes, one would expect large photoinduced fluctuations in the dominant CDW as well. Even though it is experimentally challenging to isolate the fluctuation dynamics in the $c$-axis CDW peak given the dominant photoinduced intensity change from the collapse of long-range order, there is experimental evidence that hints at the existence of its fluctuations. For example, time-resolved X-ray diffraction detects no appreciable broadening of the $c$-axis peak upon photoexcitation \cite{Moore2016,Trigo2019}. By contrast, time-resolved electron diffraction reports a significant increase in the peak width \cite{Zong2019a, Zhou2021}. The differences between X-ray and electron scattering are twofold. The former has much superior momentum resolution, so the resolution-limited peak in X-ray diffraction only captures contribution from the long-range order. In $R$Te$_3$, the long-range order peak can be as sharp as less than $3.5\times10^{-4}$~\AA$^{-1}$ in width \cite{Ru2008}, at least 100 times narrower than typical diffuse peaks \cite{Holt2001}. In the presence of the long-range order peak, the weak intensity and large momentum spread hence make diffuse peaks hard to detect in high-resolution X-ray diffraction. On the other hand, the cross section of electron diffraction at comparable energy is orders-of-magnitude larger \cite{Glaeser1985}, so it is more sensitive to the weak diffuse background. Therefore, the apparent contradiction between time-resolved X-ray and electron diffraction can be interpreted as evidence for photoinduced order parameter fluctuations. They do not affect the width of the long-range order peak but would broaden the diffuse scattering peak when CDW correlation length decreases after photoexcitation, as correctly predicted by our theoretical simulation [Fig.~\ref{fig:theory}(c)].

Another piece of evidence that hints at the proliferation of $c$-axis fluctuations is the persistence of satellite diffraction signal even at very high fluence, which is expected to completely suppress the long-range $c$-axis CDW [Figs.~\ref{fig:intro}(g), \ref{fig:double}(h), \ref{fig:kink}(b), and \ref{fig:flu}(b)] \cite{Moore2016,Trigo2019,Kogar2020}. It was previously thought that a mismatch between photon-pumped and electron-probed volumes contributes to the residual intensity at $\mathbf{q}_c$. Our findings suggest that at least part of the observed satellite intensity under strong photoexcitation originates from enhanced fluctuations. Similar to a nonvanishing residual intensity in time-resolved diffraction, the CDW gap measured in time- and angle-resolved photoemission experiments also seems to persist at high fluence \cite{Rettig2016,Zong2019a}. The finite gap size was interpreted as transient improvement of the nesting condition of the Fermi surface \cite{Rettig2016}. In light of our analysis, the residual gap could be caused by short-range $c$-axis fluctuations induced by photoexcitation, resembling the pseudogap found in equilibrium \cite{Yokoya2005,Chatterjee2015}.

\section{Comparison of photoexcited and critical states}\label{sec:compare}

In a photoexcited state where long-range order transiently vanishes, we have shown that the system and its dynamical response closely resemble the critical state near $T_c$, which is earmarked by nearly identical CDW fluctuations along both the $a$- and $c$-axis.

\begin{figure}[htb!]
	\centering
	\includegraphics[width=0.45\textwidth]{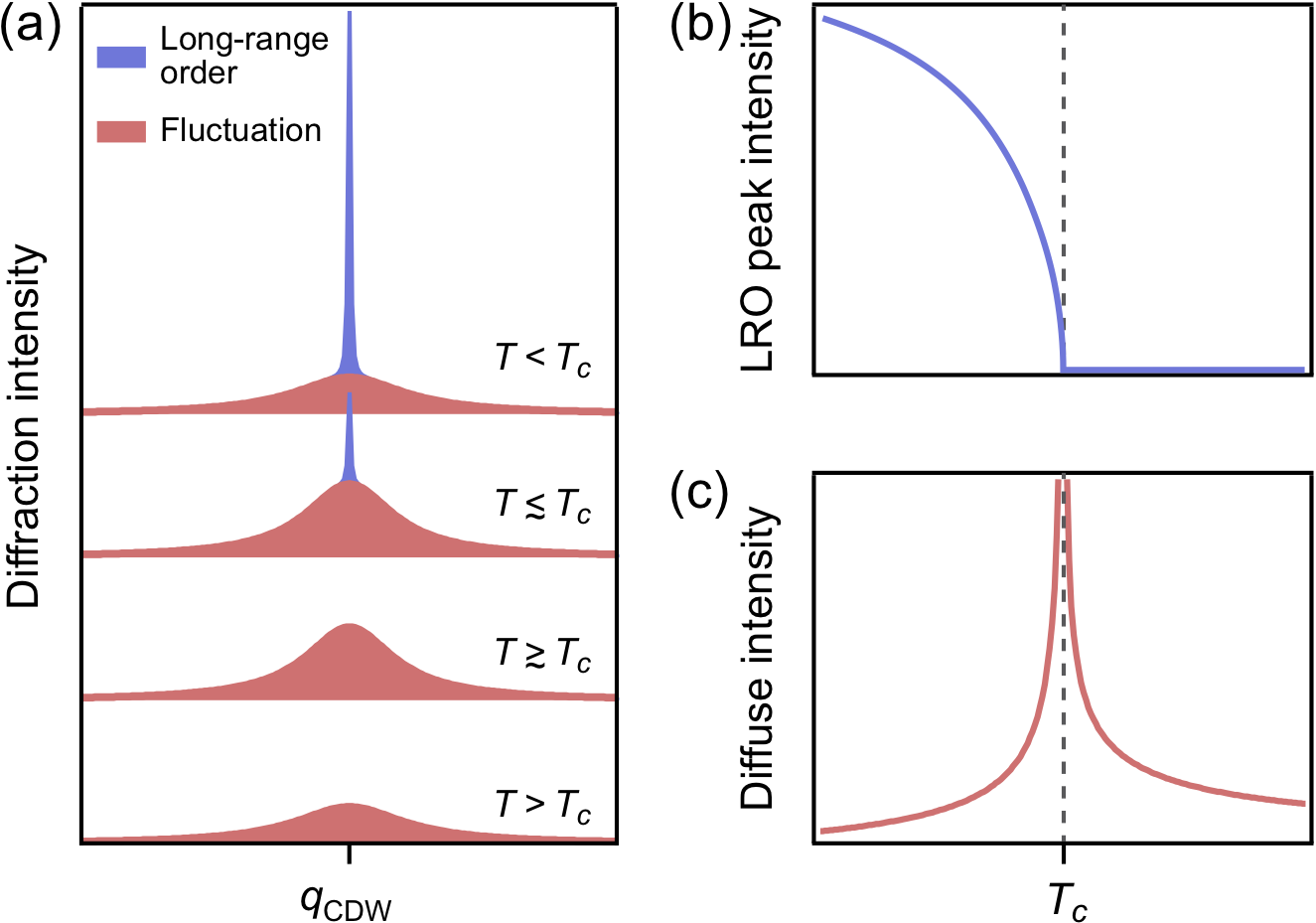}
	\caption{Evolution of long-range order and order parameter fluctuations across an equilibrium phase transition due to spontaneous symmetry breaking. (a)~Schematic intensity profile of a CDW satellite peak near $T_c$, showing a resolution-limited long-range order peak (blue) and a diffuse peak due to critical fluctuations (red). The relative peak width and height are exaggerated for illustrative purposes. (b)(c)~Schematic intensity evolutions of the sharp long-range order peak (b) and the diffuse scattering (c) across $T_c$.}
    \label{fig:lro_sro}
\end{figure}

\begin{figure}[htb!]
	\centering
	\includegraphics[width=0.48\textwidth]{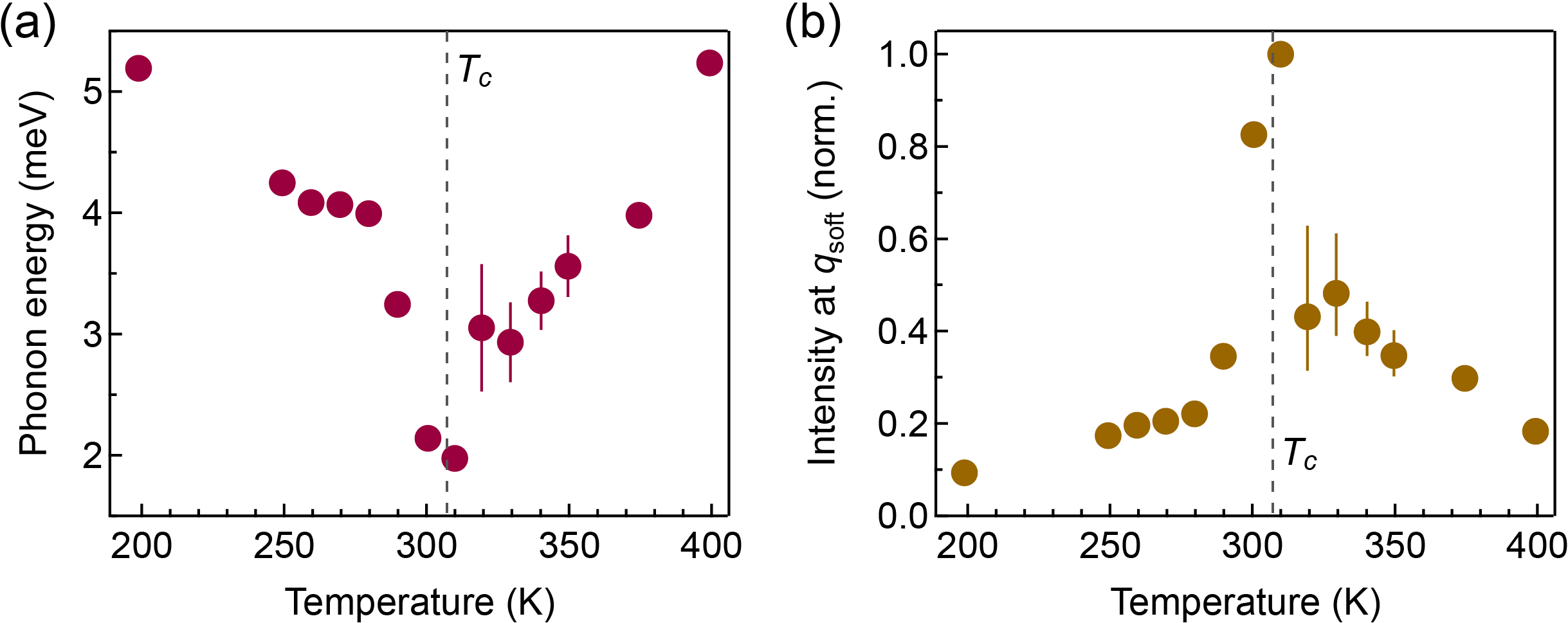}
	\caption{Equilibrium diffuse peak intensity in DyTe$_3$. (a)~Temperature dependent energy of the soft phonon for the $a$-axis CDW obtained from inelastic X-ray scattering by tracking the diffuse peak at $q_{\text{soft}} = (0.32,~7,~3)$. Adapted from ref.~\cite{Maschek2018}. (b)~Calculated diffuse peak intensity at $q_{\text{soft}}$ assuming a thermal population of the soft phonons. The diffuse $a$-axis CDW peak is the most intense at $T_c$. Intensity values are normalized so that the maximum is 1.}
    \label{fig:phonon_freq}
\end{figure}

The analogy between photoexcited and critical state motivates a connection between the respective control knob: pump laser fluence and temperature. In the $O(N)$ model, both quantities are used to adjust the phenomenological parameter $r\equiv (r_c+r_a)/2$ [see Eq.~\eqref{eq:F_i}], which determines whether the potential energy has minima away from the origin. However, the effect of a larger fluence should not be confused with a higher temperature. In equilibrium, critical fluctuations only develop close to $T_c$ and subside at either higher or lower temperatures. This behavior is summarized by the schematic of temperature-dependent long-range order and diffuse peaks in a typical charge density wave compound in Fig.~\ref{fig:lro_sro}. By leveraging previous results from inelastic X-ray scattering experiments \cite{Maschek2018}, we can explicitly compute the intensity of the diffuse peaks due to the CDW fluctuations. As shown in Fig.~\ref{fig:phonon_freq}, the energy of the soft phonon along the $a$-axis reaches a minimum at $T_c$. Correspondingly, the diffuse spot intensity is the most pronounced, calculated using the formalism in Sec.~\ref{sec:kohn}. Either above or below $T_c$, the soft phonon frequency increases significantly, leading to suppressed diffuse intensities away from $T_c$ as well.

The picture is different if we replace temperature with pump laser fluence. As long as sufficient fluence is provided to suppress the long-range order, order parameter fluctuations will continue to grow with increasing fluence until a saturation point is reached, which is set by the electron density near the Fermi level [Figs.~\ref{fig:flu}(a) and \ref{fig:theory}(b)]. This distinction can be understood by the different shapes of the free energy profile at $T_c$ and after photoexcitation. The former features a flat profile, which allows thermal fluctuations to develop. Away from $T_c$, the flat potential disappears, accounting for the weakened fluctuations. On the other hand, a photoexcited state has a fast evolving energy landscape. The amount of light-induced fluctuations are not tied to a specific shape of the free energy at a particular point in the temporal evolution. Instead, they are determined by the initial change in the free energy, which is set by the pump laser fluence.

The different origins of order parameter fluctuations in and out of equilibrium also dictate their different correlation lengths. Our calculation shows that the correlation of photoinduced fluctuations remains short-range. Larger laser fluence or an additional pump pulse only serves to further decrease the correlation length [Fig.~\ref{fig:theory}(c)]. In sharp contrast, correlation length diverges near $T_c$, which follows the universal scaling law that governs the phase transition. The notion of universality is, however, not restricted to equilibrium transitions. The close resemblance between photoexcited state and critical state also encourages one to search for universal scaling behavior in systems far from equilibrium. Theoretical simulations have made several concrete predictions of time- and momentum-scalings for light-induced order parameter fluctuations \cite{Dolgirev2020b}, which can be tested in future experiments.

\section{Experimental indicators of order parameter fluctuations}\label{sec:fluc_exp}

Based on the positive correlation between equilibrium and photoinduced fluctuations in Figs.~\ref{fig:sim} and \ref{fig:stiff}, we suggested in the main text that one should look for material classes with strong order parameter fluctuations in equilibrium as candidate platforms to search for ``hidden'' photoinduced states. In our theoretical model, the amount of fluctuations is tuned by the stiffness parameter, $K$, in Eq.~\eqref{eq:F_i}. Here, we provide some experimental guidance in characterizing the degree of fluctuations in equilibrium, which depends on the nature of the order parameter in question.

In superconductors, a common method to characterize the strength of superconducting fluctuations is the Nernst signal above $T_c$, which is used to study, for example, underdoped cuprates \cite{Xu2000} and amorphous superconducting films \cite{Pourret2006}. More recently, the Nernst effect is also evidenced in quasi-2D organic molecular metals $\kappa$-(BEDT-TTF)$_2X$ \cite{Nam2007,Nam2013}, which are found to exhibit light-induced superconductivity above their equilibrium $T_c$ \cite{Buzzi2020}. By comparing different members in this organic metal family, Buzzi and coworkers revealed that the light-induced superconductivity does not exist unless the Nernst signal is present above $T_c$ in equilibrium \cite{Buzzi2021}. This finding suggests that superconducting fluctuations in equilibrium are strong indicators of the photoinduced superconductivity in this compound, lending further support to our outlook of searching for exotic states of matter in a nonequilibrium setting.

We can further quantify the degree of fluctuations in a superconductor by comparing a few temperature scales. The fluctuations above $T_c$ represent pre-formed Cooper pairs that do not have long-range phase coherence. In a BCS-like three-dimensional superconductor with low disorder, the Ginzburg criterion dictates that fluctuations are not experimentally accessible because they are only present very close to $T_c$. For low-dimensional, disordered or non-BCS superconductors, strong fluctuations can exist above $T_c$. One practical indicator for the fluctuation strength is the ratio of $T_{\text{pair}}/T_c$, where $T_{\text{pair}}$ is the Cooper-pairing temperature and $T_c$ is the zero resistance temperature \cite{He2021}. Experimentally, $T_{\text{pair}}$ can be extracted from the spectral gap opening temperature in angle-resolved photoemission spectroscopy (ARPES) or it can be approximated by the mean-field pairing temperature from specific heat measurements \cite{Tallon2011}. A higher $T_{\text{pair}}/T_c$ ratio is therefore a strong hint for larger superconducting fluctuations.

In a charge density wave compound, the fluctuations manifest as diffuse scatterings due to the soft phonons at temperatures well above $T_c$, such as in $R$Te$_3$ studied in the present work, or other compounds including 1$T$-TiSe$_2$ \cite{Holt2001}, 2$H$-NbSe$_2$ \cite{Weber2011}, and ZrTe$_3$ \cite{Hoesch2009}. These CDW materials typically possess a quasi-1D or quasi-2D structure, where the reduced dimensionality leads to significant fluctuations in a regime above $T_c$ and below the mean-field transition temperature \cite{Gruner1994}. In this temperature range, short-range CDWs form along 1D chains or within 2D planes, but chains or planes are not strongly coupled, preventing the formation of long-range order. These short-range CDWs give rise to diffuse peaks in the diffraction pattern, whose intensity signifies the strength of fluctuations. Besides diffraction probes, these CDW fluctuations can also be detected in ARPES, where a suppression in the density of states at the Fermi level is observed even above $T_c$ \cite{Chatterjee2015}. The amount of suppression is a direct indicator of the strength of the CDW fluctuations.

For magnetically ordered systems, nuclear magnetic resonance (NMR) is a sensitive probe of the spin fluctuations, which modify the spin-lattice relaxation rate. This technique has been extensively applied to investigate, for example, the spin fluctuations in iron-based superconductors \cite{Ma2013}. For an antiferromagnet, one can also adopt a similar guideline used in CDW compounds, where antiferromagnetic fluctuations above the transition temperature are detected through the diffuse peak in magnetic neutron diffraction or X-ray scattering.\\

\section{Underlying high-energy symmetries}\label{sec:underlying}

In the main text, we envisioned that the parallels between a photoexcited state and a critical state provides a viable pathway to attain hidden symmetries of a system. Indeed, spontaneous symmetry breaking is a low-energy property of a system. When the material is excited by a strong laser pulse, the system transiently gains access to its high-energy symmetries without passing through the transition temperature. If a crystal melts or disintegrates before its equilibrium $T_c$ is reached, photoexcitation provides an alternative passage to access the high-symmetry regime. For $R$Te$_3$, the underlying symmetry between its $a$- and $c$-axis CDW is recovered: Given the experimental resolution, the two CDW fluctuations cannot be distinguished by their wavevector, diffuse intensity, or temporal evolution. It is worth noting that the unit cell of $R$Te$_3$ has a weak anisotropy between the two in-plane axes and it lacks $C_4$ symmetry due to a glide plane [Fig.~\ref{fig:intro}(a)]. The restored fourfold symmetry is hence exclusive to the CDWs, whose enhanced fluctuations overcome the anisotropy of the lattice structure.

An enticing prospect of nonthermal access to the underlying high-energy state is the discovery of symmetries corresponding to some hidden order parameters, which could in principle be manipulated and investigated with pump-probe experiments. In particular, in systems that harbor competing phases, an ultrafast pulse could be a promising route to unveil subdominant order that is suppressed in equilibrium. This mechanism has been proposed for light-induced superconductivity at the expense of CDWs in cuprates \cite{Kaiser2017} and a light-induced ferromagnetic metal out of an antiferromagnetic insulator in strained manganites \cite{Zhang2016,McLeod2020}. As experimental techniques progress, we could also test ideas like the $SO(5)$ theory of high-$T_c$ superconductivity by observing photoinduced fluctuations of both $d$-wave order parameter and antiferromagnetism~\cite{Zhang1997, Demler2004}. Ultimately, an ultrafast laser pulse would allow us to construct a ``nonequilibrium phase diagram'', which contains hidden symmetries that are otherwise inaccessible.


\begin{thebibliography}{75}%
\makeatletter
\providecommand \@ifxundefined [1]{%
 \@ifx{#1\undefined}
}%
\providecommand \@ifnum [1]{%
 \ifnum #1\expandafter \@firstoftwo
 \else \expandafter \@secondoftwo
 \fi
}%
\providecommand \@ifx [1]{%
 \ifx #1\expandafter \@firstoftwo
 \else \expandafter \@secondoftwo
 \fi
}%
\providecommand \natexlab [1]{#1}%
\providecommand \enquote  [1]{``#1''}%
\providecommand \bibnamefont  [1]{#1}%
\providecommand \bibfnamefont [1]{#1}%
\providecommand \citenamefont [1]{#1}%
\providecommand \href@noop [0]{\@secondoftwo}%
\providecommand \href [0]{\begingroup \@sanitize@url \@href}%
\providecommand \@href[1]{\@@startlink{#1}\@@href}%
\providecommand \@@href[1]{\endgroup#1\@@endlink}%
\providecommand \@sanitize@url [0]{\catcode `\\12\catcode `\$12\catcode
  `\&12\catcode `\#12\catcode `\^12\catcode `\_12\catcode `\%12\relax}%
\providecommand \@@startlink[1]{}%
\providecommand \@@endlink[0]{}%
\providecommand \url  [0]{\begingroup\@sanitize@url \@url }%
\providecommand \@url [1]{\endgroup\@href {#1}{\urlprefix }}%
\providecommand \urlprefix  [0]{URL }%
\providecommand \Eprint [0]{\href }%
\providecommand \doibase [0]{https://doi.org/}%
\providecommand \selectlanguage [0]{\@gobble}%
\providecommand \bibinfo  [0]{\@secondoftwo}%
\providecommand \bibfield  [0]{\@secondoftwo}%
\providecommand \translation [1]{[#1]}%
\providecommand \BibitemOpen [0]{}%
\providecommand \bibitemStop [0]{}%
\providecommand \bibitemNoStop [0]{.\EOS\space}%
\providecommand \EOS [0]{\spacefactor3000\relax}%
\providecommand \BibitemShut  [1]{\csname bibitem#1\endcsname}%
\let\auto@bib@innerbib\@empty
\bibitem [{\citenamefont {Kaiser}(2017)}]{Kaiser2017}%
  \BibitemOpen
  \bibfield  {author} {\bibinfo {author} {\bibfnamefont {S.}~\bibnamefont
  {Kaiser}},\ }\bibfield  {title} {\bibinfo {title} {{Light-induced
  superconductivity in high-$T_c$ cuprates}},\ }\href
  {https://doi.org/10.1088/1402-4896/aa8201} {\bibfield  {journal} {\bibinfo
  {journal} {Phys. Scr.}\ }\textbf {\bibinfo {volume} {92}},\ \bibinfo {pages}
  {103001} (\bibinfo {year} {2017})}\BibitemShut {NoStop}%
\bibitem [{\citenamefont {Mitrano}\ \emph {et~al.}(2016)\citenamefont
  {Mitrano}, \citenamefont {Cantaluppi}, \citenamefont {Nicoletti},
  \citenamefont {Kaiser}, \citenamefont {Perucchi}, \citenamefont {Lupi},
  \citenamefont {{Di Pietro}}, \citenamefont {Pontiroli}, \citenamefont
  {Ricc{\`{o}}}, \citenamefont {Clark}, \citenamefont {Jaksch},\ and\
  \citenamefont {Cavalleri}}]{Mitrano2016}%
  \BibitemOpen
  \bibfield  {author} {\bibinfo {author} {\bibfnamefont {M.}~\bibnamefont
  {Mitrano}}, \bibinfo {author} {\bibfnamefont {A.}~\bibnamefont {Cantaluppi}},
  \bibinfo {author} {\bibfnamefont {D.}~\bibnamefont {Nicoletti}}, \bibinfo
  {author} {\bibfnamefont {S.}~\bibnamefont {Kaiser}}, \bibinfo {author}
  {\bibfnamefont {A.}~\bibnamefont {Perucchi}}, \bibinfo {author}
  {\bibfnamefont {S.}~\bibnamefont {Lupi}}, \bibinfo {author} {\bibfnamefont
  {P.}~\bibnamefont {{Di Pietro}}}, \bibinfo {author} {\bibfnamefont
  {D.}~\bibnamefont {Pontiroli}}, \bibinfo {author} {\bibfnamefont
  {M.}~\bibnamefont {Ricc{\`{o}}}}, \bibinfo {author} {\bibfnamefont {S.~R.}\
  \bibnamefont {Clark}}, \bibinfo {author} {\bibfnamefont {D.}~\bibnamefont
  {Jaksch}},\ and\ \bibinfo {author} {\bibfnamefont {A.}~\bibnamefont
  {Cavalleri}},\ }\bibfield  {title} {\bibinfo {title} {{Possible light-induced
  superconductivity in K$_3$C$_{60}$ at high temperature}},\ }\href
  {https://doi.org/10.1038/nature16522} {\bibfield  {journal} {\bibinfo
  {journal} {Nature}\ }\textbf {\bibinfo {volume} {530}},\ \bibinfo {pages}
  {461} (\bibinfo {year} {2016})}\BibitemShut {NoStop}%
\bibitem [{\citenamefont {Buzzi}\ \emph {et~al.}(2020)\citenamefont {Buzzi},
  \citenamefont {Nicoletti}, \citenamefont {Fechner}, \citenamefont
  {Tancogne-Dejean}, \citenamefont {Sentef}, \citenamefont {Georges},
  \citenamefont {Biesner}, \citenamefont {Uykur}, \citenamefont {Dressel},
  \citenamefont {Henderson}, \citenamefont {Siegrist}, \citenamefont
  {Schlueter}, \citenamefont {Miyagawa}, \citenamefont {Kanoda}, \citenamefont
  {Nam}, \citenamefont {Ardavan}, \citenamefont {Coulthard}, \citenamefont
  {Tindall}, \citenamefont {Schlawin}, \citenamefont {Jaksch},\ and\
  \citenamefont {Cavalleri}}]{Buzzi2020}%
  \BibitemOpen
  \bibfield  {author} {\bibinfo {author} {\bibfnamefont {M.}~\bibnamefont
  {Buzzi}}, \bibinfo {author} {\bibfnamefont {D.}~\bibnamefont {Nicoletti}},
  \bibinfo {author} {\bibfnamefont {M.}~\bibnamefont {Fechner}}, \bibinfo
  {author} {\bibfnamefont {N.}~\bibnamefont {Tancogne-Dejean}}, \bibinfo
  {author} {\bibfnamefont {M.~A.}\ \bibnamefont {Sentef}}, \bibinfo {author}
  {\bibfnamefont {A.}~\bibnamefont {Georges}}, \bibinfo {author} {\bibfnamefont
  {T.}~\bibnamefont {Biesner}}, \bibinfo {author} {\bibfnamefont
  {E.}~\bibnamefont {Uykur}}, \bibinfo {author} {\bibfnamefont
  {M.}~\bibnamefont {Dressel}}, \bibinfo {author} {\bibfnamefont
  {A.}~\bibnamefont {Henderson}}, \bibinfo {author} {\bibfnamefont
  {T.}~\bibnamefont {Siegrist}}, \bibinfo {author} {\bibfnamefont {J.~A.}\
  \bibnamefont {Schlueter}}, \bibinfo {author} {\bibfnamefont {K.}~\bibnamefont
  {Miyagawa}}, \bibinfo {author} {\bibfnamefont {K.}~\bibnamefont {Kanoda}},
  \bibinfo {author} {\bibfnamefont {M.-S.}\ \bibnamefont {Nam}}, \bibinfo
  {author} {\bibfnamefont {A.}~\bibnamefont {Ardavan}}, \bibinfo {author}
  {\bibfnamefont {J.}~\bibnamefont {Coulthard}}, \bibinfo {author}
  {\bibfnamefont {J.}~\bibnamefont {Tindall}}, \bibinfo {author} {\bibfnamefont
  {F.}~\bibnamefont {Schlawin}}, \bibinfo {author} {\bibfnamefont
  {D.}~\bibnamefont {Jaksch}},\ and\ \bibinfo {author} {\bibfnamefont
  {A.}~\bibnamefont {Cavalleri}},\ }\bibfield  {title} {\bibinfo {title}
  {{Photomolecular high-temperature superconductivity}},\ }\href
  {https://doi.org/10.1103/PhysRevX.10.031028} {\bibfield  {journal} {\bibinfo
  {journal} {Phys. Rev. X}\ }\textbf {\bibinfo {volume} {10}},\ \bibinfo
  {pages} {031028} (\bibinfo {year} {2020})}\BibitemShut {NoStop}%
\bibitem [{\citenamefont {Kogar}\ \emph {et~al.}(2020)\citenamefont {Kogar},
  \citenamefont {Zong}, \citenamefont {Dolgirev}, \citenamefont {Shen},
  \citenamefont {Straquadine}, \citenamefont {Bie}, \citenamefont {Wang},
  \citenamefont {Rohwer}, \citenamefont {Tung}, \citenamefont {Yang},
  \citenamefont {Li}, \citenamefont {Yang}, \citenamefont {Weathersby},
  \citenamefont {Park}, \citenamefont {Kozina}, \citenamefont {Sie},
  \citenamefont {Wen}, \citenamefont {Jarillo-Herrero}, \citenamefont {Fisher},
  \citenamefont {Wang},\ and\ \citenamefont {Gedik}}]{Kogar2020}%
  \BibitemOpen
  \bibfield  {author} {\bibinfo {author} {\bibfnamefont {A.}~\bibnamefont
  {Kogar}}, \bibinfo {author} {\bibfnamefont {A.}~\bibnamefont {Zong}},
  \bibinfo {author} {\bibfnamefont {P.~E.}\ \bibnamefont {Dolgirev}}, \bibinfo
  {author} {\bibfnamefont {X.}~\bibnamefont {Shen}}, \bibinfo {author}
  {\bibfnamefont {J.}~\bibnamefont {Straquadine}}, \bibinfo {author}
  {\bibfnamefont {Y.-Q.}\ \bibnamefont {Bie}}, \bibinfo {author} {\bibfnamefont
  {X.}~\bibnamefont {Wang}}, \bibinfo {author} {\bibfnamefont {T.}~\bibnamefont
  {Rohwer}}, \bibinfo {author} {\bibfnamefont {I.-C.}\ \bibnamefont {Tung}},
  \bibinfo {author} {\bibfnamefont {Y.}~\bibnamefont {Yang}}, \bibinfo {author}
  {\bibfnamefont {R.}~\bibnamefont {Li}}, \bibinfo {author} {\bibfnamefont
  {J.}~\bibnamefont {Yang}}, \bibinfo {author} {\bibfnamefont {S.}~\bibnamefont
  {Weathersby}}, \bibinfo {author} {\bibfnamefont {S.}~\bibnamefont {Park}},
  \bibinfo {author} {\bibfnamefont {M.~E.}\ \bibnamefont {Kozina}}, \bibinfo
  {author} {\bibfnamefont {E.~J.}\ \bibnamefont {Sie}}, \bibinfo {author}
  {\bibfnamefont {H.}~\bibnamefont {Wen}}, \bibinfo {author} {\bibfnamefont
  {P.}~\bibnamefont {Jarillo-Herrero}}, \bibinfo {author} {\bibfnamefont
  {I.~R.}\ \bibnamefont {Fisher}}, \bibinfo {author} {\bibfnamefont
  {X.}~\bibnamefont {Wang}},\ and\ \bibinfo {author} {\bibfnamefont
  {N.}~\bibnamefont {Gedik}},\ }\bibfield  {title} {\bibinfo {title}
  {{Light-induced charge density wave in LaTe$_3$}},\ }\href
  {https://doi.org/10.1038/s41567-019-0705-3} {\bibfield  {journal} {\bibinfo
  {journal} {Nat. Phys.}\ }\textbf {\bibinfo {volume} {16}},\ \bibinfo {pages}
  {159} (\bibinfo {year} {2020})}\BibitemShut {NoStop}%
\bibitem [{\citenamefont {Zhou}\ \emph {et~al.}(2021)\citenamefont {Zhou},
  \citenamefont {Williams}, \citenamefont {Sun}, \citenamefont {Malliakas},
  \citenamefont {Kanatzidis}, \citenamefont {Kemper},\ and\ \citenamefont
  {Ruan}}]{Zhou2021}%
  \BibitemOpen
  \bibfield  {author} {\bibinfo {author} {\bibfnamefont {F.}~\bibnamefont
  {Zhou}}, \bibinfo {author} {\bibfnamefont {J.}~\bibnamefont {Williams}},
  \bibinfo {author} {\bibfnamefont {S.}~\bibnamefont {Sun}}, \bibinfo {author}
  {\bibfnamefont {C.~D.}\ \bibnamefont {Malliakas}}, \bibinfo {author}
  {\bibfnamefont {M.~G.}\ \bibnamefont {Kanatzidis}}, \bibinfo {author}
  {\bibfnamefont {A.~F.}\ \bibnamefont {Kemper}},\ and\ \bibinfo {author}
  {\bibfnamefont {C.-Y.}\ \bibnamefont {Ruan}},\ }\bibfield  {title} {\bibinfo
  {title} {{Nonequilibrium dynamics of spontaneous symmetry breaking into a
  hidden state of charge-density wave}},\ }\href
  {https://doi.org/10.1038/s41467-020-20834-5} {\bibfield  {journal} {\bibinfo
  {journal} {Nat. Commun.}\ }\textbf {\bibinfo {volume} {12}},\ \bibinfo
  {pages} {566} (\bibinfo {year} {2021})}\BibitemShut {NoStop}%
\bibitem [{\citenamefont {Han}\ \emph {et~al.}(2015)\citenamefont {Han},
  \citenamefont {Zhou}, \citenamefont {Malliakas}, \citenamefont {Duxbury},
  \citenamefont {Mahanti}, \citenamefont {Kanatzidis},\ and\ \citenamefont
  {Ruan}}]{Han2015}%
  \BibitemOpen
  \bibfield  {author} {\bibinfo {author} {\bibfnamefont {T.-R.~T.}\
  \bibnamefont {Han}}, \bibinfo {author} {\bibfnamefont {F.}~\bibnamefont
  {Zhou}}, \bibinfo {author} {\bibfnamefont {C.~D.}\ \bibnamefont {Malliakas}},
  \bibinfo {author} {\bibfnamefont {P.~M.}\ \bibnamefont {Duxbury}}, \bibinfo
  {author} {\bibfnamefont {S.~D.}\ \bibnamefont {Mahanti}}, \bibinfo {author}
  {\bibfnamefont {M.~G.}\ \bibnamefont {Kanatzidis}},\ and\ \bibinfo {author}
  {\bibfnamefont {C.-Y.}\ \bibnamefont {Ruan}},\ }\bibfield  {title} {\bibinfo
  {title} {{Exploration of metastability and hidden phases in correlated
  electron crystals visualized by femtosecond optical doping and electron
  crystallography}},\ }\href {https://doi.org/10.1126/sciadv.1400173}
  {\bibfield  {journal} {\bibinfo  {journal} {Sci. Adv.}\ }\textbf {\bibinfo
  {volume} {1}},\ \bibinfo {pages} {e1400173} (\bibinfo {year}
  {2015})}\BibitemShut {NoStop}%
\bibitem [{\citenamefont {Kim}\ \emph {et~al.}(2012)\citenamefont {Kim},
  \citenamefont {Pashkin}, \citenamefont {Sch{\"{a}}fer}, \citenamefont
  {Beyer}, \citenamefont {Porer}, \citenamefont {Wolf}, \citenamefont
  {Bernhard}, \citenamefont {Demsar}, \citenamefont {Huber},\ and\
  \citenamefont {Leitenstorfer}}]{Kim2012}%
  \BibitemOpen
  \bibfield  {author} {\bibinfo {author} {\bibfnamefont {K.~W.}\ \bibnamefont
  {Kim}}, \bibinfo {author} {\bibfnamefont {A.}~\bibnamefont {Pashkin}},
  \bibinfo {author} {\bibfnamefont {H.}~\bibnamefont {Sch{\"{a}}fer}}, \bibinfo
  {author} {\bibfnamefont {M.}~\bibnamefont {Beyer}}, \bibinfo {author}
  {\bibfnamefont {M.}~\bibnamefont {Porer}}, \bibinfo {author} {\bibfnamefont
  {T.}~\bibnamefont {Wolf}}, \bibinfo {author} {\bibfnamefont {C.}~\bibnamefont
  {Bernhard}}, \bibinfo {author} {\bibfnamefont {J.}~\bibnamefont {Demsar}},
  \bibinfo {author} {\bibfnamefont {R.}~\bibnamefont {Huber}},\ and\ \bibinfo
  {author} {\bibfnamefont {A.}~\bibnamefont {Leitenstorfer}},\ }\bibfield
  {title} {\bibinfo {title} {{Ultrafast transient generation of
  spin-density-wave order in the normal state of BaFe$_2$As$_2$ driven by
  coherent lattice vibrations}},\ }\href {https://doi.org/10.1038/nmat3294}
  {\bibfield  {journal} {\bibinfo  {journal} {Nat. Mater.}\ }\textbf {\bibinfo
  {volume} {11}},\ \bibinfo {pages} {497} (\bibinfo {year} {2012})}\BibitemShut
  {NoStop}%
\bibitem [{\citenamefont {Li}\ \emph {et~al.}(2019)\citenamefont {Li},
  \citenamefont {Qiu}, \citenamefont {Zhang}, \citenamefont {Baldini},
  \citenamefont {Lu}, \citenamefont {Rappe},\ and\ \citenamefont
  {Nelson}}]{Li2019}%
  \BibitemOpen
  \bibfield  {author} {\bibinfo {author} {\bibfnamefont {X.}~\bibnamefont
  {Li}}, \bibinfo {author} {\bibfnamefont {T.}~\bibnamefont {Qiu}}, \bibinfo
  {author} {\bibfnamefont {J.}~\bibnamefont {Zhang}}, \bibinfo {author}
  {\bibfnamefont {E.}~\bibnamefont {Baldini}}, \bibinfo {author} {\bibfnamefont
  {J.}~\bibnamefont {Lu}}, \bibinfo {author} {\bibfnamefont {A.~M.}\
  \bibnamefont {Rappe}},\ and\ \bibinfo {author} {\bibfnamefont {K.~A.}\
  \bibnamefont {Nelson}},\ }\bibfield  {title} {\bibinfo {title} {{Terahertz
  field-induced ferroelectricity in quantum paraelectric SrTiO$_3$}},\ }\href
  {https://doi.org/10.1126/science.aaw4913} {\bibfield  {journal} {\bibinfo
  {journal} {Science}\ }\textbf {\bibinfo {volume} {364}},\ \bibinfo {pages}
  {1079} (\bibinfo {year} {2019})}\BibitemShut {NoStop}%
\bibitem [{\citenamefont {Nova}\ \emph {et~al.}(2019)\citenamefont {Nova},
  \citenamefont {Disa}, \citenamefont {Fechner},\ and\ \citenamefont
  {Cavalleri}}]{Nova2019}%
  \BibitemOpen
  \bibfield  {author} {\bibinfo {author} {\bibfnamefont {T.~F.}\ \bibnamefont
  {Nova}}, \bibinfo {author} {\bibfnamefont {A.~S.}\ \bibnamefont {Disa}},
  \bibinfo {author} {\bibfnamefont {M.}~\bibnamefont {Fechner}},\ and\ \bibinfo
  {author} {\bibfnamefont {A.}~\bibnamefont {Cavalleri}},\ }\bibfield  {title}
  {\bibinfo {title} {{Metastable ferroelectricity in optically strained
  SrTiO$_3$}},\ }\href {https://doi.org/10.1126/science.aaw4911} {\bibfield
  {journal} {\bibinfo  {journal} {Science}\ }\textbf {\bibinfo {volume}
  {364}},\ \bibinfo {pages} {1075} (\bibinfo {year} {2019})}\BibitemShut
  {NoStop}%
\bibitem [{\citenamefont {Fausti}\ \emph {et~al.}(2011)\citenamefont {Fausti},
  \citenamefont {Tobey}, \citenamefont {Dean}, \citenamefont {Kaiser},
  \citenamefont {Dienst}, \citenamefont {Hoffmann}, \citenamefont {Pyon},
  \citenamefont {Takayama}, \citenamefont {Takagi},\ and\ \citenamefont
  {Cavalleri}}]{Fausti2011}%
  \BibitemOpen
  \bibfield  {author} {\bibinfo {author} {\bibfnamefont {D.}~\bibnamefont
  {Fausti}}, \bibinfo {author} {\bibfnamefont {R.~I.}\ \bibnamefont {Tobey}},
  \bibinfo {author} {\bibfnamefont {N.}~\bibnamefont {Dean}}, \bibinfo {author}
  {\bibfnamefont {S.}~\bibnamefont {Kaiser}}, \bibinfo {author} {\bibfnamefont
  {A.}~\bibnamefont {Dienst}}, \bibinfo {author} {\bibfnamefont {M.~C.}\
  \bibnamefont {Hoffmann}}, \bibinfo {author} {\bibfnamefont {S.}~\bibnamefont
  {Pyon}}, \bibinfo {author} {\bibfnamefont {T.}~\bibnamefont {Takayama}},
  \bibinfo {author} {\bibfnamefont {H.}~\bibnamefont {Takagi}},\ and\ \bibinfo
  {author} {\bibfnamefont {A.}~\bibnamefont {Cavalleri}},\ }\bibfield  {title}
  {\bibinfo {title} {{Light-induced superconductivity in a stripe-ordered
  cuprate}},\ }\href {https://doi.org/10.1126/science.1197294} {\bibfield
  {journal} {\bibinfo  {journal} {Science}\ }\textbf {\bibinfo {volume}
  {331}},\ \bibinfo {pages} {189} (\bibinfo {year} {2011})}\BibitemShut
  {NoStop}%
\bibitem [{\citenamefont {Kaiser}\ \emph {et~al.}(2014)\citenamefont {Kaiser},
  \citenamefont {Hunt}, \citenamefont {Nicoletti}, \citenamefont {Hu},
  \citenamefont {Gierz}, \citenamefont {Liu}, \citenamefont {{Le Tacon}},
  \citenamefont {Loew}, \citenamefont {Haug}, \citenamefont {Keimer},\ and\
  \citenamefont {Cavalleri}}]{Kaiser2014}%
  \BibitemOpen
  \bibfield  {author} {\bibinfo {author} {\bibfnamefont {S.}~\bibnamefont
  {Kaiser}}, \bibinfo {author} {\bibfnamefont {C.~R.}\ \bibnamefont {Hunt}},
  \bibinfo {author} {\bibfnamefont {D.}~\bibnamefont {Nicoletti}}, \bibinfo
  {author} {\bibfnamefont {W.}~\bibnamefont {Hu}}, \bibinfo {author}
  {\bibfnamefont {I.}~\bibnamefont {Gierz}}, \bibinfo {author} {\bibfnamefont
  {H.~Y.}\ \bibnamefont {Liu}}, \bibinfo {author} {\bibfnamefont
  {M.}~\bibnamefont {{Le Tacon}}}, \bibinfo {author} {\bibfnamefont
  {T.}~\bibnamefont {Loew}}, \bibinfo {author} {\bibfnamefont {D.}~\bibnamefont
  {Haug}}, \bibinfo {author} {\bibfnamefont {B.}~\bibnamefont {Keimer}},\ and\
  \bibinfo {author} {\bibfnamefont {A.}~\bibnamefont {Cavalleri}},\ }\bibfield
  {title} {\bibinfo {title} {{Optically induced coherent transport far above
  $T_c$ in underdoped YBa$_2$Cu$_3$O$_{6+\delta}$}},\ }\href
  {https://doi.org/10.1103/PhysRevB.89.184516} {\bibfield  {journal} {\bibinfo
  {journal} {Phys. Rev. B}\ }\textbf {\bibinfo {volume} {89}},\ \bibinfo
  {pages} {184516} (\bibinfo {year} {2014})}\BibitemShut {NoStop}%
\bibitem [{\citenamefont {Hu}\ \emph {et~al.}(2014{\natexlab{a}})\citenamefont
  {Hu}, \citenamefont {Kaiser}, \citenamefont {Nicoletti}, \citenamefont
  {Hunt}, \citenamefont {Gierz}, \citenamefont {Hoffmann}, \citenamefont {{Le
  Tacon}}, \citenamefont {Loew}, \citenamefont {Keimer},\ and\ \citenamefont
  {Cavalleri}}]{Hu2014}%
  \BibitemOpen
  \bibfield  {author} {\bibinfo {author} {\bibfnamefont {W.}~\bibnamefont
  {Hu}}, \bibinfo {author} {\bibfnamefont {S.}~\bibnamefont {Kaiser}}, \bibinfo
  {author} {\bibfnamefont {D.}~\bibnamefont {Nicoletti}}, \bibinfo {author}
  {\bibfnamefont {C.~R.}\ \bibnamefont {Hunt}}, \bibinfo {author}
  {\bibfnamefont {I.}~\bibnamefont {Gierz}}, \bibinfo {author} {\bibfnamefont
  {M.~C.}\ \bibnamefont {Hoffmann}}, \bibinfo {author} {\bibfnamefont
  {M.}~\bibnamefont {{Le Tacon}}}, \bibinfo {author} {\bibfnamefont
  {T.}~\bibnamefont {Loew}}, \bibinfo {author} {\bibfnamefont {B.}~\bibnamefont
  {Keimer}},\ and\ \bibinfo {author} {\bibfnamefont {A.}~\bibnamefont
  {Cavalleri}},\ }\bibfield  {title} {\bibinfo {title} {{Optically enhanced
  coherent transport in YBa$_2$Cu$_3$O$_{6.5}$ by ultrafast redistribution of
  interlayer coupling}},\ }\href {https://doi.org/10.1038/nmat3963} {\bibfield
  {journal} {\bibinfo  {journal} {Nat. Mater.}\ }\textbf {\bibinfo {volume}
  {13}},\ \bibinfo {pages} {705} (\bibinfo {year}
  {2014}{\natexlab{a}})}\BibitemShut {NoStop}%
\bibitem [{\citenamefont {Nicoletti}\ \emph {et~al.}(2014)\citenamefont
  {Nicoletti}, \citenamefont {Casandruc}, \citenamefont {Laplace},
  \citenamefont {Khanna}, \citenamefont {Hunt}, \citenamefont {Kaiser},
  \citenamefont {Dhesi}, \citenamefont {Gu}, \citenamefont {Hill},\ and\
  \citenamefont {Cavalleri}}]{Nicoletti2014}%
  \BibitemOpen
  \bibfield  {author} {\bibinfo {author} {\bibfnamefont {D.}~\bibnamefont
  {Nicoletti}}, \bibinfo {author} {\bibfnamefont {E.}~\bibnamefont
  {Casandruc}}, \bibinfo {author} {\bibfnamefont {Y.}~\bibnamefont {Laplace}},
  \bibinfo {author} {\bibfnamefont {V.}~\bibnamefont {Khanna}}, \bibinfo
  {author} {\bibfnamefont {C.~R.}\ \bibnamefont {Hunt}}, \bibinfo {author}
  {\bibfnamefont {S.}~\bibnamefont {Kaiser}}, \bibinfo {author} {\bibfnamefont
  {S.~S.}\ \bibnamefont {Dhesi}}, \bibinfo {author} {\bibfnamefont {G.~D.}\
  \bibnamefont {Gu}}, \bibinfo {author} {\bibfnamefont {J.~P.}\ \bibnamefont
  {Hill}},\ and\ \bibinfo {author} {\bibfnamefont {A.}~\bibnamefont
  {Cavalleri}},\ }\bibfield  {title} {\bibinfo {title} {{Optically induced
  superconductivity in striped La$_{2-x}$Ba$_x$CuO$_4$ by
  polarization-selective excitation in the near infrared}},\ }\href
  {https://doi.org/10.1103/PhysRevB.90.100503} {\bibfield  {journal} {\bibinfo
  {journal} {Phys. Rev. B}\ }\textbf {\bibinfo {volume} {90}},\ \bibinfo
  {pages} {100503} (\bibinfo {year} {2014})}\BibitemShut {NoStop}%
\bibitem [{\citenamefont {Borroni}\ \emph {et~al.}(2017)\citenamefont
  {Borroni}, \citenamefont {Baldini}, \citenamefont {Katukuri}, \citenamefont
  {Mann}, \citenamefont {Parlinski}, \citenamefont {Legut}, \citenamefont
  {Arrell}, \citenamefont {van Mourik}, \citenamefont {Teyssier}, \citenamefont
  {Kozlowski}, \citenamefont {Piekarz}, \citenamefont {Yazyev}, \citenamefont
  {Ole\ifmmode~\acute{s}\else \'{s}\fi{}}, \citenamefont {Lorenzana},\ and\
  \citenamefont {Carbone}}]{Borroni2017}%
  \BibitemOpen
  \bibfield  {author} {\bibinfo {author} {\bibfnamefont {S.}~\bibnamefont
  {Borroni}}, \bibinfo {author} {\bibfnamefont {E.}~\bibnamefont {Baldini}},
  \bibinfo {author} {\bibfnamefont {V.~M.}\ \bibnamefont {Katukuri}}, \bibinfo
  {author} {\bibfnamefont {A.}~\bibnamefont {Mann}}, \bibinfo {author}
  {\bibfnamefont {K.}~\bibnamefont {Parlinski}}, \bibinfo {author}
  {\bibfnamefont {D.}~\bibnamefont {Legut}}, \bibinfo {author} {\bibfnamefont
  {C.}~\bibnamefont {Arrell}}, \bibinfo {author} {\bibfnamefont
  {F.}~\bibnamefont {van Mourik}}, \bibinfo {author} {\bibfnamefont
  {J.}~\bibnamefont {Teyssier}}, \bibinfo {author} {\bibfnamefont
  {A.}~\bibnamefont {Kozlowski}}, \bibinfo {author} {\bibfnamefont
  {P.}~\bibnamefont {Piekarz}}, \bibinfo {author} {\bibfnamefont {O.~V.}\
  \bibnamefont {Yazyev}}, \bibinfo {author} {\bibfnamefont {A.~M.}\
  \bibnamefont {Ole\ifmmode~\acute{s}\else \'{s}\fi{}}}, \bibinfo {author}
  {\bibfnamefont {J.}~\bibnamefont {Lorenzana}},\ and\ \bibinfo {author}
  {\bibfnamefont {F.}~\bibnamefont {Carbone}},\ }\bibfield  {title} {\bibinfo
  {title} {{Coherent generation of symmetry-forbidden phonons by light-induced
  electron-phonon interactions in magnetite}},\ }\href
  {https://doi.org/10.1103/PhysRevB.96.104308} {\bibfield  {journal} {\bibinfo
  {journal} {Phys. Rev. B}\ }\textbf {\bibinfo {volume} {96}},\ \bibinfo
  {pages} {104308} (\bibinfo {year} {2017})}\BibitemShut {NoStop}%
\bibitem [{\citenamefont {Emery}\ and\ \citenamefont
  {Kivelson}(1995)}]{Emery1995}%
  \BibitemOpen
  \bibfield  {author} {\bibinfo {author} {\bibfnamefont {V.~J.}\ \bibnamefont
  {Emery}}\ and\ \bibinfo {author} {\bibfnamefont {S.~A.}\ \bibnamefont
  {Kivelson}},\ }\bibfield  {title} {\bibinfo {title} {{Importance of phase
  fluctuations in superconductors with small superfluid density}},\ }\href@noop
  {} {\bibfield  {journal} {\bibinfo  {journal} {Nature}\ }\textbf {\bibinfo
  {volume} {374}},\ \bibinfo {pages} {434} (\bibinfo {year}
  {1995})}\BibitemShut {NoStop}%
\bibitem [{\citenamefont {Nam}\ \emph {et~al.}(2007)\citenamefont {Nam},
  \citenamefont {Ardavan}, \citenamefont {Blundell},\ and\ \citenamefont
  {Schlueter}}]{Nam2007}%
  \BibitemOpen
  \bibfield  {author} {\bibinfo {author} {\bibfnamefont {M.-S.}\ \bibnamefont
  {Nam}}, \bibinfo {author} {\bibfnamefont {A.}~\bibnamefont {Ardavan}},
  \bibinfo {author} {\bibfnamefont {S.~J.}\ \bibnamefont {Blundell}},\ and\
  \bibinfo {author} {\bibfnamefont {J.~A.}\ \bibnamefont {Schlueter}},\
  }\bibfield  {title} {\bibinfo {title} {{Fluctuating superconductivity in
  organic molecular metals close to the Mott transition}},\ }\href
  {https://doi.org/10.1038/nature06182} {\bibfield  {journal} {\bibinfo
  {journal} {Nature}\ }\textbf {\bibinfo {volume} {449}},\ \bibinfo {pages}
  {584} (\bibinfo {year} {2007})}\BibitemShut {NoStop}%
\bibitem [{\citenamefont {Kagawa}\ \emph {et~al.}(2009)\citenamefont {Kagawa},
  \citenamefont {Miyagawa},\ and\ \citenamefont {Kanoda}}]{Kagawa2009}%
  \BibitemOpen
  \bibfield  {author} {\bibinfo {author} {\bibfnamefont {F.}~\bibnamefont
  {Kagawa}}, \bibinfo {author} {\bibfnamefont {K.}~\bibnamefont {Miyagawa}},\
  and\ \bibinfo {author} {\bibfnamefont {K.}~\bibnamefont {Kanoda}},\
  }\bibfield  {title} {\bibinfo {title} {{Magnetic Mott criticality in a
  $\kappa$-type organic salt probed by NMR}},\ }\href
  {https://doi.org/10.1038/nphys1428} {\bibfield  {journal} {\bibinfo
  {journal} {Nat. Phys.}\ }\textbf {\bibinfo {volume} {5}},\ \bibinfo {pages}
  {880} (\bibinfo {year} {2009})}\BibitemShut {NoStop}%
\bibitem [{\citenamefont {M{\"{u}}ller}\ and\ \citenamefont
  {Burkard}(1979)}]{Muller1979}%
  \BibitemOpen
  \bibfield  {author} {\bibinfo {author} {\bibfnamefont {K.~A.}\ \bibnamefont
  {M{\"{u}}ller}}\ and\ \bibinfo {author} {\bibfnamefont {H.}~\bibnamefont
  {Burkard}},\ }\bibfield  {title} {\bibinfo {title} {{SrTiO$_3$: An intrinsic
  quantum paraelectric below 4~K}},\ }\href
  {https://doi.org/10.1103/PhysRevB.19.3593} {\bibfield  {journal} {\bibinfo
  {journal} {Phys. Rev. B}\ }\textbf {\bibinfo {volume} {19}},\ \bibinfo
  {pages} {3593} (\bibinfo {year} {1979})}\BibitemShut {NoStop}%
\bibitem [{\citenamefont {Ru}(2008)}]{RuThesis}%
  \BibitemOpen
  \bibfield  {author} {\bibinfo {author} {\bibfnamefont {N.}~\bibnamefont
  {Ru}},\ }\emph {\bibinfo {title} {{Charge Density Wave Formation in
  Rare-earth Tellurides}}},\ \href
  {https://web.stanford.edu/group/fisher/people/Nancy_Ru_thesis.pdf} {\bibinfo
  {type} {{Ph.D. thesis}}},\ \bibinfo  {school} {Stanford University}, \bibinfo
  {address} {Stanford} (\bibinfo {year} {2008})\BibitemShut {NoStop}%
\bibitem [{\citenamefont {Eiter}\ \emph {et~al.}(2013)\citenamefont {Eiter},
  \citenamefont {Lavagnini}, \citenamefont {Hackl}, \citenamefont {Nowadnick},
  \citenamefont {Kemper}, \citenamefont {Devereaux}, \citenamefont {Chu},
  \citenamefont {Analytis}, \citenamefont {Fisher},\ and\ \citenamefont
  {Degiorgi}}]{Eiter2013}%
  \BibitemOpen
  \bibfield  {author} {\bibinfo {author} {\bibfnamefont {H.-M.}\ \bibnamefont
  {Eiter}}, \bibinfo {author} {\bibfnamefont {M.}~\bibnamefont {Lavagnini}},
  \bibinfo {author} {\bibfnamefont {R.}~\bibnamefont {Hackl}}, \bibinfo
  {author} {\bibfnamefont {E.~A.}\ \bibnamefont {Nowadnick}}, \bibinfo {author}
  {\bibfnamefont {A.~F.}\ \bibnamefont {Kemper}}, \bibinfo {author}
  {\bibfnamefont {T.~P.}\ \bibnamefont {Devereaux}}, \bibinfo {author}
  {\bibfnamefont {J.-H.}\ \bibnamefont {Chu}}, \bibinfo {author} {\bibfnamefont
  {J.~G.}\ \bibnamefont {Analytis}}, \bibinfo {author} {\bibfnamefont {I.~R.}\
  \bibnamefont {Fisher}},\ and\ \bibinfo {author} {\bibfnamefont
  {L.}~\bibnamefont {Degiorgi}},\ }\bibfield  {title} {\bibinfo {title}
  {{Alternative route to charge density wave formation in multiband systems}},\
  }\href {https://doi.org/10.1073/pnas.1214745110} {\bibfield  {journal}
  {\bibinfo  {journal} {Proc. Natl. Acad. Sci. U.S.A.}\ }\textbf {\bibinfo
  {volume} {110}},\ \bibinfo {pages} {64} (\bibinfo {year} {2013})}\BibitemShut
  {NoStop}%
\bibitem [{\citenamefont {Maschek}\ \emph {et~al.}(2018)\citenamefont
  {Maschek}, \citenamefont {Zocco}, \citenamefont {Rosenkranz}, \citenamefont
  {Heid}, \citenamefont {Said}, \citenamefont {Alatas}, \citenamefont
  {Walmsley}, \citenamefont {Fisher},\ and\ \citenamefont
  {Weber}}]{Maschek2018}%
  \BibitemOpen
  \bibfield  {author} {\bibinfo {author} {\bibfnamefont {M.}~\bibnamefont
  {Maschek}}, \bibinfo {author} {\bibfnamefont {D.~A.}\ \bibnamefont {Zocco}},
  \bibinfo {author} {\bibfnamefont {S.}~\bibnamefont {Rosenkranz}}, \bibinfo
  {author} {\bibfnamefont {R.}~\bibnamefont {Heid}}, \bibinfo {author}
  {\bibfnamefont {A.~H.}\ \bibnamefont {Said}}, \bibinfo {author}
  {\bibfnamefont {A.}~\bibnamefont {Alatas}}, \bibinfo {author} {\bibfnamefont
  {P.}~\bibnamefont {Walmsley}}, \bibinfo {author} {\bibfnamefont {I.~R.}\
  \bibnamefont {Fisher}},\ and\ \bibinfo {author} {\bibfnamefont
  {F.}~\bibnamefont {Weber}},\ }\bibfield  {title} {\bibinfo {title}
  {{Competing soft phonon modes at the charge-density-wave transitions in
  DyTe$_3$}},\ }\href {https://doi.org/10.1103/PhysRevB.98.094304} {\bibfield
  {journal} {\bibinfo  {journal} {Phys. Rev. B}\ }\textbf {\bibinfo {volume}
  {98}},\ \bibinfo {pages} {094304} (\bibinfo {year} {2018})}\BibitemShut
  {NoStop}%
\bibitem [{\citenamefont {Ru}\ \emph {et~al.}(2008)\citenamefont {Ru},
  \citenamefont {Condron}, \citenamefont {Margulis}, \citenamefont {Shin},
  \citenamefont {Laverock}, \citenamefont {Dugdale}, \citenamefont {Toney},\
  and\ \citenamefont {Fisher}}]{Ru2008}%
  \BibitemOpen
  \bibfield  {author} {\bibinfo {author} {\bibfnamefont {N.}~\bibnamefont
  {Ru}}, \bibinfo {author} {\bibfnamefont {C.~L.}\ \bibnamefont {Condron}},
  \bibinfo {author} {\bibfnamefont {G.~Y.}\ \bibnamefont {Margulis}}, \bibinfo
  {author} {\bibfnamefont {K.~Y.}\ \bibnamefont {Shin}}, \bibinfo {author}
  {\bibfnamefont {J.}~\bibnamefont {Laverock}}, \bibinfo {author}
  {\bibfnamefont {S.~B.}\ \bibnamefont {Dugdale}}, \bibinfo {author}
  {\bibfnamefont {M.~F.}\ \bibnamefont {Toney}},\ and\ \bibinfo {author}
  {\bibfnamefont {I.~R.}\ \bibnamefont {Fisher}},\ }\bibfield  {title}
  {\bibinfo {title} {{Effect of chemical pressure on the charge density wave
  transition in rare-earth tritellurides $R$Te$_3$}},\ }\href
  {https://doi.org/10.1103/PhysRevB.77.035114} {\bibfield  {journal} {\bibinfo
  {journal} {Phys. Rev. B}\ }\textbf {\bibinfo {volume} {77}},\ \bibinfo
  {pages} {035114} (\bibinfo {year} {2008})}\BibitemShut {NoStop}%
\bibitem [{\citenamefont {Hu}\ \emph {et~al.}(2014{\natexlab{b}})\citenamefont
  {Hu}, \citenamefont {Cheng}, \citenamefont {Yuan}, \citenamefont {Dong},\
  and\ \citenamefont {Wang}}]{Hu2014b}%
  \BibitemOpen
  \bibfield  {author} {\bibinfo {author} {\bibfnamefont {B.~F.}\ \bibnamefont
  {Hu}}, \bibinfo {author} {\bibfnamefont {B.}~\bibnamefont {Cheng}}, \bibinfo
  {author} {\bibfnamefont {R.~H.}\ \bibnamefont {Yuan}}, \bibinfo {author}
  {\bibfnamefont {T.}~\bibnamefont {Dong}},\ and\ \bibinfo {author}
  {\bibfnamefont {N.~L.}\ \bibnamefont {Wang}},\ }\bibfield  {title} {\bibinfo
  {title} {{Coexistence and competition of multiple charge-density-wave orders
  in rare-earth tritellurides}},\ }\href
  {https://doi.org/10.1103/PhysRevB.90.085105} {\bibfield  {journal} {\bibinfo
  {journal} {Phys. Rev. B}\ }\textbf {\bibinfo {volume} {90}},\ \bibinfo
  {pages} {085105} (\bibinfo {year} {2014}{\natexlab{b}})}\BibitemShut
  {NoStop}%
\bibitem [{Note1()}]{Note1}%
  \BibitemOpen
  \bibinfo {note} {Unlike LaTe$_3$ that only develops a unidirectional CDW
  under ambient pressure, there are two CDW transitions in DyTe$_3$ \cite
  {RuThesis}. Starting from its normal metallic state, DyTe$_3$ first develops
  a unidirectional CDW at $T_{c1}$; the second transition into a bidirectional
  CDW state occurs at a lower temperature $T_{c2}$. Here, we are only concerned
  with the high-temperature transition and we denote $T_{c1}$ by $T_c$ for
  brevity.}\BibitemShut {Stop}%
\bibitem [{Rem()}]{Remark1}%
  \BibitemOpen
  \href@noop {} {}\bibinfo {note} {See the Supplemental Material for more
  details.}\BibitemShut {Stop}%
\bibitem [{Note2()}]{Note2}%
  \BibitemOpen
  \bibinfo {note} {The image was taken above $T_{c2}=68$~K for DyTe$_3$ \cite
  {Maschek2018}.}\BibitemShut {Stop}%
\bibitem [{\citenamefont {Straquadine}\ \emph {et~al.}(2019)\citenamefont
  {Straquadine}, \citenamefont {Weber}, \citenamefont {Rosenkranz},
  \citenamefont {Said},\ and\ \citenamefont {Fisher}}]{Straquadine2019}%
  \BibitemOpen
  \bibfield  {author} {\bibinfo {author} {\bibfnamefont {J.~A.~W.}\
  \bibnamefont {Straquadine}}, \bibinfo {author} {\bibfnamefont
  {F.}~\bibnamefont {Weber}}, \bibinfo {author} {\bibfnamefont
  {S.}~\bibnamefont {Rosenkranz}}, \bibinfo {author} {\bibfnamefont {A.~H.}\
  \bibnamefont {Said}},\ and\ \bibinfo {author} {\bibfnamefont {I.~R.}\
  \bibnamefont {Fisher}},\ }\bibfield  {title} {\bibinfo {title} {{Suppression
  of charge density wave order by disorder in Pd-intercalated ErTe$_3$}},\
  }\href {https://doi.org/10.1103/PhysRevB.99.235138} {\bibfield  {journal}
  {\bibinfo  {journal} {Phys. Rev. B}\ }\textbf {\bibinfo {volume} {99}},\
  \bibinfo {pages} {235138} (\bibinfo {year} {2019})}\BibitemShut {NoStop}%
\bibitem [{Note3()}]{Note3}%
  \BibitemOpen
  \bibinfo {note} {The speed of sound is deduced from the phonon dispersion in
  DyTe$_3$ calculated by density functional perturbation theory and verified by
  inelastic X-ray scattering \cite {Maschek2018}. The speed of sound associated
  with the phason excitation may be an alternative choice for this correlation
  length estimate. The phason dispersion is unavailable for $R$Te$_3$ but we
  take note of values in other incommensurate CDWs. The phason speed ranges
  from $4\times 10^2$~m/s in 1$T$-TaS$_2$ \cite {Minor1989} to $2\times
  10^4$~m/s in K$_{0.3}$MoO$_3$ \cite {Pouget1991}, hence not changing the
  conclusion that the largest possible correlation length of the transient CDW
  in LaTe$_3$ is still orders of magnitude smaller compared to its dominant CDW
  in equilibrium.}\BibitemShut {Stop}%
\bibitem [{\citenamefont {Otto}\ \emph {et~al.}(2021)\citenamefont {Otto},
  \citenamefont {P{\"{o}}hls}, \citenamefont {{Ren{\'{e}} de Cotret}},
  \citenamefont {Stern}, \citenamefont {Sutton},\ and\ \citenamefont
  {Siwick}}]{Otto2019b}%
  \BibitemOpen
  \bibfield  {author} {\bibinfo {author} {\bibfnamefont {M.~R.}\ \bibnamefont
  {Otto}}, \bibinfo {author} {\bibfnamefont {J.-H.}\ \bibnamefont
  {P{\"{o}}hls}}, \bibinfo {author} {\bibfnamefont {L.~P.}\ \bibnamefont
  {{Ren{\'{e}} de Cotret}}}, \bibinfo {author} {\bibfnamefont {M.~J.}\
  \bibnamefont {Stern}}, \bibinfo {author} {\bibfnamefont {M.}~\bibnamefont
  {Sutton}},\ and\ \bibinfo {author} {\bibfnamefont {B.~J.}\ \bibnamefont
  {Siwick}},\ }\bibfield  {title} {\bibinfo {title} {{Mechanisms of
  electron-phonon coupling unraveled in momentum and time: The case of soft
  phonons in TiSe$_2$}},\ }\href {https://doi.org/10.1126/sciadv.abf2810}
  {\bibfield  {journal} {\bibinfo  {journal} {Sci. Adv.}\ }\textbf {\bibinfo
  {volume} {7}},\ \bibinfo {pages} {eabf2810} (\bibinfo {year}
  {2021})}\BibitemShut {NoStop}%
\bibitem [{\citenamefont {Hellmann}\ \emph {et~al.}(2012)\citenamefont
  {Hellmann}, \citenamefont {Rohwer}, \citenamefont {Kall{\"{a}}ne},
  \citenamefont {Hanff}, \citenamefont {Sohrt}, \citenamefont {Stange},
  \citenamefont {Carr}, \citenamefont {Murnane}, \citenamefont {Kapteyn},
  \citenamefont {Kipp}, \citenamefont {Bauer},\ and\ \citenamefont
  {Rossnagel}}]{Hellmann2012}%
  \BibitemOpen
  \bibfield  {author} {\bibinfo {author} {\bibfnamefont {S.}~\bibnamefont
  {Hellmann}}, \bibinfo {author} {\bibfnamefont {T.}~\bibnamefont {Rohwer}},
  \bibinfo {author} {\bibfnamefont {M.}~\bibnamefont {Kall{\"{a}}ne}}, \bibinfo
  {author} {\bibfnamefont {K.}~\bibnamefont {Hanff}}, \bibinfo {author}
  {\bibfnamefont {C.}~\bibnamefont {Sohrt}}, \bibinfo {author} {\bibfnamefont
  {A.}~\bibnamefont {Stange}}, \bibinfo {author} {\bibfnamefont
  {A.}~\bibnamefont {Carr}}, \bibinfo {author} {\bibfnamefont {M.~M.}\
  \bibnamefont {Murnane}}, \bibinfo {author} {\bibfnamefont {H.~C.}\
  \bibnamefont {Kapteyn}}, \bibinfo {author} {\bibfnamefont {L.}~\bibnamefont
  {Kipp}}, \bibinfo {author} {\bibfnamefont {M.}~\bibnamefont {Bauer}},\ and\
  \bibinfo {author} {\bibfnamefont {K.}~\bibnamefont {Rossnagel}},\ }\bibfield
  {title} {\bibinfo {title} {{Time-domain classification of charge-density-wave
  insulators}},\ }\href {https://doi.org/10.1038/ncomms2078} {\bibfield
  {journal} {\bibinfo  {journal} {Nat. Commun.}\ }\textbf {\bibinfo {volume}
  {3}},\ \bibinfo {pages} {1069} (\bibinfo {year} {2012})}\BibitemShut
  {NoStop}%
\bibitem [{\citenamefont {Zong}\ \emph
  {et~al.}(2019{\natexlab{a}})\citenamefont {Zong}, \citenamefont {Dolgirev},
  \citenamefont {Kogar}, \citenamefont {Erge{\c{c}}en}, \citenamefont {Yilmaz},
  \citenamefont {Bie}, \citenamefont {Rohwer}, \citenamefont {Tung},
  \citenamefont {Straquadine}, \citenamefont {Wang}, \citenamefont {Yang},
  \citenamefont {Shen}, \citenamefont {Li}, \citenamefont {Yang}, \citenamefont
  {Park}, \citenamefont {Hoffmann}, \citenamefont {Ofori-Okai}, \citenamefont
  {Kozina}, \citenamefont {Wen}, \citenamefont {Wang}, \citenamefont {Fisher},
  \citenamefont {Jarillo-Herrero},\ and\ \citenamefont {Gedik}}]{Zong2019b}%
  \BibitemOpen
  \bibfield  {author} {\bibinfo {author} {\bibfnamefont {A.}~\bibnamefont
  {Zong}}, \bibinfo {author} {\bibfnamefont {P.~E.}\ \bibnamefont {Dolgirev}},
  \bibinfo {author} {\bibfnamefont {A.}~\bibnamefont {Kogar}}, \bibinfo
  {author} {\bibfnamefont {E.}~\bibnamefont {Erge{\c{c}}en}}, \bibinfo {author}
  {\bibfnamefont {M.~B.}\ \bibnamefont {Yilmaz}}, \bibinfo {author}
  {\bibfnamefont {Y.-Q.}\ \bibnamefont {Bie}}, \bibinfo {author} {\bibfnamefont
  {T.}~\bibnamefont {Rohwer}}, \bibinfo {author} {\bibfnamefont {I.-C.}\
  \bibnamefont {Tung}}, \bibinfo {author} {\bibfnamefont {J.}~\bibnamefont
  {Straquadine}}, \bibinfo {author} {\bibfnamefont {X.}~\bibnamefont {Wang}},
  \bibinfo {author} {\bibfnamefont {Y.}~\bibnamefont {Yang}}, \bibinfo {author}
  {\bibfnamefont {X.}~\bibnamefont {Shen}}, \bibinfo {author} {\bibfnamefont
  {R.}~\bibnamefont {Li}}, \bibinfo {author} {\bibfnamefont {J.}~\bibnamefont
  {Yang}}, \bibinfo {author} {\bibfnamefont {S.}~\bibnamefont {Park}}, \bibinfo
  {author} {\bibfnamefont {M.~C.}\ \bibnamefont {Hoffmann}}, \bibinfo {author}
  {\bibfnamefont {B.~K.}\ \bibnamefont {Ofori-Okai}}, \bibinfo {author}
  {\bibfnamefont {M.~E.}\ \bibnamefont {Kozina}}, \bibinfo {author}
  {\bibfnamefont {H.}~\bibnamefont {Wen}}, \bibinfo {author} {\bibfnamefont
  {X.}~\bibnamefont {Wang}}, \bibinfo {author} {\bibfnamefont {I.~R.}\
  \bibnamefont {Fisher}}, \bibinfo {author} {\bibfnamefont {P.}~\bibnamefont
  {Jarillo-Herrero}},\ and\ \bibinfo {author} {\bibfnamefont {N.}~\bibnamefont
  {Gedik}},\ }\bibfield  {title} {\bibinfo {title} {{Dynamical slowing-down in
  an ultrafast photoinduced phase transition}},\ }\href
  {https://doi.org/10.1103/PhysRevLett.123.097601} {\bibfield  {journal}
  {\bibinfo  {journal} {Phys. Rev. Lett.}\ }\textbf {\bibinfo {volume} {123}},\
  \bibinfo {pages} {097601} (\bibinfo {year} {2019}{\natexlab{a}})}\BibitemShut
  {NoStop}%
\bibitem [{\citenamefont {Goldenfeld}(1992)}]{Goldenfeld1992}%
  \BibitemOpen
  \bibfield  {author} {\bibinfo {author} {\bibfnamefont {N.}~\bibnamefont
  {Goldenfeld}},\ }\href {https://doi.org/10.1201/9780429493492} {\emph
  {\bibinfo {title} {{Lectures on Phase Transitions and the Renormalization
  Group}}}}\ (\bibinfo  {publisher} {Westview},\ \bibinfo {address} {Boulder},\
  \bibinfo {year} {1992})\BibitemShut {NoStop}%
\bibitem [{\citenamefont {Dolgirev}\ \emph
  {et~al.}(2020{\natexlab{a}})\citenamefont {Dolgirev}, \citenamefont
  {Michael}, \citenamefont {Zong}, \citenamefont {Gedik},\ and\ \citenamefont
  {Demler}}]{Dolgirev2020b}%
  \BibitemOpen
  \bibfield  {author} {\bibinfo {author} {\bibfnamefont {P.~E.}\ \bibnamefont
  {Dolgirev}}, \bibinfo {author} {\bibfnamefont {M.~H.}\ \bibnamefont
  {Michael}}, \bibinfo {author} {\bibfnamefont {A.}~\bibnamefont {Zong}},
  \bibinfo {author} {\bibfnamefont {N.}~\bibnamefont {Gedik}},\ and\ \bibinfo
  {author} {\bibfnamefont {E.}~\bibnamefont {Demler}},\ }\bibfield  {title}
  {\bibinfo {title} {{Self-similar dynamics of order parameter fluctuations in
  pump-probe experiments}},\ }\href
  {https://doi.org/10.1103/PhysRevB.101.174306} {\bibfield  {journal} {\bibinfo
   {journal} {Phys. Rev. B}\ }\textbf {\bibinfo {volume} {101}},\ \bibinfo
  {pages} {174306} (\bibinfo {year} {2020}{\natexlab{a}})}\BibitemShut
  {NoStop}%
\bibitem [{\citenamefont {Dolgirev}\ \emph
  {et~al.}(2020{\natexlab{b}})\citenamefont {Dolgirev}, \citenamefont
  {Rozhkov}, \citenamefont {Zong}, \citenamefont {Kogar}, \citenamefont
  {Gedik},\ and\ \citenamefont {Fine}}]{Dolgirev2020a}%
  \BibitemOpen
  \bibfield  {author} {\bibinfo {author} {\bibfnamefont {P.~E.}\ \bibnamefont
  {Dolgirev}}, \bibinfo {author} {\bibfnamefont {A.~V.}\ \bibnamefont
  {Rozhkov}}, \bibinfo {author} {\bibfnamefont {A.}~\bibnamefont {Zong}},
  \bibinfo {author} {\bibfnamefont {A.}~\bibnamefont {Kogar}}, \bibinfo
  {author} {\bibfnamefont {N.}~\bibnamefont {Gedik}},\ and\ \bibinfo {author}
  {\bibfnamefont {B.~V.}\ \bibnamefont {Fine}},\ }\bibfield  {title} {\bibinfo
  {title} {{Amplitude dynamics of charge density wave in LaTe$_3$: Theoretical
  description of pump-probe experiments}},\ }\href
  {https://doi.org/10.1103/PhysRevB.101.054203} {\bibfield  {journal} {\bibinfo
   {journal} {Phys. Rev. B}\ }\textbf {\bibinfo {volume} {101}},\ \bibinfo
  {pages} {054203} (\bibinfo {year} {2020}{\natexlab{b}})}\BibitemShut
  {NoStop}%
\bibitem [{\citenamefont {Laulh{\'{e}}}\ \emph {et~al.}(2017)\citenamefont
  {Laulh{\'{e}}}, \citenamefont {Huber}, \citenamefont {Lantz}, \citenamefont
  {Ferrer}, \citenamefont {Mariager}, \citenamefont {Gr{\"{u}}bel},
  \citenamefont {Rittmann}, \citenamefont {Johnson}, \citenamefont {Esposito},
  \citenamefont {L{\"{u}}bcke}, \citenamefont {Huber}, \citenamefont {Kubli},
  \citenamefont {Savoini}, \citenamefont {Jacques}, \citenamefont {Cario},
  \citenamefont {Corraze}, \citenamefont {Janod}, \citenamefont {Ingold},
  \citenamefont {Beaud}, \citenamefont {Johnson},\ and\ \citenamefont
  {Ravy}}]{Laulhe2017}%
  \BibitemOpen
  \bibfield  {author} {\bibinfo {author} {\bibfnamefont {C.}~\bibnamefont
  {Laulh{\'{e}}}}, \bibinfo {author} {\bibfnamefont {T.}~\bibnamefont {Huber}},
  \bibinfo {author} {\bibfnamefont {G.}~\bibnamefont {Lantz}}, \bibinfo
  {author} {\bibfnamefont {A.}~\bibnamefont {Ferrer}}, \bibinfo {author}
  {\bibfnamefont {S.~O.}\ \bibnamefont {Mariager}}, \bibinfo {author}
  {\bibfnamefont {S.}~\bibnamefont {Gr{\"{u}}bel}}, \bibinfo {author}
  {\bibfnamefont {J.}~\bibnamefont {Rittmann}}, \bibinfo {author}
  {\bibfnamefont {J.~A.}\ \bibnamefont {Johnson}}, \bibinfo {author}
  {\bibfnamefont {V.}~\bibnamefont {Esposito}}, \bibinfo {author}
  {\bibfnamefont {A.}~\bibnamefont {L{\"{u}}bcke}}, \bibinfo {author}
  {\bibfnamefont {L.}~\bibnamefont {Huber}}, \bibinfo {author} {\bibfnamefont
  {M.}~\bibnamefont {Kubli}}, \bibinfo {author} {\bibfnamefont
  {M.}~\bibnamefont {Savoini}}, \bibinfo {author} {\bibfnamefont {V.~L.~R.}\
  \bibnamefont {Jacques}}, \bibinfo {author} {\bibfnamefont {L.}~\bibnamefont
  {Cario}}, \bibinfo {author} {\bibfnamefont {B.}~\bibnamefont {Corraze}},
  \bibinfo {author} {\bibfnamefont {E.}~\bibnamefont {Janod}}, \bibinfo
  {author} {\bibfnamefont {G.}~\bibnamefont {Ingold}}, \bibinfo {author}
  {\bibfnamefont {P.}~\bibnamefont {Beaud}}, \bibinfo {author} {\bibfnamefont
  {S.~L.}\ \bibnamefont {Johnson}},\ and\ \bibinfo {author} {\bibfnamefont
  {S.}~\bibnamefont {Ravy}},\ }\bibfield  {title} {\bibinfo {title} {{Ultrafast
  formation of a charge density wave state in 1$T$-TaS$_2$: Observation at
  nanometer scales using time-resolved x-ray diffraction}},\ }\href
  {https://doi.org/10.1103/PhysRevLett.118.247401} {\bibfield  {journal}
  {\bibinfo  {journal} {Phys. Rev. Lett.}\ }\textbf {\bibinfo {volume} {118}},\
  \bibinfo {pages} {247401} (\bibinfo {year} {2017})}\BibitemShut {NoStop}%
\bibitem [{\citenamefont {Vogelgesang}\ \emph {et~al.}(2018)\citenamefont
  {Vogelgesang}, \citenamefont {Storeck}, \citenamefont {Horstmann},
  \citenamefont {Diekmann}, \citenamefont {Sivis}, \citenamefont {Schramm},
  \citenamefont {Rossnagel}, \citenamefont {Sch{\"a}fer},\ and\ \citenamefont
  {Ropers}}]{Vogelgesang2018}%
  \BibitemOpen
  \bibfield  {author} {\bibinfo {author} {\bibfnamefont {S.}~\bibnamefont
  {Vogelgesang}}, \bibinfo {author} {\bibfnamefont {G.}~\bibnamefont
  {Storeck}}, \bibinfo {author} {\bibfnamefont {J.~G.}\ \bibnamefont
  {Horstmann}}, \bibinfo {author} {\bibfnamefont {T.}~\bibnamefont {Diekmann}},
  \bibinfo {author} {\bibfnamefont {M.}~\bibnamefont {Sivis}}, \bibinfo
  {author} {\bibfnamefont {S.}~\bibnamefont {Schramm}}, \bibinfo {author}
  {\bibfnamefont {K.}~\bibnamefont {Rossnagel}}, \bibinfo {author}
  {\bibfnamefont {S.}~\bibnamefont {Sch{\"a}fer}},\ and\ \bibinfo {author}
  {\bibfnamefont {C.}~\bibnamefont {Ropers}},\ }\bibfield  {title} {\bibinfo
  {title} {{Phase ordering of charge density waves traced by ultrafast
  low-energy electron diffraction}},\ }\href
  {https://doi.org/10.1038/nphys4309} {\bibfield  {journal} {\bibinfo
  {journal} {Nat. Phys.}\ }\textbf {\bibinfo {volume} {14}},\ \bibinfo {pages}
  {184} (\bibinfo {year} {2018})}\BibitemShut {NoStop}%
\bibitem [{\citenamefont {Mitrano}\ \emph {et~al.}(2019)\citenamefont
  {Mitrano}, \citenamefont {Lee}, \citenamefont {Husain}, \citenamefont
  {Delacretaz}, \citenamefont {Zhu}, \citenamefont {{de la Pe{\~{n}}a Munoz}},
  \citenamefont {Sun}, \citenamefont {Joe}, \citenamefont {Reid}, \citenamefont
  {Wandel}, \citenamefont {Coslovich}, \citenamefont {Schlotter}, \citenamefont
  {van Driel}, \citenamefont {Schneeloch}, \citenamefont {Gu}, \citenamefont
  {Hartnoll}, \citenamefont {Goldenfeld},\ and\ \citenamefont
  {Abbamonte}}]{Mitrano2019}%
  \BibitemOpen
  \bibfield  {author} {\bibinfo {author} {\bibfnamefont {M.}~\bibnamefont
  {Mitrano}}, \bibinfo {author} {\bibfnamefont {S.}~\bibnamefont {Lee}},
  \bibinfo {author} {\bibfnamefont {A.~A.}\ \bibnamefont {Husain}}, \bibinfo
  {author} {\bibfnamefont {L.}~\bibnamefont {Delacretaz}}, \bibinfo {author}
  {\bibfnamefont {M.}~\bibnamefont {Zhu}}, \bibinfo {author} {\bibfnamefont
  {G.}~\bibnamefont {{de la Pe{\~{n}}a Munoz}}}, \bibinfo {author}
  {\bibfnamefont {S.~X.-L.}\ \bibnamefont {Sun}}, \bibinfo {author}
  {\bibfnamefont {Y.~I.}\ \bibnamefont {Joe}}, \bibinfo {author} {\bibfnamefont
  {A.~H.}\ \bibnamefont {Reid}}, \bibinfo {author} {\bibfnamefont {S.~F.}\
  \bibnamefont {Wandel}}, \bibinfo {author} {\bibfnamefont {G.}~\bibnamefont
  {Coslovich}}, \bibinfo {author} {\bibfnamefont {W.}~\bibnamefont
  {Schlotter}}, \bibinfo {author} {\bibfnamefont {T.}~\bibnamefont {van
  Driel}}, \bibinfo {author} {\bibfnamefont {J.}~\bibnamefont {Schneeloch}},
  \bibinfo {author} {\bibfnamefont {G.~D.}\ \bibnamefont {Gu}}, \bibinfo
  {author} {\bibfnamefont {S.}~\bibnamefont {Hartnoll}}, \bibinfo {author}
  {\bibfnamefont {N.}~\bibnamefont {Goldenfeld}},\ and\ \bibinfo {author}
  {\bibfnamefont {P.}~\bibnamefont {Abbamonte}},\ }\bibfield  {title} {\bibinfo
  {title} {{Ultrafast time-resolved x-ray scattering reveals diffusive charge
  order dynamics in La$_{2-x}$Ba$_x$CuO$_4$}},\ }\href
  {https://doi.org/10.1126/sciadv.aax3346} {\bibfield  {journal} {\bibinfo
  {journal} {Sci. Adv.}\ }\textbf {\bibinfo {volume} {5}},\ \bibinfo {pages}
  {eaax3346} (\bibinfo {year} {2019})}\BibitemShut {NoStop}%
\bibitem [{\citenamefont {Minor}\ \emph {et~al.}(1989)\citenamefont {Minor},
  \citenamefont {Chapman}, \citenamefont {Ehrlich},\ and\ \citenamefont
  {Colella}}]{Minor1989}%
  \BibitemOpen
  \bibfield  {author} {\bibinfo {author} {\bibfnamefont {W.}~\bibnamefont
  {Minor}}, \bibinfo {author} {\bibfnamefont {L.~D.}\ \bibnamefont {Chapman}},
  \bibinfo {author} {\bibfnamefont {S.~N.}\ \bibnamefont {Ehrlich}},\ and\
  \bibinfo {author} {\bibfnamefont {R.}~\bibnamefont {Colella}},\ }\bibfield
  {title} {\bibinfo {title} {{Phason velocities in TaS$_2$ by X-ray diffuse
  scattering}},\ }\href {https://doi.org/10.1103/PhysRevB.39.1360} {\bibfield
  {journal} {\bibinfo  {journal} {Phys. Rev. B}\ }\textbf {\bibinfo {volume}
  {39}},\ \bibinfo {pages} {1360} (\bibinfo {year} {1989})}\BibitemShut
  {NoStop}%
\bibitem [{\citenamefont {Pouget}\ \emph {et~al.}(1991)\citenamefont {Pouget},
  \citenamefont {Hennion}, \citenamefont {Escribe-Filippini},\ and\
  \citenamefont {Sato}}]{Pouget1991}%
  \BibitemOpen
  \bibfield  {author} {\bibinfo {author} {\bibfnamefont {J.~P.}\ \bibnamefont
  {Pouget}}, \bibinfo {author} {\bibfnamefont {B.}~\bibnamefont {Hennion}},
  \bibinfo {author} {\bibfnamefont {C.}~\bibnamefont {Escribe-Filippini}},\
  and\ \bibinfo {author} {\bibfnamefont {M.}~\bibnamefont {Sato}},\ }\bibfield
  {title} {\bibinfo {title} {{Neutron-scattering investigations of the Kohn
  anomaly and of the phase and amplitude charge-density-wave excitations of the
  blue bronze K$_{0.3}$MoO$_3$}},\ }\href
  {https://doi.org/10.1103/PhysRevB.43.8421} {\bibfield  {journal} {\bibinfo
  {journal} {Phys. Rev. B}\ }\textbf {\bibinfo {volume} {43}},\ \bibinfo
  {pages} {8421} (\bibinfo {year} {1991})}\BibitemShut {NoStop}%
\bibitem [{\citenamefont {Ru}\ and\ \citenamefont {Fisher}(2006)}]{Ru2006}%
  \BibitemOpen
  \bibfield  {author} {\bibinfo {author} {\bibfnamefont {N.}~\bibnamefont
  {Ru}}\ and\ \bibinfo {author} {\bibfnamefont {I.~R.}\ \bibnamefont
  {Fisher}},\ }\bibfield  {title} {\bibinfo {title} {{Thermodynamic and
  transport properties of YTe$_3$, LaTe$_3$, CeTe$_3$}},\ }\href
  {https://doi.org/10.1103/PhysRevB.73.033101} {\bibfield  {journal} {\bibinfo
  {journal} {Phys. Rev. B}\ }\textbf {\bibinfo {volume} {73}},\ \bibinfo
  {pages} {033101} (\bibinfo {year} {2006})}\BibitemShut {NoStop}%
\bibitem [{\citenamefont {Bie}\ \emph {et~al.}(2021)\citenamefont {Bie},
  \citenamefont {Zong}, \citenamefont {Wang}, \citenamefont {Jarillo-Herrero},\
  and\ \citenamefont {Gedik}}]{Bie2021}%
  \BibitemOpen
  \bibfield  {author} {\bibinfo {author} {\bibfnamefont {Y.-Q.}\ \bibnamefont
  {Bie}}, \bibinfo {author} {\bibfnamefont {A.}~\bibnamefont {Zong}}, \bibinfo
  {author} {\bibfnamefont {X.}~\bibnamefont {Wang}}, \bibinfo {author}
  {\bibfnamefont {P.}~\bibnamefont {Jarillo-Herrero}},\ and\ \bibinfo {author}
  {\bibfnamefont {N.}~\bibnamefont {Gedik}},\ }\bibfield  {title} {\bibinfo
  {title} {{A versatile sample fabrication method for ultrafast electron
  diffraction}},\ }\href {https://doi.org/10.1016/j.ultramic.2021.113389}
  {\bibfield  {journal} {\bibinfo  {journal} {Ultramicroscopy}\ }\textbf
  {\bibinfo {volume} {230}},\ \bibinfo {pages} {113389} (\bibinfo {year}
  {2021})}\BibitemShut {NoStop}%
\bibitem [{\citenamefont {Weathersby}\ \emph {et~al.}(2015)\citenamefont
  {Weathersby}, \citenamefont {Brown}, \citenamefont {Centurion}, \citenamefont
  {Chase}, \citenamefont {Coffee}, \citenamefont {Corbett}, \citenamefont
  {Eichner}, \citenamefont {Frisch}, \citenamefont {Fry}, \citenamefont
  {G{\"{u}}hr}, \citenamefont {Hartmann}, \citenamefont {Hast}, \citenamefont
  {Hettel}, \citenamefont {Jobe}, \citenamefont {Jongewaard}, \citenamefont
  {Lewandowski}, \citenamefont {Li}, \citenamefont {Lindenberg}, \citenamefont
  {Makasyuk}, \citenamefont {May}, \citenamefont {McCormick}, \citenamefont
  {Nguyen}, \citenamefont {Reid}, \citenamefont {Shen}, \citenamefont
  {Sokolowski-Tinten}, \citenamefont {Vecchione}, \citenamefont {Vetter},
  \citenamefont {Wu}, \citenamefont {Yang}, \citenamefont {D{\"{u}}rr},\ and\
  \citenamefont {Wang}}]{Weathersby2015}%
  \BibitemOpen
  \bibfield  {author} {\bibinfo {author} {\bibfnamefont {S.~P.}\ \bibnamefont
  {Weathersby}}, \bibinfo {author} {\bibfnamefont {G.}~\bibnamefont {Brown}},
  \bibinfo {author} {\bibfnamefont {M.}~\bibnamefont {Centurion}}, \bibinfo
  {author} {\bibfnamefont {T.~F.}\ \bibnamefont {Chase}}, \bibinfo {author}
  {\bibfnamefont {R.}~\bibnamefont {Coffee}}, \bibinfo {author} {\bibfnamefont
  {J.}~\bibnamefont {Corbett}}, \bibinfo {author} {\bibfnamefont {J.~P.}\
  \bibnamefont {Eichner}}, \bibinfo {author} {\bibfnamefont {J.~C.}\
  \bibnamefont {Frisch}}, \bibinfo {author} {\bibfnamefont {A.~R.}\
  \bibnamefont {Fry}}, \bibinfo {author} {\bibfnamefont {M.}~\bibnamefont
  {G{\"{u}}hr}}, \bibinfo {author} {\bibfnamefont {N.}~\bibnamefont
  {Hartmann}}, \bibinfo {author} {\bibfnamefont {C.}~\bibnamefont {Hast}},
  \bibinfo {author} {\bibfnamefont {R.}~\bibnamefont {Hettel}}, \bibinfo
  {author} {\bibfnamefont {R.~K.}\ \bibnamefont {Jobe}}, \bibinfo {author}
  {\bibfnamefont {E.~N.}\ \bibnamefont {Jongewaard}}, \bibinfo {author}
  {\bibfnamefont {J.~R.}\ \bibnamefont {Lewandowski}}, \bibinfo {author}
  {\bibfnamefont {R.~K.}\ \bibnamefont {Li}}, \bibinfo {author} {\bibfnamefont
  {A.~M.}\ \bibnamefont {Lindenberg}}, \bibinfo {author} {\bibfnamefont
  {I.}~\bibnamefont {Makasyuk}}, \bibinfo {author} {\bibfnamefont {J.~E.}\
  \bibnamefont {May}}, \bibinfo {author} {\bibfnamefont {D.}~\bibnamefont
  {McCormick}}, \bibinfo {author} {\bibfnamefont {M.~N.}\ \bibnamefont
  {Nguyen}}, \bibinfo {author} {\bibfnamefont {A.~H.}\ \bibnamefont {Reid}},
  \bibinfo {author} {\bibfnamefont {X.}~\bibnamefont {Shen}}, \bibinfo {author}
  {\bibfnamefont {K.}~\bibnamefont {Sokolowski-Tinten}}, \bibinfo {author}
  {\bibfnamefont {T.}~\bibnamefont {Vecchione}}, \bibinfo {author}
  {\bibfnamefont {S.~L.}\ \bibnamefont {Vetter}}, \bibinfo {author}
  {\bibfnamefont {J.}~\bibnamefont {Wu}}, \bibinfo {author} {\bibfnamefont
  {J.}~\bibnamefont {Yang}}, \bibinfo {author} {\bibfnamefont {H.~A.}\
  \bibnamefont {D{\"{u}}rr}},\ and\ \bibinfo {author} {\bibfnamefont {X.~J.}\
  \bibnamefont {Wang}},\ }\bibfield  {title} {\bibinfo {title}
  {{Mega-electron-volt ultrafast electron diffraction at SLAC National
  Accelerator Laboratory}},\ }\href {https://doi.org/10.1063/1.4926994}
  {\bibfield  {journal} {\bibinfo  {journal} {Rev. Sci. Instrum.}\ }\textbf
  {\bibinfo {volume} {86}},\ \bibinfo {pages} {073702} (\bibinfo {year}
  {2015})}\BibitemShut {NoStop}%
\bibitem [{\citenamefont {Shen}\ \emph {et~al.}(2018)\citenamefont {Shen},
  \citenamefont {Li}, \citenamefont {Lundstr{\"{o}}m}, \citenamefont {Lane},
  \citenamefont {Reid}, \citenamefont {Weathersby},\ and\ \citenamefont
  {Wang}}]{Shen2018}%
  \BibitemOpen
  \bibfield  {author} {\bibinfo {author} {\bibfnamefont {X.}~\bibnamefont
  {Shen}}, \bibinfo {author} {\bibfnamefont {R.~K.}\ \bibnamefont {Li}},
  \bibinfo {author} {\bibfnamefont {U.}~\bibnamefont {Lundstr{\"{o}}m}},
  \bibinfo {author} {\bibfnamefont {T.~J.}\ \bibnamefont {Lane}}, \bibinfo
  {author} {\bibfnamefont {A.~H.}\ \bibnamefont {Reid}}, \bibinfo {author}
  {\bibfnamefont {S.~P.}\ \bibnamefont {Weathersby}},\ and\ \bibinfo {author}
  {\bibfnamefont {X.~J.}\ \bibnamefont {Wang}},\ }\bibfield  {title} {\bibinfo
  {title} {{Femtosecond mega-electron-volt electron microdiffraction}},\ }\href
  {https://doi.org/10.1016/j.ultramic.2017.08.019} {\bibfield  {journal}
  {\bibinfo  {journal} {Ultramicroscopy}\ }\textbf {\bibinfo {volume} {184}},\
  \bibinfo {pages} {172} (\bibinfo {year} {2018})}\BibitemShut {NoStop}%
\bibitem [{\citenamefont {Moore}\ \emph {et~al.}(2016)\citenamefont {Moore},
  \citenamefont {Lee}, \citenamefont {Kirchman}, \citenamefont {Chuang},
  \citenamefont {Kemper}, \citenamefont {Trigo}, \citenamefont {Patthey},
  \citenamefont {Lu}, \citenamefont {Krupin}, \citenamefont {Yi}, \citenamefont
  {Reis}, \citenamefont {Doering}, \citenamefont {Denes}, \citenamefont
  {Schlotter}, \citenamefont {Turner}, \citenamefont {Hays}, \citenamefont
  {Hering}, \citenamefont {Benson}, \citenamefont {Chu}, \citenamefont
  {Devereaux}, \citenamefont {Fisher}, \citenamefont {Hussain},\ and\
  \citenamefont {Shen}}]{Moore2016}%
  \BibitemOpen
  \bibfield  {author} {\bibinfo {author} {\bibfnamefont {R.~G.}\ \bibnamefont
  {Moore}}, \bibinfo {author} {\bibfnamefont {W.~S.}\ \bibnamefont {Lee}},
  \bibinfo {author} {\bibfnamefont {P.~S.}\ \bibnamefont {Kirchman}}, \bibinfo
  {author} {\bibfnamefont {Y.~D.}\ \bibnamefont {Chuang}}, \bibinfo {author}
  {\bibfnamefont {A.~F.}\ \bibnamefont {Kemper}}, \bibinfo {author}
  {\bibfnamefont {M.}~\bibnamefont {Trigo}}, \bibinfo {author} {\bibfnamefont
  {L.}~\bibnamefont {Patthey}}, \bibinfo {author} {\bibfnamefont {D.~H.}\
  \bibnamefont {Lu}}, \bibinfo {author} {\bibfnamefont {O.}~\bibnamefont
  {Krupin}}, \bibinfo {author} {\bibfnamefont {M.}~\bibnamefont {Yi}}, \bibinfo
  {author} {\bibfnamefont {D.~A.}\ \bibnamefont {Reis}}, \bibinfo {author}
  {\bibfnamefont {D.}~\bibnamefont {Doering}}, \bibinfo {author} {\bibfnamefont
  {P.}~\bibnamefont {Denes}}, \bibinfo {author} {\bibfnamefont {W.~F.}\
  \bibnamefont {Schlotter}}, \bibinfo {author} {\bibfnamefont {J.~J.}\
  \bibnamefont {Turner}}, \bibinfo {author} {\bibfnamefont {G.}~\bibnamefont
  {Hays}}, \bibinfo {author} {\bibfnamefont {P.}~\bibnamefont {Hering}},
  \bibinfo {author} {\bibfnamefont {T.}~\bibnamefont {Benson}}, \bibinfo
  {author} {\bibfnamefont {J.-H.}\ \bibnamefont {Chu}}, \bibinfo {author}
  {\bibfnamefont {T.~P.}\ \bibnamefont {Devereaux}}, \bibinfo {author}
  {\bibfnamefont {I.~R.}\ \bibnamefont {Fisher}}, \bibinfo {author}
  {\bibfnamefont {Z.}~\bibnamefont {Hussain}},\ and\ \bibinfo {author}
  {\bibfnamefont {Z.-X.}\ \bibnamefont {Shen}},\ }\bibfield  {title} {\bibinfo
  {title} {{Ultrafast resonant soft X-ray diffraction dynamics of the charge
  density wave in TbTe$_3$}},\ }\href
  {https://doi.org/10.1103/PhysRevB.93.024304} {\bibfield  {journal} {\bibinfo
  {journal} {Phys. Rev. B}\ }\textbf {\bibinfo {volume} {93}},\ \bibinfo
  {pages} {024304} (\bibinfo {year} {2016})}\BibitemShut {NoStop}%
\bibitem [{\citenamefont {Sacchetti}\ \emph {et~al.}(2006)\citenamefont
  {Sacchetti}, \citenamefont {Degiorgi}, \citenamefont {Giamarchi},
  \citenamefont {Ru},\ and\ \citenamefont {Fisher}}]{Sacchetti2006}%
  \BibitemOpen
  \bibfield  {author} {\bibinfo {author} {\bibfnamefont {A.}~\bibnamefont
  {Sacchetti}}, \bibinfo {author} {\bibfnamefont {L.}~\bibnamefont {Degiorgi}},
  \bibinfo {author} {\bibfnamefont {T.}~\bibnamefont {Giamarchi}}, \bibinfo
  {author} {\bibfnamefont {N.}~\bibnamefont {Ru}},\ and\ \bibinfo {author}
  {\bibfnamefont {I.~R.}\ \bibnamefont {Fisher}},\ }\bibfield  {title}
  {\bibinfo {title} {{Chemical pressure and hidden one-dimensional behavior in
  rare-earth tri-telluride charge-density wave compounds}},\ }\href
  {https://doi.org/10.1103/PhysRevB.74.125115} {\bibfield  {journal} {\bibinfo
  {journal} {Phys. Rev. B}\ }\textbf {\bibinfo {volume} {74}},\ \bibinfo
  {pages} {125115} (\bibinfo {year} {2006})}\BibitemShut {NoStop}%
\bibitem [{\citenamefont {Ramsey}\ \emph {et~al.}(1965)\citenamefont {Ramsey},
  \citenamefont {Steinfink},\ and\ \citenamefont {Weiss}}]{Ramsey1965}%
  \BibitemOpen
  \bibfield  {author} {\bibinfo {author} {\bibfnamefont {T.~H.}\ \bibnamefont
  {Ramsey}}, \bibinfo {author} {\bibfnamefont {H.}~\bibnamefont {Steinfink}},\
  and\ \bibinfo {author} {\bibfnamefont {E.~J.}\ \bibnamefont {Weiss}},\
  }\bibfield  {title} {\bibinfo {title} {{Thermoelectric and electrical
  measurements in the La-Te System}},\ }\href
  {https://doi.org/10.1063/1.1714028} {\bibfield  {journal} {\bibinfo
  {journal} {J. Appl. Phys.}\ }\textbf {\bibinfo {volume} {36}},\ \bibinfo
  {pages} {548} (\bibinfo {year} {1965})}\BibitemShut {NoStop}%
\bibitem [{\citenamefont {{Ren{\'{e}} de Cotret}}\ \emph
  {et~al.}(2019)\citenamefont {{Ren{\'{e}} de Cotret}}, \citenamefont
  {P{\"{o}}hls}, \citenamefont {Stern}, \citenamefont {Otto}, \citenamefont
  {Sutton},\ and\ \citenamefont {Siwick}}]{RenedeCotret2019}%
  \BibitemOpen
  \bibfield  {author} {\bibinfo {author} {\bibfnamefont {L.~P.}\ \bibnamefont
  {{Ren{\'{e}} de Cotret}}}, \bibinfo {author} {\bibfnamefont {J.-H.}\
  \bibnamefont {P{\"{o}}hls}}, \bibinfo {author} {\bibfnamefont {M.~J.}\
  \bibnamefont {Stern}}, \bibinfo {author} {\bibfnamefont {M.~R.}\ \bibnamefont
  {Otto}}, \bibinfo {author} {\bibfnamefont {M.}~\bibnamefont {Sutton}},\ and\
  \bibinfo {author} {\bibfnamefont {B.~J.}\ \bibnamefont {Siwick}},\ }\bibfield
   {title} {\bibinfo {title} {{Time- and momentum-resolved phonon population
  dynamics with ultrafast electron diffuse scattering}},\ }\href
  {https://doi.org/10.1103/PhysRevB.100.214115} {\bibfield  {journal} {\bibinfo
   {journal} {Phys. Rev. B}\ }\textbf {\bibinfo {volume} {100}},\ \bibinfo
  {pages} {214115} (\bibinfo {year} {2019})}\BibitemShut {NoStop}%
\bibitem [{\citenamefont {Maschek}\ \emph {et~al.}(2015)\citenamefont
  {Maschek}, \citenamefont {Rosenkranz}, \citenamefont {Heid}, \citenamefont
  {Said}, \citenamefont {Giraldo-Gallo}, \citenamefont {Fisher},\ and\
  \citenamefont {Weber}}]{Maschek2015}%
  \BibitemOpen
  \bibfield  {author} {\bibinfo {author} {\bibfnamefont {M.}~\bibnamefont
  {Maschek}}, \bibinfo {author} {\bibfnamefont {S.}~\bibnamefont {Rosenkranz}},
  \bibinfo {author} {\bibfnamefont {R.}~\bibnamefont {Heid}}, \bibinfo {author}
  {\bibfnamefont {A.~H.}\ \bibnamefont {Said}}, \bibinfo {author}
  {\bibfnamefont {P.}~\bibnamefont {Giraldo-Gallo}}, \bibinfo {author}
  {\bibfnamefont {I.~R.}\ \bibnamefont {Fisher}},\ and\ \bibinfo {author}
  {\bibfnamefont {F.}~\bibnamefont {Weber}},\ }\bibfield  {title} {\bibinfo
  {title} {{Wave-vector-dependent electron-phonon coupling and the
  charge-density-wave transition in TbTe$_3$}},\ }\href
  {https://doi.org/10.1103/PhysRevB.91.235146} {\bibfield  {journal} {\bibinfo
  {journal} {Phys. Rev. B}\ }\textbf {\bibinfo {volume} {91}},\ \bibinfo
  {pages} {235146} (\bibinfo {year} {2015})}\BibitemShut {NoStop}%
\bibitem [{\citenamefont {Stern}\ \emph {et~al.}(2018)\citenamefont {Stern},
  \citenamefont {{Ren{\'{e}} de Cotret}}, \citenamefont {Otto}, \citenamefont
  {Chatelain}, \citenamefont {Boisvert}, \citenamefont {Sutton},\ and\
  \citenamefont {Siwick}}]{Stern2018}%
  \BibitemOpen
  \bibfield  {author} {\bibinfo {author} {\bibfnamefont {M.~J.}\ \bibnamefont
  {Stern}}, \bibinfo {author} {\bibfnamefont {L.~P.}\ \bibnamefont {{Ren{\'{e}}
  de Cotret}}}, \bibinfo {author} {\bibfnamefont {M.~R.}\ \bibnamefont {Otto}},
  \bibinfo {author} {\bibfnamefont {R.~P.}\ \bibnamefont {Chatelain}}, \bibinfo
  {author} {\bibfnamefont {J.-P.}\ \bibnamefont {Boisvert}}, \bibinfo {author}
  {\bibfnamefont {M.}~\bibnamefont {Sutton}},\ and\ \bibinfo {author}
  {\bibfnamefont {B.~J.}\ \bibnamefont {Siwick}},\ }\bibfield  {title}
  {\bibinfo {title} {{Mapping momentum-dependent electron-phonon coupling and
  nonequilibrium phonon dynamics with ultrafast electron diffuse scattering}},\
  }\href {https://doi.org/10.1103/PhysRevB.97.165416} {\bibfield  {journal}
  {\bibinfo  {journal} {Phys. Rev. B}\ }\textbf {\bibinfo {volume} {97}},\
  \bibinfo {pages} {165416} (\bibinfo {year} {2018})}\BibitemShut {NoStop}%
\bibitem [{\citenamefont {Mazenko}\ and\ \citenamefont
  {Zannetti}(1985)}]{Mazenko1985}%
  \BibitemOpen
  \bibfield  {author} {\bibinfo {author} {\bibfnamefont {G.~F.}\ \bibnamefont
  {Mazenko}}\ and\ \bibinfo {author} {\bibfnamefont {M.}~\bibnamefont
  {Zannetti}},\ }\bibfield  {title} {\bibinfo {title} {{Instability, spinodal
  decomposition, and nucleation in a system with continuous symmetry}},\ }\href
  {https://doi.org/10.1103/PhysRevB.32.4565} {\bibfield  {journal} {\bibinfo
  {journal} {Phys. Rev. B}\ }\textbf {\bibinfo {volume} {32}},\ \bibinfo
  {pages} {4565} (\bibinfo {year} {1985})}\BibitemShut {NoStop}%
\bibitem [{\citenamefont {Bray}(1994)}]{Bray1994}%
  \BibitemOpen
  \bibfield  {author} {\bibinfo {author} {\bibfnamefont {A.~J.}\ \bibnamefont
  {Bray}},\ }\bibfield  {title} {\bibinfo {title} {{Theory of phase-ordering
  kinetics}},\ }\href {https://doi.org/10.1080/00018739400101505} {\bibfield
  {journal} {\bibinfo  {journal} {Adv. Phys.}\ }\textbf {\bibinfo {volume}
  {43}},\ \bibinfo {pages} {357} (\bibinfo {year} {1994})}\BibitemShut
  {NoStop}%
\bibitem [{\citenamefont {Zong}\ \emph
  {et~al.}(2019{\natexlab{b}})\citenamefont {Zong}, \citenamefont {Kogar},
  \citenamefont {Bie}, \citenamefont {Rohwer}, \citenamefont {Lee},
  \citenamefont {Baldini}, \citenamefont {Erge{\c{c}}en}, \citenamefont
  {Yilmaz}, \citenamefont {Freelon}, \citenamefont {Sie}, \citenamefont {Zhou},
  \citenamefont {Straquadine}, \citenamefont {Walmsley}, \citenamefont
  {Dolgirev}, \citenamefont {Rozhkov}, \citenamefont {Fisher}, \citenamefont
  {Jarillo-Herrero}, \citenamefont {Fine},\ and\ \citenamefont
  {Gedik}}]{Zong2019a}%
  \BibitemOpen
  \bibfield  {author} {\bibinfo {author} {\bibfnamefont {A.}~\bibnamefont
  {Zong}}, \bibinfo {author} {\bibfnamefont {A.}~\bibnamefont {Kogar}},
  \bibinfo {author} {\bibfnamefont {Y.-Q.}\ \bibnamefont {Bie}}, \bibinfo
  {author} {\bibfnamefont {T.}~\bibnamefont {Rohwer}}, \bibinfo {author}
  {\bibfnamefont {C.}~\bibnamefont {Lee}}, \bibinfo {author} {\bibfnamefont
  {E.}~\bibnamefont {Baldini}}, \bibinfo {author} {\bibfnamefont
  {E.}~\bibnamefont {Erge{\c{c}}en}}, \bibinfo {author} {\bibfnamefont {M.~B.}\
  \bibnamefont {Yilmaz}}, \bibinfo {author} {\bibfnamefont {B.}~\bibnamefont
  {Freelon}}, \bibinfo {author} {\bibfnamefont {E.~J.}\ \bibnamefont {Sie}},
  \bibinfo {author} {\bibfnamefont {H.}~\bibnamefont {Zhou}}, \bibinfo {author}
  {\bibfnamefont {J.}~\bibnamefont {Straquadine}}, \bibinfo {author}
  {\bibfnamefont {P.}~\bibnamefont {Walmsley}}, \bibinfo {author}
  {\bibfnamefont {P.~E.}\ \bibnamefont {Dolgirev}}, \bibinfo {author}
  {\bibfnamefont {A.~V.}\ \bibnamefont {Rozhkov}}, \bibinfo {author}
  {\bibfnamefont {I.~R.}\ \bibnamefont {Fisher}}, \bibinfo {author}
  {\bibfnamefont {P.}~\bibnamefont {Jarillo-Herrero}}, \bibinfo {author}
  {\bibfnamefont {B.~V.}\ \bibnamefont {Fine}},\ and\ \bibinfo {author}
  {\bibfnamefont {N.}~\bibnamefont {Gedik}},\ }\bibfield  {title} {\bibinfo
  {title} {{Evidence for topological defects in a photoinduced phase
  transition}},\ }\href {https://doi.org/10.1038/s41567-018-0311-9} {\bibfield
  {journal} {\bibinfo  {journal} {Nat. Phys.}\ }\textbf {\bibinfo {volume}
  {15}},\ \bibinfo {pages} {27} (\bibinfo {year}
  {2019}{\natexlab{b}})}\BibitemShut {NoStop}%
\bibitem [{\citenamefont {Chaikin}\ and\ \citenamefont
  {Lubensky}(1995)}]{Chaikin1995}%
  \BibitemOpen
  \bibfield  {author} {\bibinfo {author} {\bibfnamefont {P.~M.}\ \bibnamefont
  {Chaikin}}\ and\ \bibinfo {author} {\bibfnamefont {T.~C.}\ \bibnamefont
  {Lubensky}},\ }\bibfield  {title} {\bibinfo {title} {{Mean-field theory}},\
  }in\ \href {https://doi.org/10.1017/CBO9780511813467.005} {\emph {\bibinfo
  {booktitle} {Principles of Condensed Matter Physics}}}\ (\bibinfo
  {publisher} {Cambridge University Press},\ \bibinfo {address} {Cambridge},\
  \bibinfo {year} {1995})\ pp.\ \bibinfo {pages} {144--212}\BibitemShut
  {NoStop}%
\bibitem [{\citenamefont {Hohenberg}\ and\ \citenamefont
  {Halperin}(1977)}]{Hohenberg1977}%
  \BibitemOpen
  \bibfield  {author} {\bibinfo {author} {\bibfnamefont {P.~C.}\ \bibnamefont
  {Hohenberg}}\ and\ \bibinfo {author} {\bibfnamefont {B.~I.}\ \bibnamefont
  {Halperin}},\ }\bibfield  {title} {\bibinfo {title} {{Theory of dynamic
  critical phenomena}},\ }\href {https://doi.org/10.1103/RevModPhys.49.435}
  {\bibfield  {journal} {\bibinfo  {journal} {Rev. Mod. Phys.}\ }\textbf
  {\bibinfo {volume} {49}},\ \bibinfo {pages} {435} (\bibinfo {year}
  {1977})}\BibitemShut {NoStop}%
\bibitem [{\citenamefont {Sun}\ and\ \citenamefont {Millis}(2020)}]{Sun2020}%
  \BibitemOpen
  \bibfield  {author} {\bibinfo {author} {\bibfnamefont {Z.}~\bibnamefont
  {Sun}}\ and\ \bibinfo {author} {\bibfnamefont {A.~J.}\ \bibnamefont
  {Millis}},\ }\bibfield  {title} {\bibinfo {title} {{Transient trapping into
  metastable states in systems with competing orders}},\ }\href
  {https://doi.org/10.1103/PhysRevX.10.021028} {\bibfield  {journal} {\bibinfo
  {journal} {Phys. Rev. X}\ }\textbf {\bibinfo {volume} {10}},\ \bibinfo
  {pages} {021028} (\bibinfo {year} {2020})}\BibitemShut {NoStop}%
\bibitem [{\citenamefont {Trigo}\ \emph {et~al.}(2019)\citenamefont {Trigo},
  \citenamefont {Giraldo-Gallo}, \citenamefont {Kozina}, \citenamefont
  {Henighan}, \citenamefont {Jiang}, \citenamefont {Liu}, \citenamefont
  {Clark}, \citenamefont {Chollet}, \citenamefont {Glownia}, \citenamefont
  {Zhu}, \citenamefont {Katayama}, \citenamefont {Leuenberger}, \citenamefont
  {Kirchmann}, \citenamefont {Fisher}, \citenamefont {Shen},\ and\
  \citenamefont {Reis}}]{Trigo2019}%
  \BibitemOpen
  \bibfield  {author} {\bibinfo {author} {\bibfnamefont {M.}~\bibnamefont
  {Trigo}}, \bibinfo {author} {\bibfnamefont {P.}~\bibnamefont
  {Giraldo-Gallo}}, \bibinfo {author} {\bibfnamefont {M.~E.}\ \bibnamefont
  {Kozina}}, \bibinfo {author} {\bibfnamefont {T.}~\bibnamefont {Henighan}},
  \bibinfo {author} {\bibfnamefont {M.~P.}\ \bibnamefont {Jiang}}, \bibinfo
  {author} {\bibfnamefont {H.}~\bibnamefont {Liu}}, \bibinfo {author}
  {\bibfnamefont {J.~N.}\ \bibnamefont {Clark}}, \bibinfo {author}
  {\bibfnamefont {M.}~\bibnamefont {Chollet}}, \bibinfo {author} {\bibfnamefont
  {J.~M.}\ \bibnamefont {Glownia}}, \bibinfo {author} {\bibfnamefont
  {D.}~\bibnamefont {Zhu}}, \bibinfo {author} {\bibfnamefont {T.}~\bibnamefont
  {Katayama}}, \bibinfo {author} {\bibfnamefont {D.}~\bibnamefont
  {Leuenberger}}, \bibinfo {author} {\bibfnamefont {P.~S.}\ \bibnamefont
  {Kirchmann}}, \bibinfo {author} {\bibfnamefont {I.~R.}\ \bibnamefont
  {Fisher}}, \bibinfo {author} {\bibfnamefont {Z.~X.}\ \bibnamefont {Shen}},\
  and\ \bibinfo {author} {\bibfnamefont {D.~A.}\ \bibnamefont {Reis}},\
  }\bibfield  {title} {\bibinfo {title} {{Coherent order parameter dynamics in
  SmTe$_3$}},\ }\href {https://doi.org/10.1103/PhysRevB.99.104111} {\bibfield
  {journal} {\bibinfo  {journal} {Phys. Rev. B}\ }\textbf {\bibinfo {volume}
  {99}},\ \bibinfo {pages} {104111} (\bibinfo {year} {2019})}\BibitemShut
  {NoStop}%
\bibitem [{\citenamefont {Holt}\ \emph {et~al.}(2001)\citenamefont {Holt},
  \citenamefont {Zschack}, \citenamefont {Hong}, \citenamefont {Chou},\ and\
  \citenamefont {Chiang}}]{Holt2001}%
  \BibitemOpen
  \bibfield  {author} {\bibinfo {author} {\bibfnamefont {M.}~\bibnamefont
  {Holt}}, \bibinfo {author} {\bibfnamefont {P.}~\bibnamefont {Zschack}},
  \bibinfo {author} {\bibfnamefont {H.}~\bibnamefont {Hong}}, \bibinfo {author}
  {\bibfnamefont {M.~Y.}\ \bibnamefont {Chou}},\ and\ \bibinfo {author}
  {\bibfnamefont {T.-C.}\ \bibnamefont {Chiang}},\ }\bibfield  {title}
  {\bibinfo {title} {{X-ray studies of phonon softening in TiSe$_2$}},\ }\href
  {https://doi.org/10.1103/PhysRevLett.86.3799} {\bibfield  {journal} {\bibinfo
   {journal} {Phys. Rev. Lett.}\ }\textbf {\bibinfo {volume} {86}},\ \bibinfo
  {pages} {3799} (\bibinfo {year} {2001})}\BibitemShut {NoStop}%
\bibitem [{\citenamefont {Glaeser}(1985)}]{Glaeser1985}%
  \BibitemOpen
  \bibfield  {author} {\bibinfo {author} {\bibfnamefont {R.~M.}\ \bibnamefont
  {Glaeser}},\ }\bibfield  {title} {\bibinfo {title} {{Electron crystallography
  of biological macromolecules}},\ }\href
  {https://doi.org/10.1146/annurev.pc.36.100185.001331} {\bibfield  {journal}
  {\bibinfo  {journal} {Annu. Rev. Phys. Chem.}\ }\textbf {\bibinfo {volume}
  {36}},\ \bibinfo {pages} {243} (\bibinfo {year} {1985})}\BibitemShut
  {NoStop}%
\bibitem [{\citenamefont {Rettig}\ \emph {et~al.}(2016)\citenamefont {Rettig},
  \citenamefont {Cort{\'{e}}s}, \citenamefont {Chu}, \citenamefont {Fisher},
  \citenamefont {Schmitt}, \citenamefont {Moore}, \citenamefont {Shen},
  \citenamefont {Kirchmann}, \citenamefont {Wolf},\ and\ \citenamefont
  {Bovensiepen}}]{Rettig2016}%
  \BibitemOpen
  \bibfield  {author} {\bibinfo {author} {\bibfnamefont {L.}~\bibnamefont
  {Rettig}}, \bibinfo {author} {\bibfnamefont {R.}~\bibnamefont
  {Cort{\'{e}}s}}, \bibinfo {author} {\bibfnamefont {J.-H.}\ \bibnamefont
  {Chu}}, \bibinfo {author} {\bibfnamefont {I.~R.}\ \bibnamefont {Fisher}},
  \bibinfo {author} {\bibfnamefont {F.}~\bibnamefont {Schmitt}}, \bibinfo
  {author} {\bibfnamefont {R.~G.}\ \bibnamefont {Moore}}, \bibinfo {author}
  {\bibfnamefont {Z.-X.}\ \bibnamefont {Shen}}, \bibinfo {author}
  {\bibfnamefont {P.~S.}\ \bibnamefont {Kirchmann}}, \bibinfo {author}
  {\bibfnamefont {M.}~\bibnamefont {Wolf}},\ and\ \bibinfo {author}
  {\bibfnamefont {U.}~\bibnamefont {Bovensiepen}},\ }\bibfield  {title}
  {\bibinfo {title} {{Persistent order due to transiently enhanced nesting in
  an electronically excited charge density wave}},\ }\href
  {https://doi.org/10.1038/ncomms10459} {\bibfield  {journal} {\bibinfo
  {journal} {Nat. Commun.}\ }\textbf {\bibinfo {volume} {7}},\ \bibinfo {pages}
  {10459} (\bibinfo {year} {2016})}\BibitemShut {NoStop}%
\bibitem [{\citenamefont {Yokoya}\ \emph {et~al.}(2005)\citenamefont {Yokoya},
  \citenamefont {Kiss}, \citenamefont {Chainani}, \citenamefont {Shin},\ and\
  \citenamefont {Yamaya}}]{Yokoya2005}%
  \BibitemOpen
  \bibfield  {author} {\bibinfo {author} {\bibfnamefont {T.}~\bibnamefont
  {Yokoya}}, \bibinfo {author} {\bibfnamefont {T.}~\bibnamefont {Kiss}},
  \bibinfo {author} {\bibfnamefont {A.}~\bibnamefont {Chainani}}, \bibinfo
  {author} {\bibfnamefont {S.}~\bibnamefont {Shin}},\ and\ \bibinfo {author}
  {\bibfnamefont {K.}~\bibnamefont {Yamaya}},\ }\bibfield  {title} {\bibinfo
  {title} {{Role of charge-density-wave fluctuations on the spectral function
  in a metallic charge-density-wave system}},\ }\href
  {https://doi.org/10.1103/PhysRevB.71.140504} {\bibfield  {journal} {\bibinfo
  {journal} {Phys. Rev. B}\ }\textbf {\bibinfo {volume} {71}},\ \bibinfo
  {pages} {140504(R)} (\bibinfo {year} {2005})}\BibitemShut {NoStop}%
\bibitem [{\citenamefont {Chatterjee}\ \emph {et~al.}(2015)\citenamefont
  {Chatterjee}, \citenamefont {Zhao}, \citenamefont {Iavarone}, \citenamefont
  {{Di Capua}}, \citenamefont {Castellan}, \citenamefont {Karapetrov},
  \citenamefont {Malliakas}, \citenamefont {Kanatzidis}, \citenamefont {Claus},
  \citenamefont {Ruff}, \citenamefont {Weber}, \citenamefont {{van Wezel}},
  \citenamefont {Campuzano}, \citenamefont {Osborn}, \citenamefont {Randeria},
  \citenamefont {Trivedi}, \citenamefont {Norman},\ and\ \citenamefont
  {Rosenkranz}}]{Chatterjee2015}%
  \BibitemOpen
  \bibfield  {author} {\bibinfo {author} {\bibfnamefont {U.}~\bibnamefont
  {Chatterjee}}, \bibinfo {author} {\bibfnamefont {J.}~\bibnamefont {Zhao}},
  \bibinfo {author} {\bibfnamefont {M.}~\bibnamefont {Iavarone}}, \bibinfo
  {author} {\bibfnamefont {R.}~\bibnamefont {{Di Capua}}}, \bibinfo {author}
  {\bibfnamefont {J.~P.}\ \bibnamefont {Castellan}}, \bibinfo {author}
  {\bibfnamefont {G.}~\bibnamefont {Karapetrov}}, \bibinfo {author}
  {\bibfnamefont {C.~D.}\ \bibnamefont {Malliakas}}, \bibinfo {author}
  {\bibfnamefont {M.~G.}\ \bibnamefont {Kanatzidis}}, \bibinfo {author}
  {\bibfnamefont {H.}~\bibnamefont {Claus}}, \bibinfo {author} {\bibfnamefont
  {J.~P.~C.}\ \bibnamefont {Ruff}}, \bibinfo {author} {\bibfnamefont
  {F.}~\bibnamefont {Weber}}, \bibinfo {author} {\bibfnamefont
  {J.}~\bibnamefont {{van Wezel}}}, \bibinfo {author} {\bibfnamefont {J.~C.}\
  \bibnamefont {Campuzano}}, \bibinfo {author} {\bibfnamefont {R.}~\bibnamefont
  {Osborn}}, \bibinfo {author} {\bibfnamefont {M.}~\bibnamefont {Randeria}},
  \bibinfo {author} {\bibfnamefont {N.}~\bibnamefont {Trivedi}}, \bibinfo
  {author} {\bibfnamefont {M.~R.}\ \bibnamefont {Norman}},\ and\ \bibinfo
  {author} {\bibfnamefont {S.}~\bibnamefont {Rosenkranz}},\ }\bibfield  {title}
  {\bibinfo {title} {{Emergence of coherence in the charge-density wave state
  of 2$H$-NbSe$_2$}},\ }\href {https://doi.org/10.1038/ncomms7313} {\bibfield
  {journal} {\bibinfo  {journal} {Nat. Commun.}\ }\textbf {\bibinfo {volume}
  {6}},\ \bibinfo {pages} {6313} (\bibinfo {year} {2015})}\BibitemShut
  {NoStop}%
\bibitem [{\citenamefont {Xu}\ \emph {et~al.}(2000)\citenamefont {Xu},
  \citenamefont {Ong}, \citenamefont {Wang}, \citenamefont {Kakeshita},\ and\
  \citenamefont {Uchida}}]{Xu2000}%
  \BibitemOpen
  \bibfield  {author} {\bibinfo {author} {\bibfnamefont {Z.~A.}\ \bibnamefont
  {Xu}}, \bibinfo {author} {\bibfnamefont {N.~P.}\ \bibnamefont {Ong}},
  \bibinfo {author} {\bibfnamefont {Y.}~\bibnamefont {Wang}}, \bibinfo {author}
  {\bibfnamefont {T.}~\bibnamefont {Kakeshita}},\ and\ \bibinfo {author}
  {\bibfnamefont {S.}~\bibnamefont {Uchida}},\ }\bibfield  {title} {\bibinfo
  {title} {{Vortex-like excitations and the onset of superconducting phase
  fluctuation in underdoped La$_{2-x}$Sr$_x$CuO$_4$}},\ }\href
  {https://doi.org/10.1038/35020016} {\bibfield  {journal} {\bibinfo  {journal}
  {Nature}\ }\textbf {\bibinfo {volume} {406}},\ \bibinfo {pages} {486}
  (\bibinfo {year} {2000})}\BibitemShut {NoStop}%
\bibitem [{\citenamefont {Pourret}\ \emph {et~al.}(2006)\citenamefont
  {Pourret}, \citenamefont {Aubin}, \citenamefont {Lesueur}, \citenamefont
  {Marrache-Kikuchi}, \citenamefont {Berg{\'{e}}}, \citenamefont {Dumoulin},\
  and\ \citenamefont {Behnia}}]{Pourret2006}%
  \BibitemOpen
  \bibfield  {author} {\bibinfo {author} {\bibfnamefont {A.}~\bibnamefont
  {Pourret}}, \bibinfo {author} {\bibfnamefont {H.}~\bibnamefont {Aubin}},
  \bibinfo {author} {\bibfnamefont {J.}~\bibnamefont {Lesueur}}, \bibinfo
  {author} {\bibfnamefont {C.~A.}\ \bibnamefont {Marrache-Kikuchi}}, \bibinfo
  {author} {\bibfnamefont {L.}~\bibnamefont {Berg{\'{e}}}}, \bibinfo {author}
  {\bibfnamefont {L.}~\bibnamefont {Dumoulin}},\ and\ \bibinfo {author}
  {\bibfnamefont {K.}~\bibnamefont {Behnia}},\ }\bibfield  {title} {\bibinfo
  {title} {{Observation of the Nernst signal generated by fluctuating Cooper
  pairs}},\ }\href {https://doi.org/10.1038/nphys413} {\bibfield  {journal}
  {\bibinfo  {journal} {Nat. Phys.}\ }\textbf {\bibinfo {volume} {2}},\
  \bibinfo {pages} {683} (\bibinfo {year} {2006})}\BibitemShut {NoStop}%
\bibitem [{\citenamefont {Nam}\ \emph {et~al.}(2013)\citenamefont {Nam},
  \citenamefont {M{\'{e}}zi{\`{e}}re}, \citenamefont {Batail}, \citenamefont
  {Zorina}, \citenamefont {Simonov},\ and\ \citenamefont {Ardavan}}]{Nam2013}%
  \BibitemOpen
  \bibfield  {author} {\bibinfo {author} {\bibfnamefont {M.-S.}\ \bibnamefont
  {Nam}}, \bibinfo {author} {\bibfnamefont {C.}~\bibnamefont
  {M{\'{e}}zi{\`{e}}re}}, \bibinfo {author} {\bibfnamefont {P.}~\bibnamefont
  {Batail}}, \bibinfo {author} {\bibfnamefont {L.}~\bibnamefont {Zorina}},
  \bibinfo {author} {\bibfnamefont {S.}~\bibnamefont {Simonov}},\ and\ \bibinfo
  {author} {\bibfnamefont {A.}~\bibnamefont {Ardavan}},\ }\bibfield  {title}
  {\bibinfo {title} {{Superconducting fluctuations in organic molecular metals
  enhanced by Mott criticality}},\ }\href {https://doi.org/10.1038/srep03390}
  {\bibfield  {journal} {\bibinfo  {journal} {Sci. Rep.}\ }\textbf {\bibinfo
  {volume} {3}},\ \bibinfo {pages} {3390} (\bibinfo {year} {2013})}\BibitemShut
  {NoStop}%
\bibitem [{\citenamefont {Buzzi}\ \emph {et~al.}(2021)\citenamefont {Buzzi},
  \citenamefont {Nicoletti}, \citenamefont {Fava}, \citenamefont {Jotzu},
  \citenamefont {Miyagawa}, \citenamefont {Kanoda}, \citenamefont {Henderson},
  \citenamefont {Siegrist}, \citenamefont {Schlueter}, \citenamefont {Nam},
  \citenamefont {Ardavan},\ and\ \citenamefont {Cavalleri}}]{Buzzi2021}%
  \BibitemOpen
  \bibfield  {author} {\bibinfo {author} {\bibfnamefont {M.}~\bibnamefont
  {Buzzi}}, \bibinfo {author} {\bibfnamefont {D.}~\bibnamefont {Nicoletti}},
  \bibinfo {author} {\bibfnamefont {S.}~\bibnamefont {Fava}}, \bibinfo {author}
  {\bibfnamefont {G.}~\bibnamefont {Jotzu}}, \bibinfo {author} {\bibfnamefont
  {K.}~\bibnamefont {Miyagawa}}, \bibinfo {author} {\bibfnamefont
  {K.}~\bibnamefont {Kanoda}}, \bibinfo {author} {\bibfnamefont
  {A.}~\bibnamefont {Henderson}}, \bibinfo {author} {\bibfnamefont
  {T.}~\bibnamefont {Siegrist}}, \bibinfo {author} {\bibfnamefont {J.~A.}\
  \bibnamefont {Schlueter}}, \bibinfo {author} {\bibfnamefont {M.-S.}\
  \bibnamefont {Nam}}, \bibinfo {author} {\bibfnamefont {A.}~\bibnamefont
  {Ardavan}},\ and\ \bibinfo {author} {\bibfnamefont {A.}~\bibnamefont
  {Cavalleri}},\ }\href@noop {} {\bibinfo {title} {{A phase diagram for
  light-induced superconductivity in $\kappa$-(ET)$_2$-X}}} (\bibinfo {year}
  {2021}),\ \Eprint {https://arxiv.org/abs/2106.14244} {arXiv:2106.14244}
  \BibitemShut {NoStop}%
\bibitem [{\citenamefont {He}\ \emph {et~al.}(2021)\citenamefont {He},
  \citenamefont {Chen}, \citenamefont {Li}, \citenamefont {Zhao}, \citenamefont
  {Song}, \citenamefont {Yoshida}, \citenamefont {Eisaki}, \citenamefont {Wu},
  \citenamefont {Chen}, \citenamefont {Lu}, \citenamefont {Meingast},
  \citenamefont {Devereaux}, \citenamefont {Birgeneau}, \citenamefont
  {Hashimoto}, \citenamefont {Lee},\ and\ \citenamefont {Shen}}]{He2021}%
  \BibitemOpen
  \bibfield  {author} {\bibinfo {author} {\bibfnamefont {Y.}~\bibnamefont
  {He}}, \bibinfo {author} {\bibfnamefont {S.-D.}\ \bibnamefont {Chen}},
  \bibinfo {author} {\bibfnamefont {Z.-X.}\ \bibnamefont {Li}}, \bibinfo
  {author} {\bibfnamefont {D.}~\bibnamefont {Zhao}}, \bibinfo {author}
  {\bibfnamefont {D.}~\bibnamefont {Song}}, \bibinfo {author} {\bibfnamefont
  {Y.}~\bibnamefont {Yoshida}}, \bibinfo {author} {\bibfnamefont
  {H.}~\bibnamefont {Eisaki}}, \bibinfo {author} {\bibfnamefont
  {T.}~\bibnamefont {Wu}}, \bibinfo {author} {\bibfnamefont {X.-H.}\
  \bibnamefont {Chen}}, \bibinfo {author} {\bibfnamefont {D.-H.}\ \bibnamefont
  {Lu}}, \bibinfo {author} {\bibfnamefont {C.}~\bibnamefont {Meingast}},
  \bibinfo {author} {\bibfnamefont {T.~P.}\ \bibnamefont {Devereaux}}, \bibinfo
  {author} {\bibfnamefont {R.~J.}\ \bibnamefont {Birgeneau}}, \bibinfo {author}
  {\bibfnamefont {M.}~\bibnamefont {Hashimoto}}, \bibinfo {author}
  {\bibfnamefont {D.-H.}\ \bibnamefont {Lee}},\ and\ \bibinfo {author}
  {\bibfnamefont {Z.-X.}\ \bibnamefont {Shen}},\ }\bibfield  {title} {\bibinfo
  {title} {{Superconducting fluctuations in overdoped
  Bi$_2$Sr$_2$CaCu$_2$O$_{8+\delta}$}},\ }\href
  {https://doi.org/10.1103/PhysRevX.11.031068} {\bibfield  {journal} {\bibinfo
  {journal} {Phys. Rev. X}\ }\textbf {\bibinfo {volume} {11}},\ \bibinfo
  {pages} {031068} (\bibinfo {year} {2021})}\BibitemShut {NoStop}%
\bibitem [{\citenamefont {Tallon}\ \emph {et~al.}(2011)\citenamefont {Tallon},
  \citenamefont {Storey},\ and\ \citenamefont {Loram}}]{Tallon2011}%
  \BibitemOpen
  \bibfield  {author} {\bibinfo {author} {\bibfnamefont {J.~L.}\ \bibnamefont
  {Tallon}}, \bibinfo {author} {\bibfnamefont {J.~G.}\ \bibnamefont {Storey}},\
  and\ \bibinfo {author} {\bibfnamefont {J.~W.}\ \bibnamefont {Loram}},\
  }\bibfield  {title} {\bibinfo {title} {{Fluctuations and critical temperature
  reduction in cuprate superconductors}},\ }\href
  {https://doi.org/10.1103/PhysRevB.83.092502} {\bibfield  {journal} {\bibinfo
  {journal} {Phys. Rev. B}\ }\textbf {\bibinfo {volume} {83}},\ \bibinfo
  {pages} {092502} (\bibinfo {year} {2011})}\BibitemShut {NoStop}%
\bibitem [{\citenamefont {Weber}\ \emph {et~al.}(2011)\citenamefont {Weber},
  \citenamefont {Rosenkranz}, \citenamefont {Castellan}, \citenamefont
  {Osborn}, \citenamefont {Hott}, \citenamefont {Heid}, \citenamefont {Bohnen},
  \citenamefont {Egami}, \citenamefont {Said},\ and\ \citenamefont
  {Reznik}}]{Weber2011}%
  \BibitemOpen
  \bibfield  {author} {\bibinfo {author} {\bibfnamefont {F.}~\bibnamefont
  {Weber}}, \bibinfo {author} {\bibfnamefont {S.}~\bibnamefont {Rosenkranz}},
  \bibinfo {author} {\bibfnamefont {J.-P.}\ \bibnamefont {Castellan}}, \bibinfo
  {author} {\bibfnamefont {R.}~\bibnamefont {Osborn}}, \bibinfo {author}
  {\bibfnamefont {R.}~\bibnamefont {Hott}}, \bibinfo {author} {\bibfnamefont
  {R.}~\bibnamefont {Heid}}, \bibinfo {author} {\bibfnamefont {K.-P.}\
  \bibnamefont {Bohnen}}, \bibinfo {author} {\bibfnamefont {T.}~\bibnamefont
  {Egami}}, \bibinfo {author} {\bibfnamefont {A.~H.}\ \bibnamefont {Said}},\
  and\ \bibinfo {author} {\bibfnamefont {D.}~\bibnamefont {Reznik}},\
  }\bibfield  {title} {\bibinfo {title} {{Extended phonon collapse and the
  origin of the charge-density wave in 2$H$-NbSe$_2$}},\ }\href
  {https://doi.org/10.1103/PhysRevLett.107.107403} {\bibfield  {journal}
  {\bibinfo  {journal} {Phys. Rev. Lett.}\ }\textbf {\bibinfo {volume} {107}},\
  \bibinfo {pages} {107403} (\bibinfo {year} {2011})}\BibitemShut {NoStop}%
\bibitem [{\citenamefont {Hoesch}\ \emph {et~al.}(2009)\citenamefont {Hoesch},
  \citenamefont {Bosak}, \citenamefont {Chernyshov}, \citenamefont {Berger},\
  and\ \citenamefont {Krisch}}]{Hoesch2009}%
  \BibitemOpen
  \bibfield  {author} {\bibinfo {author} {\bibfnamefont {M.}~\bibnamefont
  {Hoesch}}, \bibinfo {author} {\bibfnamefont {A.}~\bibnamefont {Bosak}},
  \bibinfo {author} {\bibfnamefont {D.}~\bibnamefont {Chernyshov}}, \bibinfo
  {author} {\bibfnamefont {H.}~\bibnamefont {Berger}},\ and\ \bibinfo {author}
  {\bibfnamefont {M.}~\bibnamefont {Krisch}},\ }\bibfield  {title} {\bibinfo
  {title} {{Giant Kohn anomaly and the phase transition in charge density wave
  ZrTe$_3$}},\ }\href {https://doi.org/10.1103/PhysRevLett.102.086402}
  {\bibfield  {journal} {\bibinfo  {journal} {Phys. Rev. Lett.}\ }\textbf
  {\bibinfo {volume} {102}},\ \bibinfo {pages} {086402} (\bibinfo {year}
  {2009})}\BibitemShut {NoStop}%
\bibitem [{\citenamefont {Gr{\"{u}}ner}(1994)}]{Gruner1994}%
  \BibitemOpen
  \bibfield  {author} {\bibinfo {author} {\bibfnamefont {G.}~\bibnamefont
  {Gr{\"{u}}ner}},\ }\href {https://doi.org/10.1201/9780429501012} {\emph
  {\bibinfo {title} {{Density Waves in Solids}}}}\ (\bibinfo  {publisher}
  {Addison-Wesley},\ \bibinfo {address} {Boston},\ \bibinfo {year}
  {1994})\BibitemShut {NoStop}%
\bibitem [{\citenamefont {Ma}\ and\ \citenamefont {Yu}(2013)}]{Ma2013}%
  \BibitemOpen
  \bibfield  {author} {\bibinfo {author} {\bibfnamefont {L.}~\bibnamefont
  {Ma}}\ and\ \bibinfo {author} {\bibfnamefont {W.-Q.}\ \bibnamefont {Yu}},\
  }\bibfield  {title} {\bibinfo {title} {{Review of nuclear magnetic resonance
  studies on iron-based superconductors}},\ }\href
  {https://doi.org/10.1088/1674-1056/22/8/087414} {\bibfield  {journal}
  {\bibinfo  {journal} {Chinese Phys. B}\ }\textbf {\bibinfo {volume} {22}},\
  \bibinfo {pages} {087414} (\bibinfo {year} {2013})}\BibitemShut {NoStop}%
\bibitem [{\citenamefont {Zhang}\ \emph {et~al.}(2016)\citenamefont {Zhang},
  \citenamefont {Tan}, \citenamefont {Liu}, \citenamefont {Teitelbaum},
  \citenamefont {Post}, \citenamefont {Jin}, \citenamefont {Nelson},
  \citenamefont {Basov}, \citenamefont {Wu},\ and\ \citenamefont
  {Averitt}}]{Zhang2016}%
  \BibitemOpen
  \bibfield  {author} {\bibinfo {author} {\bibfnamefont {J.}~\bibnamefont
  {Zhang}}, \bibinfo {author} {\bibfnamefont {X.}~\bibnamefont {Tan}}, \bibinfo
  {author} {\bibfnamefont {M.}~\bibnamefont {Liu}}, \bibinfo {author}
  {\bibfnamefont {S.~W.}\ \bibnamefont {Teitelbaum}}, \bibinfo {author}
  {\bibfnamefont {K.~W.}\ \bibnamefont {Post}}, \bibinfo {author}
  {\bibfnamefont {F.}~\bibnamefont {Jin}}, \bibinfo {author} {\bibfnamefont
  {K.~A.}\ \bibnamefont {Nelson}}, \bibinfo {author} {\bibfnamefont {D.~N.}\
  \bibnamefont {Basov}}, \bibinfo {author} {\bibfnamefont {W.}~\bibnamefont
  {Wu}},\ and\ \bibinfo {author} {\bibfnamefont {R.~D.}\ \bibnamefont
  {Averitt}},\ }\bibfield  {title} {\bibinfo {title} {{Cooperative photoinduced
  metastable phase control in strained manganite films}},\ }\href
  {https://doi.org/10.1038/nmat4695} {\bibfield  {journal} {\bibinfo  {journal}
  {Nat. Mater.}\ }\textbf {\bibinfo {volume} {15}},\ \bibinfo {pages} {956}
  (\bibinfo {year} {2016})}\BibitemShut {NoStop}%
\bibitem [{\citenamefont {McLeod}\ \emph {et~al.}(2020)\citenamefont {McLeod},
  \citenamefont {Zhang}, \citenamefont {Gu}, \citenamefont {Jin}, \citenamefont
  {Zhang}, \citenamefont {Post}, \citenamefont {Zhao}, \citenamefont {Millis},
  \citenamefont {Wu}, \citenamefont {Rondinelli}, \citenamefont {Averitt},\
  and\ \citenamefont {Basov}}]{McLeod2020}%
  \BibitemOpen
  \bibfield  {author} {\bibinfo {author} {\bibfnamefont {A.~S.}\ \bibnamefont
  {McLeod}}, \bibinfo {author} {\bibfnamefont {J.}~\bibnamefont {Zhang}},
  \bibinfo {author} {\bibfnamefont {M.~Q.}\ \bibnamefont {Gu}}, \bibinfo
  {author} {\bibfnamefont {F.}~\bibnamefont {Jin}}, \bibinfo {author}
  {\bibfnamefont {G.}~\bibnamefont {Zhang}}, \bibinfo {author} {\bibfnamefont
  {K.~W.}\ \bibnamefont {Post}}, \bibinfo {author} {\bibfnamefont {X.~G.}\
  \bibnamefont {Zhao}}, \bibinfo {author} {\bibfnamefont {A.~J.}\ \bibnamefont
  {Millis}}, \bibinfo {author} {\bibfnamefont {W.~B.}\ \bibnamefont {Wu}},
  \bibinfo {author} {\bibfnamefont {J.~M.}\ \bibnamefont {Rondinelli}},
  \bibinfo {author} {\bibfnamefont {R.~D.}\ \bibnamefont {Averitt}},\ and\
  \bibinfo {author} {\bibfnamefont {D.~N.}\ \bibnamefont {Basov}},\ }\bibfield
  {title} {\bibinfo {title} {{Multi-messenger nanoprobes of hidden magnetism in
  a strained manganite}},\ }\href {https://doi.org/10.1038/s41563-019-0533-y}
  {\bibfield  {journal} {\bibinfo  {journal} {Nat. Mater.}\ }\textbf {\bibinfo
  {volume} {19}},\ \bibinfo {pages} {397} (\bibinfo {year} {2020})}\BibitemShut
  {NoStop}%
\bibitem [{\citenamefont {Zhang}(1997)}]{Zhang1997}%
  \BibitemOpen
  \bibfield  {author} {\bibinfo {author} {\bibfnamefont {S.-C.}\ \bibnamefont
  {Zhang}},\ }\bibfield  {title} {\bibinfo {title} {{A unified theory based on
  $SO(5)$ symmetry of superconductivity and antiferromagnetism}},\ }\href
  {https://doi.org/10.1126/science.275.5303.1089} {\bibfield  {journal}
  {\bibinfo  {journal} {Science}\ }\textbf {\bibinfo {volume} {275}},\ \bibinfo
  {pages} {1089} (\bibinfo {year} {1997})}\BibitemShut {NoStop}%
\bibitem [{\citenamefont {Demler}\ \emph {et~al.}(2004)\citenamefont {Demler},
  \citenamefont {Hanke},\ and\ \citenamefont {Zhang}}]{Demler2004}%
  \BibitemOpen
  \bibfield  {author} {\bibinfo {author} {\bibfnamefont {E.}~\bibnamefont
  {Demler}}, \bibinfo {author} {\bibfnamefont {W.}~\bibnamefont {Hanke}},\ and\
  \bibinfo {author} {\bibfnamefont {S.-C.}\ \bibnamefont {Zhang}},\ }\bibfield
  {title} {\bibinfo {title} {{$SO(5)$ theory of antiferromagnetism and
  superconductivity}},\ }\href {https://doi.org/10.1103/RevModPhys.76.909}
  {\bibfield  {journal} {\bibinfo  {journal} {Rev. Mod. Phys.}\ }\textbf
  {\bibinfo {volume} {76}},\ \bibinfo {pages} {909} (\bibinfo {year}
  {2004})}\BibitemShut {NoStop}%
\end{thebibliography}
\end{document}